\documentclass[12pt,a4paper]{article}

\usepackage{jcappub}

\usepackage[english]{babel}
\usepackage{amsmath,amsfonts,amssymb}
\usepackage{comment}
\usepackage{bbold}
\usepackage{color}
\usepackage{slashed}

\usepackage[top=30mm,bottom=30mm,left=20mm,right=20mm]{geometry}

\usepackage{hyperref}
\usepackage{appendix}

\usepackage{booktabs}
\usepackage{graphicx}
\usepackage[font=normalsize]{caption}
\usepackage{ulem}
\usepackage{enumitem}

\usepackage{multirow}
\usepackage{arydshln}
\usepackage{afterpage}

\usepackage{tabu}
\usepackage{float}

\usepackage{tikz-feynman}
\tikzfeynmanset{compat=1.0.0}
\usetikzlibrary{shapes.misc}
\tikzset{cross/.style={cross out, draw=black, minimum size=2*(#1-\pgflinewidth), inner sep=0pt, outer sep=0pt},cross/.default={3pt}}

\newcommand{\GeV}[0]{\,\mathrm{GeV}}

\DeclareMathOperator{\diag}{diag}

\def\BOX{\hbox to 3cm}

\def\MGUT{M_{\text{GUT}}}

\def\MZ{M_{Z}}

\def\CKM{^{\text{CKM}}}
\def\PMNS{^{\text{PMNS}}}

\begin{document}

\begin{titlepage}
\vspace*{0.7cm}

\begin{center}

{\LARGE {\bf Nucleon decay in a minimal non-SUSY GUT  \\[2pt] with predicted quark-lepton Yukawa ratios}}\\[8mm]

	Stefan Antusch$^{\star}$\footnote{Email: \texttt{stefan.antusch@unibas.ch}}	
	and Kevin Hinze$^\star$\footnote{Email: \texttt{kevin.hinze@unibas.ch}}

\end{center}

\vspace*{0.20cm}

\centerline{$^{\star}$ \it
Department of Physics, University of Basel,}
\centerline{\it
Klingelbergstr.\ 82, CH-4056 Basel, Switzerland}

\vspace*{1.2cm}

\begin{abstract}
\noindent 
We investigate the predictions for various nucleon decay rates and their ratios in non-supersymmetric SU(5) Grand Unified Theories (GUTs) where the masses of the third and second family down-type quarks and charged leptons each stem dominantly from single GUT operators. 
Extending the Georgi-Glashow SU(5) model by a 45-dimensional GUT-Higgs representation, the gauge couplings can meet  at the high scale $M_\mathrm{GUT}$ and the GUT scale predictions $y_\tau/y_b = 3/2$ and $y_\mu/y_s = 9/2$ can emerge within \textit{single operator dominance}. 
Explaining the observed neutrino masses via the type I seesaw mechanism by adding SU(5) singlet fermion representations and taking their renormalization group effects into account, we show that these predictions can lead to viable low scale second and third family down-type quark and charged lepton masses. 
To investigate nucleon decay predictions, we extend the minimal scenario to two ``toy models'' towards explaining quark and lepton masses and mixings with different flavor structure, and perform Markov Chain Monte Carlo (MCMC) analyses confronting the models with the available experimental data. We show that if several nucleon decay channels are observed, the ratios between their partial decay rates can serve as ``fingerprints'', allowing to separate between GUT models with different flavor structure.   
\end{abstract}

\end{titlepage}

\newpage


\section{Introduction}

Grand Unified Theories (GUTs) \cite{Georgi:1974sy,Fritzsch:1974nn,Georgi:1974my} provide an interesting framework to study physics beyond the Standard Model (SM). The (partial) unification of matter into larger GUT representations, together with the unification of the gauge couplings, helps to address the flavor puzzle, i.e.\ the question about the origin of the observed fermion masses, mixing angles and CP violating phases, while explaining the quantization of the hypercharge in the SM.

One of the major predictions of the Georgi-Glashow model \cite{Georgi:1974sy} of SU(5) unification is the GUT scale relation 
\begin{equation}\label{eq: Ye=Yd}
\textbf{Y}_e=\textbf{Y}_d^T
\end{equation}
between the charged lepton and down-type Yukawa matrices. Denoting the diagonal elements of the Yukawa matrices after transformation to the mass basis as $y_i$, with $i$ indicating the fermion flavor, this relation implies a unification of the tau and bottom Yukawa couplings, as well as of the muon and strange quark Yukawa couplings, i.e.\ 
\begin{equation}
y_\tau/y_b=1 \quad \mbox{and} \quad  y_\mu/y_s=1 \:. 
\end{equation}
However, these Yukawa unification relations are challenged by the (low energy) experimental data for the quark and lepton masses, at least when only the known SM particles are considered in the renormalization group (RG) evolution to high energies \cite{Arason:1991ic}. 

There are two typical approaches to overcome this shortcoming. In the first approach non-renormalizable operators are added, while the same minimal field content is used (see e.g.\ \cite{Ellis:1979fg, Dorsner:2006hw, Bajc:2006ia, Bajc:2007zf}). For the second approach new fields are added to the minimal field content, while the renormalizability of the model is preserved. These added fields either allow the introduction of additional new renormalizable operators (see e.g. \cite{Dorsner:2014wva, Tsuyuki:2014xja}) such that Eq.~\eqref{eq: Ye=Yd} no-longer holds, or change the renormalization group evolution of $\textbf{Y}_e$ and $\textbf{Y}_d$, while Eq.~\eqref{eq: Ye=Yd} is maintained at the GUT scale. The latter is very challenging, especially for the second family. In the former cases extra degrees of freedom are added and thus predictivity for the GUT scale Yukawa ratios is lost.

In this paper we consider a different approach, employing the concept of \textit{single operator dominance} (cf.\ \cite{Antusch:2009gu, Antusch:2013rxa}) where each Yukawa entry is dominated by a different single GUT operator.\footnote{The concept of \textit{single operator dominance} has so far mainly been studied in the context of SUSY GUTs, cf.\ e.g.~\cite{Antusch:2013kna, Antusch:2019avd, Antusch:2018gnu, Antusch:2017ano, Antusch:2012fb}, but is rarely considered in the context of non-SUSY GUTs.}
 These GUT operators may be renormalizable or non-renormalizable, and allow to consider different GUT scale  ratios between different entries of the lepton and down-type quark Yukawa matrices. This way a high predictivity with respect to fermion masses can be maintained, but alternative GUT scale rations between charged lepton and down-type quark Yukawa couplings can be realized that are potentially more promising w.r.t.\ the experimental data. As we will argue below, the ratios 
 \begin{equation}
 y_\tau/y_b=3/2 \quad \mbox{and} \quad  y_\mu/y_s=9/2 
 \end{equation}
 are particularly interesting in this context when the light neutrino masses are explained by the type I seesaw mechanism~\cite{Minkowski:1977sc}. The GUT operators generating these ratios can be realized in a minimal extension of the Georgi-Glashow SU(5) model where the Higgs sector consists of the representations \textbf{5}, \textbf{24} and \textbf{45}. 

A typical prediction of GUTs and probably the most promising way to test grand unification is nucleon decay. Any grand unification group based on SU(5) (or which contains SU(5) as a subgroup) includes in the adjoint representation of SU(5) gauge bosons transforming as (\textbf{3}, \textbf{2})$_{-\frac{5}{6}}$ under the SM gauge group SU(3)$\,\times\,$SU(2)$\,\times\,$U(1). These gauge bosons mediate interactions which violate the baryon and lepton number ($B$ and $L$) symmetries and generate below their mass scale dimension six nucleon decay operators. From these operators the total nucleon decay rate can be roughly estimated as
\begin{equation}\label{eq: nucleon decay estimation rough}
\Gamma\approx \frac{g^4\, m_N^5}{M_X^2},
\end{equation} 
where $g$ is the unified gauge coupling, $m_N\approx 0.94\,$GeV is the nucleon mass, and $M_X$ is the mass of the gauge bosons which mediate interactions violating the $B$ and $L$ symmetries. 

There are GUT models which also contain other sources of nucleon decay: (i) scalar leptoquarks can also mediate dimension six nucleon decay operators, (ii) in supersymmetric (SUSY) theories there are additional dimension five nucleon decay operators and if no R-parity conservation is assumed even dimension four nucleon decay is allowed. But in this paper only the dimension six gauge boson mediated nucleon decay will be relevant.

The procedure for a more accurate calculation of the decay rate predictions within a specific GUT model (replacing the rough estimate in Eq.~\eqref{eq: nucleon decay estimation rough}) is well-known, cf.\ e.g.\ \cite{Nath:2006ut}, and has been implemented in the recently released Mathematica package \texttt{ProtonDecay}~\cite{Antusch:2020ztu} which (amongst its dimension five nucleon decay features) takes the dimension six nucleon decay operators as input and uses them to compute the nucleon decay rates of 13 different decay channels. With these predictions for the decay rates, a specific GUT model can be confronted with the current constraints \cite{Brock:2012ogj, Mine:2016mxy, Takenaka:2020vqy, Bajc:2016qcc, Miura:2016krn, Abe:2014mwa, 1205.6538} and the future experimental sensitivities \cite{Acciarri:2015uup, Abe:2018uyc, An:2015jdp}. 

In this work, using the \texttt{ProtonDecay} package \cite{Antusch:2020ztu},  we investigate the nucleon decay rates for these 13 decay channels in two ``toy models'' (following \cite{Antusch:2014poa, Antusch:2013kna}) which both realize the GUT scale ratios $y_\tau/y_b=3/2$ and $y_\mu/y_s=9/2$, but feature a different flavor structure in the Yukawa matrices. Both models are SU(5) GUT scenarios with a Higgs sector consisting of the representations \textbf{5}, \textbf{24} and \textbf{45},\footnote{A model with this field content but different predictions for the GUT scale Yukawa ratios has been proposed in \cite{Georgi:1979df}. A lower bound of $1\times10^{-68}$~GeV for the total proton decay rate was later given in \cite{Dorsner:2006dj, Perez:2007rm}, 
however not in the here considered scenario with predicted Yukawa ratios and specified flavor structures. 
} such that the GUT operators leading to both relations can be realized. In addition, to explain the observed neutrino masses via the type I seesaw mechanism,  extra fermionic SU(5) singlet representations are added.
We perform Markov Chain Monte Carlo (MCMC) analyses in order to confront the models with the available experimental data. As we will show, if several nucleon decay channels are observed, the ratios between their partial decay rates can serve as ``fingerprints'', allowing to separate between the two ``toy models''.

The paper is organized as follows. We introduce in Section~\ref{sec: A minimal non-SUSY} our minimal GUT scenario which predicts the GUT scale relations $y_\tau/y_b=3/2$ and $y_\mu/y_s=9/2$ and discuss their viability w.r.t.\ the experimental data. In Section~\ref{sec: nucleon decay fingerprints of two toy models} we extend our minimal scenario to two ``toy models'' with different Yukawa textures and perform an MCMC analysis to compute their predictions for the nucleon decay rates as well as the fermion masses and mixings, before discussing our results and concluding in Section~\ref{sec:conclusions}. In Appendix~\ref{sec: running of neutrino Yukawa coupling} we approximate the running of the neutrino couplings above the GUT scale and discuss the perturbativity of our model. 

\newpage
\section{A minimal non-SUSY SU(5) GUT scenario with predicted \\ Yukawa ratios}\label{sec: A minimal non-SUSY}

We consider an SU(5) GUT with the Higgs sector fields $\textbf{5}_H$, $\textbf{24}_H$, and $\textbf{45}_H$ (in the respective representations) and the fermion fields 
$\boldsymbol{\overline{5}}_{Fi}$, $\textbf{10}_{Fi}$, with $i=1,2,3$, plus additional singlets $\textbf{1}_{Fi}$ with $i=1, ... , n$ and $n\ge 2$ to explain the observed massive neutrinos via the type I seesaw mechanism~\cite{Minkowski:1977sc}.
As we will show, this rather minimal GUT scenario allows for unification of the gauge SM couplings as well as to realize the GUT scale predictions $y_\tau/y_b = 3/2$ and $y_\mu/y_s = 9/2$ within \textit{single operator dominance}. 

The SM fermions are embedded in the usual way into three generations of $\boldsymbol{\overline{5}}_{Fi}$, $\textbf{10}_{Fi}$,
\begin{equation}
\begin{split}
\boldsymbol{\overline{5}}_{Fi}&=d_i^c\oplus L_i \: ,\\
\textbf{10}_{Fi}&=Q_i\oplus u_i^c\oplus e_i^c \:,
\end{split}
\end{equation}
and the $\textbf{1}_{Fi}$ contain the right-chiral neutrinos,
\begin{equation}
\textbf{1}_{Fi}=\nu^c_i\:.
\end{equation}
The GUT Higgs fields decompose under the SM gauge group SU(3)$\,\times\,$SU(2)$_\mathrm{L}\,\times\,$U(1)$_\mathrm{Y}$ as 
\begin{equation}
\begin{split}
\textbf{5}_H\;&=\;T\oplus H=(\textbf{3},\textbf{1})_{-\frac{1}{3}}\oplus(\textbf{1},\textbf{2})_{\frac{1}{2}},\\
\textbf{24}_H\;&=\;\Sigma_8\oplus \Sigma_3\oplus\Sigma_{(3,2)}\oplus\Sigma_{\overline{3},2}\oplus\Sigma_1\\
&=\;(\textbf{8},\textbf{1})_0\oplus(\textbf{1},\textbf{3})_0\oplus(\textbf{3},\textbf{2})_{-\frac{5}{6}}\oplus(\boldsymbol{\overline{3}},\textbf{2})_{\frac{5}{6}}\oplus(\textbf{1},\textbf{1})_0,\\
\textbf{45}_H\;&=\;\Phi_8\oplus \Phi_{\overline{6}}\oplus\Phi_{(3,3)}\oplus\Phi_{(\overline{3},2)}\oplus\Phi_3\oplus\Phi_{\overline{3}}\oplus\Phi_2\\
&=\;(\textbf{8},\textbf{2})_{\frac{1}{2}}\oplus(\boldsymbol{\overline{6}},\textbf{1})_{-\frac{1}{3}}\oplus(\textbf{3},\textbf{3})_{-\frac{1}{3}}\oplus(\boldsymbol{\overline{3}},\textbf{2})_{-\frac{7}{6}}\oplus(\textbf{3},\textbf{1})_{-\frac{1}{3}}\oplus(\boldsymbol{\overline{3}},\textbf{1})_\frac{4}{3}\oplus(\textbf{1},\textbf{2})_{\frac{1}{2}}.
\end{split}
\end{equation}
The SM Higgs doublet $h$ is then a linear combination of the fields $H$ and $\Phi_2$. We will in this paper assume that the mass of the second Higgs doublet $h^\perp$, which is given by a linear combination of $H$ and $\Phi_2$ orthogonal to $h$, is at the GUT scale $\MGUT$.

\subsection{Predicted Yukawa ratios $\boldsymbol{y_\tau/y_b = 3/2}$ and $\boldsymbol{y_\mu/y_s = 9/2}$}
For the Yukawa sector we will assume the concept of \textit{single operator dominance} (cf.\ \cite{Antusch:2009gu, Antusch:2013rxa}), i.e.\ we will assume that each entry of the Yukawa matrices is governed by one singlet GUT operator giving the dominant contribution to this entry. Furthermore, we will assume that the Yukawa matrices have hierarchical structure. Focusing on the GUT scale Yukawa matrix $\boldsymbol{Y_{\overline{5}}}$ leading simultaneously to the masses for the down-type quarks and charged leptons, it follows that the second and third family masses stem dominantly from the GUT operators ${\cal O}_2$ and ${\cal O}_3$ dominating the 22 and 33 positions of the hierarchical Yukawa matrix, i.e.:
\begin{equation}
\boldsymbol{Y_{\overline{5}}}=
\begin{pmatrix}
0&& 0 && 0 \\
0 && {\cal O}_2 && 0 \\
0 && 0 && {\cal O}_3
\end{pmatrix} + \dots \:,
\end{equation}
with the dots indicating terms that are subdominant for the second and third family masses. With the particle content specified above, it is possible to construct the two operators
\begin{equation}\label{eq: operator 9/2}
\begin{split}
{\cal O}_2  &= (\textbf{10}_{F3}\boldsymbol{\overline{45}}_H)_{\textbf{5}}(\boldsymbol{\overline{5}}_{F3}\textbf{24}_H)_{\boldsymbol{\overline{5}}} \\
{\cal O}_3 &= (\textbf{10}_{F3}\boldsymbol{\overline{5}}_H)_{\textbf{45}}(\boldsymbol{\overline{5}}_{F3}\textbf{24}_H)_{\boldsymbol{\overline{45}}}
\end{split}
\end{equation}
which give rise to the GUT scale Yukawa ratios $y_\tau/y_b = 3/2$ and  $y_\mu/y_s = 9/2$ \cite{Antusch:2009gu}.\footnote{Alternatively, instead of ${\cal O}_3$, one could also use a 5-plet messenger field, which then leads to a predicted ratio of  $y_\tau/y_b = - 3/2$. For the phenomenology, the minus sign of course makes no difference.} In these operators, the representations $X$ given as index for the brackets $(.)_X$ specify the contractions of the GUT representations' indices inside the bracket. 
We thus arrive at the following leading order structure of the GUT scale Yukawa matrices $\textbf{Y}_e$ and $\textbf{Y}_d$: 
\begin{equation}
\textbf{Y}_e=
\begin{pmatrix}
0 && 0 && 0 \\
0 && \frac{9}{2} && 0 \\
0 && 0 && \frac{3}{2}
\end{pmatrix}\cdot \textbf{Y}_d^T \:,
\end{equation}
where by the dot $\cdot$ entry-wise multiplication of matrices is specified. We note that arriving at the situations that the operators ${\cal O}_2$ and ${\cal O}_3$ are indeed dominating the 22 and 33 positions of the hierarchical Yukawa matrix typically requires additional family symmetries. Examples for models realizing \textit{single operator dominance} this way can e.g.\ be found in \cite{Antusch:2012fb, Antusch:2013kna, Antusch:2017ano, Antusch:2018gnu, Antusch:2019avd}. In the following we will further assume that the applied family symmetry also makes sure that $\textbf{45}_H$ only couples to the 22 entry of $\boldsymbol{Y_{\overline{5}}}$.

\subsection{Gauge coupling unification}\label{sec: gauge coupling unification}
\begin{figure}
\includegraphics[width=7.5cm]{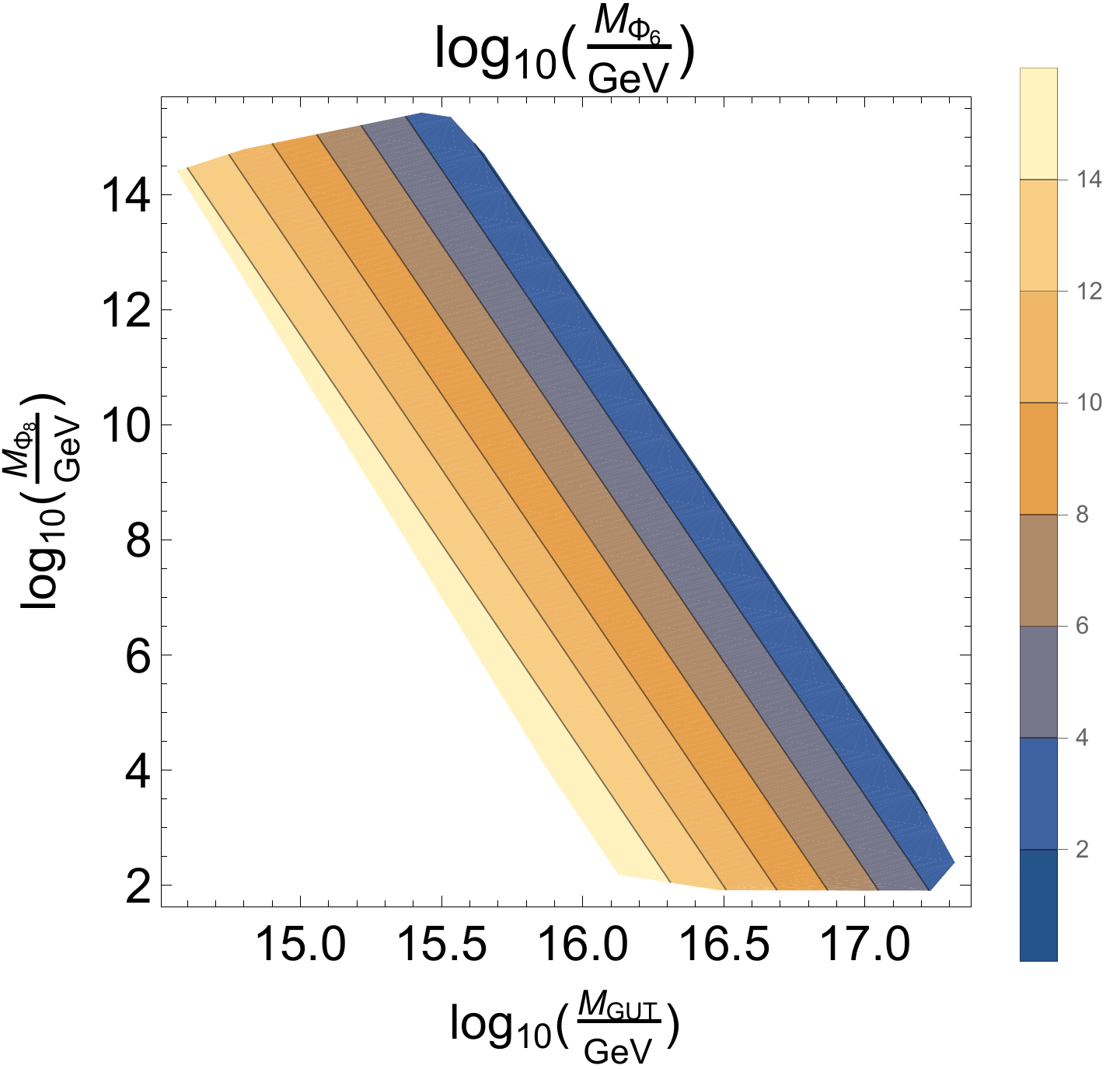}\hspace{10mm}
\includegraphics[width=7.5cm]{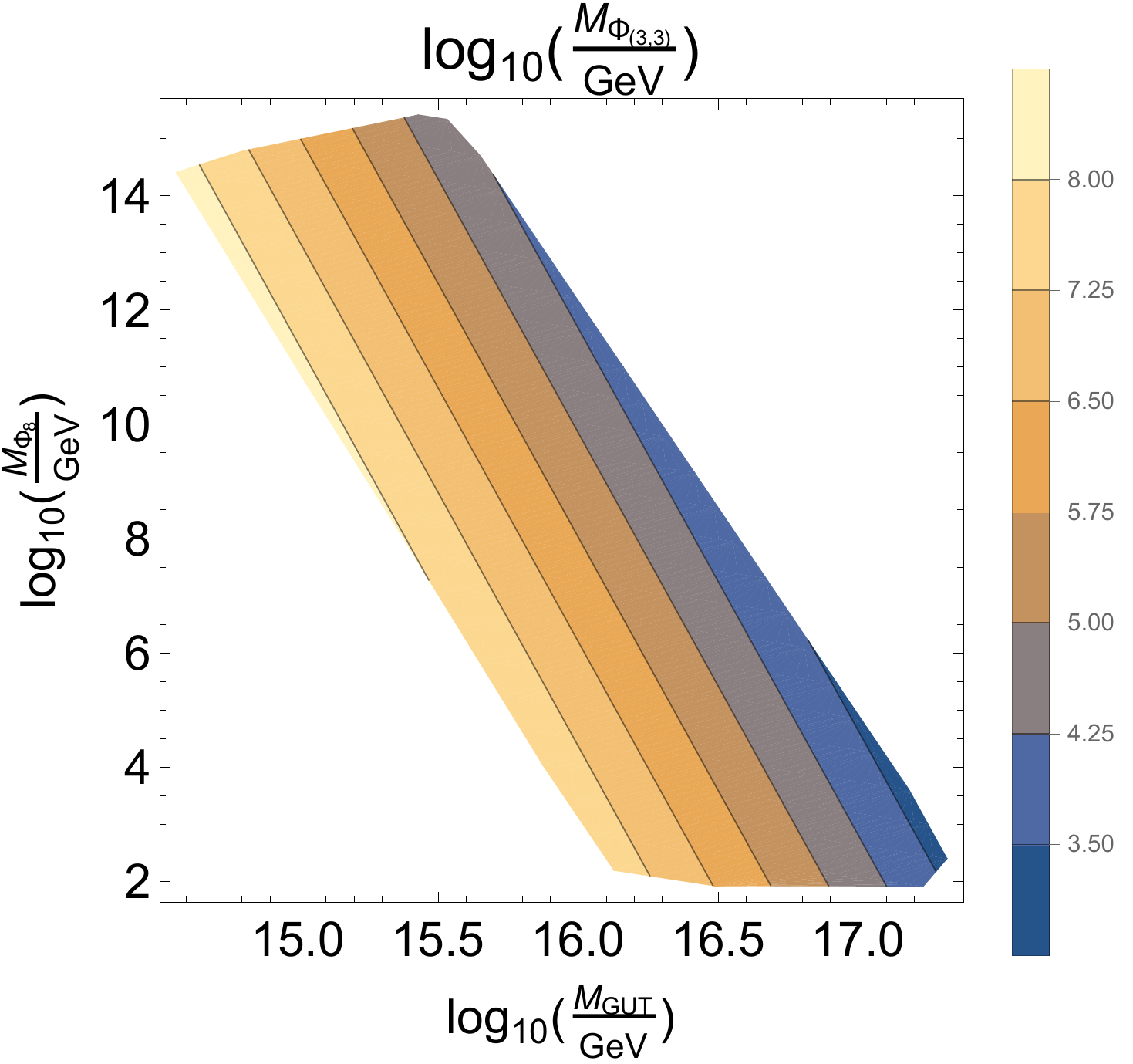}
\caption{Contour plot of the values of $M_{\Phi_6}$ and $M_{\Phi_{(3,3)}}$ in the $\MGUT$--$M_{\Phi_8}$ plane which allow for gauge coupling unification at one-loop level. 
}\label{fig: contourplot unification}
\end{figure}

To demonstrate that gauge coupling unification can be achieved in the above-specified scenario, we consider here as a ``first look'' the RG evolution of the gauge couplings at the one-loop level\footnote{We note that below, when we perform MCMC analyses, we will of course consider 2-loop running.} (cf.\ \cite{Cheng:1973nv})
\begin{equation}
\alpha_i^{-1}(\MZ)=\alpha_i^{-1}(\MGUT)+\frac{1}{2\pi}\left(b_i\ln\left(\frac{\MGUT}{\MZ}\right)+\sum_I b_{iI}\ln\left(\frac{\MGUT}{M_I}\right)\right),
\end{equation}
where $i=1,2,3,$ stands for U(1)$_\mathrm{Y}$, SU(2)$_\mathrm{L}$ and SU(3)$_\mathrm{C}$ and the $b_i$ denote the SM one-loop coefficients (with one Higgs doublet), i.e.\ $b_i=(41/10,\,-19/6,\,-7)$. The $b_{iI}$ are the one-loop coefficients of the particles with intermediate-scale masses $M_I$, i.e.\ $\MZ\leq M_I\leq \MGUT$. The constraints $\alpha_1(\MGUT)=\alpha_2(\MGUT)=\alpha_3(\MGUT)$ then yield the following two conditions which have to be satisfied by the masses $M_I$ in order to allow for gauge coupling unification at the one-loop level:
\begin{equation}\label{eq: unification constraints}
\begin{split}
\exp\left[2\pi\left(\alpha_2^{-1}(\MZ)-\alpha_1^{-1}(\MZ)\right)\right]=\left(\frac{\MGUT}{\MZ}\right)^{b_2-b_1}\prod_I \left(\frac{\MGUT}{M_I}\right)^{b_{2I}-b_{1I}},\\
\exp\left[2\pi\left(\alpha_3^{-1}(\MZ)-\alpha_2^{-1}(\MZ)\right)\right]=\left(\frac{\MGUT}{\MZ}\right)^{b_3-b_2}\prod_I \left(\frac{\MGUT}{M_I}\right)^{b_{3I}-b_{2I}}.
\end{split}
\end{equation}
Being interested in a minimal model which allows for gauge coupling unification, we choose the masses $M_{\Phi_8}$, $M_{\Phi_6}$ and $M_{\Phi_{(3,3)}}$ of $\Phi_8$, $\Phi_6$ and $\Phi_{(3,3)}$ (as well as the mass of the SM Higgs doublet $h$) to be below the GUT scale $\MGUT$ and set the masses of the remaining scalar fields to $\MGUT$. In particular, only one of the Higgs doublets $h$ and $h^\perp$ has a mass below $\MGUT$. While the mass of the Higgs doublet $h$ is fixed by experiment, the masses $M_{\Phi_8}$, $M_{\Phi_6}$ and $M_{\Phi_{(3,3)}}$ have to satisfy Eq.~\eqref{eq: unification constraints}. Using Eq.~\eqref{eq: unification constraints} as well as the constraints $\MZ\leq M_I\leq\MGUT$ we compute the allowed values of the mass parameters $\MGUT$, $M_{\Phi_8}$, $M_{\Phi_6}$ and $M_{\Phi_{(3,3)}}$ which we show in contourplots in the $\MGUT$-$M_{\Phi_8}$ plane in Figure~\ref{fig: contourplot unification}. The legend to the right of each plot indicates which values are represented by which color --- note that in different plots the same color represents different values. 

We observe that gauge coupling unification is indeed possible and that the GUT scale $\MGUT$ can be between $10^{14.5}$ and $10^{17.4}$~GeV. The mass $M_{\Phi_8}$ varies between $\MZ$ and $10^{15.5}$~GeV --- in particular, gauge coupling unification is still possible if we set $M_{\Phi_8}=\MGUT$, but then $\MGUT$ is restricted to be below $10^{15.5}$~GeV which would lead to proton decay rates close to the current experimental limits (cf.~Section~\ref{sec: results}). Similarly, $M_{\Phi_6}$ is bounded between $M_Z$ and $10^{16.1}$~GeV, so gauge coupling unification is still possible if $M_{\Phi_6}$ is set to $\MGUT$. But, we still allow that this mass is below $\MGUT$, so that we have more freedom for the range of $\MGUT$. Finally, the mass $M_{\Phi_{(3,3)}}$ lies between $10^{3.1}$ and $10^{8.4}$~GeV. Hence, $M_{\Phi_{(3,3)}}$ needs to be significantly below the GUT scale for a successful gauge coupling unification. We note that in a general set-up such a low mass of $\Phi_{(3,3)}$ may cause problems, since this multiplet can mediate too fast nucleon decay. However since, as mentioned above, $\textbf{45}_H$ only couples to the 22 entry of $\boldsymbol{Y_{\overline{5}}}$, this does not cause a problem in the here proposed scenario.

\subsection{Viability of the GUT scale Yukawa ratios $\boldsymbol{y_\tau/y_b = 3/2}$ and $\boldsymbol{y_\mu/y_s = 9/2}$}\label{sec: Viability of the GUT scale Yukawa ratios}
In order to investigate the viability of the GUT scale predictions $y_\tau/y_b = 3/2$ and $y_\mu/y_s = 9/2$ in the context of the above-described scenario, we study the 2-loop RG evolution taking into account the impact of the neutrino Yukawa couplings, as well as performing an (MCMC) analysis to account for the uncertainties in the low energy experimental data. To be specific (and to give a representative example), we will assume that in the mass basis of the right-chiral neutrinos (with $n=3$), the neutrino Yukawa matrix is hierarchical and dominated by its 33 element $y_{\nu_3}$, and that together with the right-chiral neutrino mass $M_3$ it gives rise to the light neutrino mass $m_{\nu_3}$ with a normal neutrino mass ordering (i.e.\ using $m_{\nu_3}=\sqrt{\Delta m_{31}^2} \approx 50.14 $\,meV~\cite{Minkowski:1977sc}).

For illustration, let us look at the RGEs of $\textbf{Y}_d$ and $\textbf{Y}_e$ with neutrino Yukawa couplings included (cf.\ e.g.\ \cite{Antusch:2002rr}), which read 
\begin{equation}\label{eq: RGE Yd Ye}
\begin{split}
16\pi^2\,\mu\frac{d}{d\mu}\textbf{Y}_d=\textbf{Y}_d\bigg\lbrace&\frac{3}{2}\textbf{Y}_d^\dagger \textbf{Y}_d -\frac{3}{2}\textbf{Y}_u^\dagger \textbf{Y}_u -\frac{1}{4}g_1^2-\frac{9}{4}g_2^2-8g_3^2\\
&+\text{Tr}\left[\textbf{Y}_e^\dagger \textbf{Y}_e+\textbf{Y}_\nu^\dagger \textbf{Y}_\nu+3\textbf{Y}_d^\dagger \textbf{Y}_d+3\textbf{Y}_u^\dagger \textbf{Y}_u\right]\bigg\rbrace,\\
16\pi^2\,\mu\frac{d}{d\mu}\textbf{Y}_e=\textbf{Y}_e\bigg\lbrace&\frac{3}{2}\textbf{Y}_e^\dagger \textbf{Y}_e -\frac{3}{2}\textbf{Y}_\nu^\dagger \textbf{Y}_\nu -\frac{9}{4}g_1^2-\frac{9}{4}g_2^2\\
&+\text{Tr}\left[\textbf{Y}_e^\dagger \textbf{Y}_e+\textbf{Y}_\nu^\dagger \textbf{Y}_\nu+3\textbf{Y}_d^\dagger \textbf{Y}_d+3\textbf{Y}_u^\dagger \textbf{Y}_u\right]\bigg\rbrace.
\end{split}
\end{equation}
We note that the positive term $(\textbf{Y}_d\,\text{Tr}\textbf{Y}_\nu^\dagger \textbf{Y}_\nu)_{33}$ in the RGE of $\textbf{Y}_d$ ``speeds up" the running of $y_b$, while the negative term $-\frac{3}{2}(\textbf{Y}_e\textbf{Y}_\nu^\dagger \textbf{Y}_\nu)_{33}$ in the RGE of $\textbf{Y}_e$ ``slows down" the running of $y_\tau$. Therefore, with nonzero contributions from $\textbf{Y}_\nu$, for a fixed GUT scale a smaller GUT scale ratio $\frac{y_\tau}{y_b}$ can be viable than for the running in the SM (without $\textbf{Y}_\nu$). Or, equivalently, the predicted given ratio $y_\tau/y_b = 3/2$ can be viable for a larger GUT scale. Assuming hierarchical $\textbf{Y}_\nu$, this effect is only relevant for the third family, but negligible for the second family. 

\begin{figure}
\includegraphics[width=7.5cm]{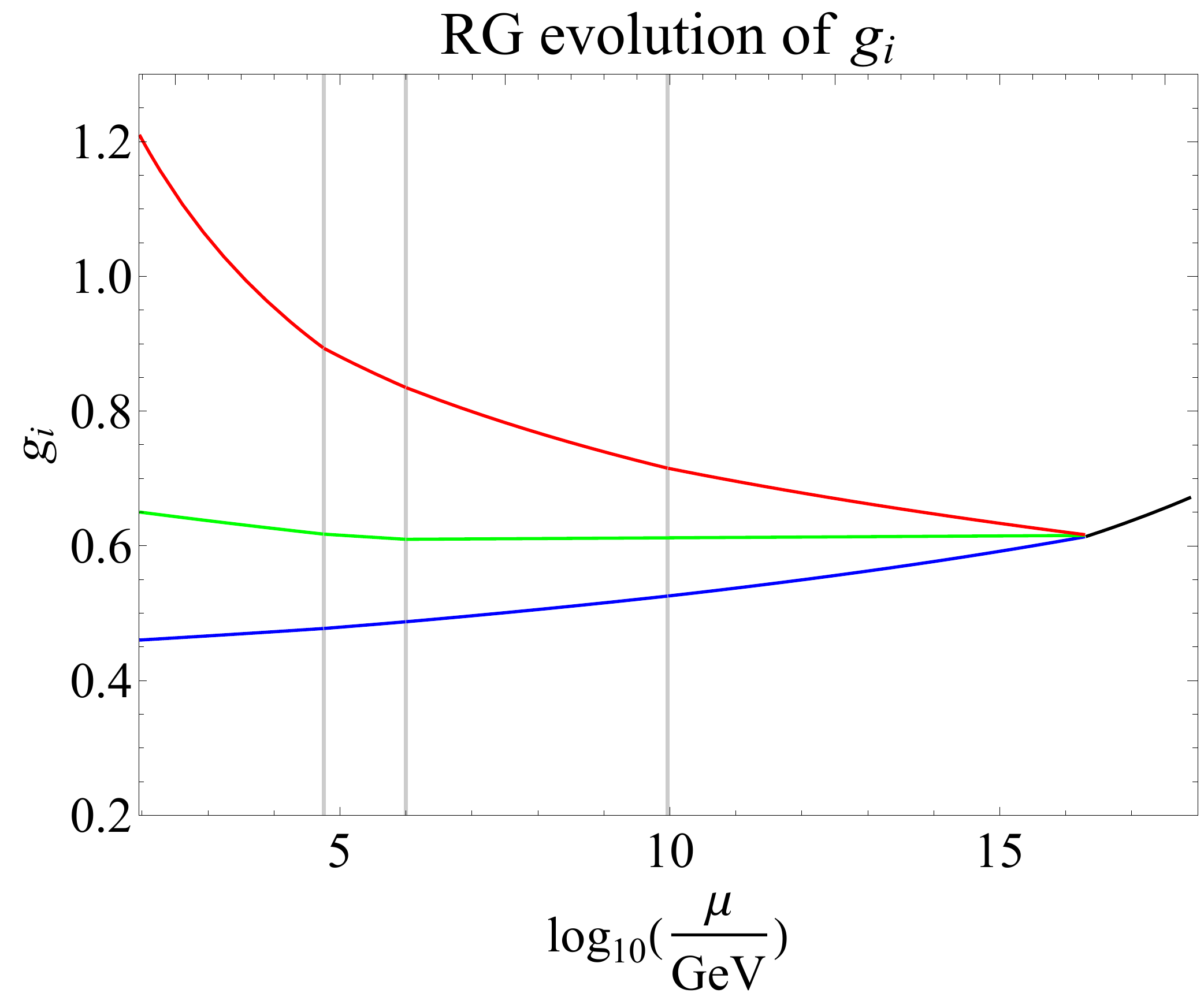}\hspace{1cm}
\includegraphics[width=7.5cm]{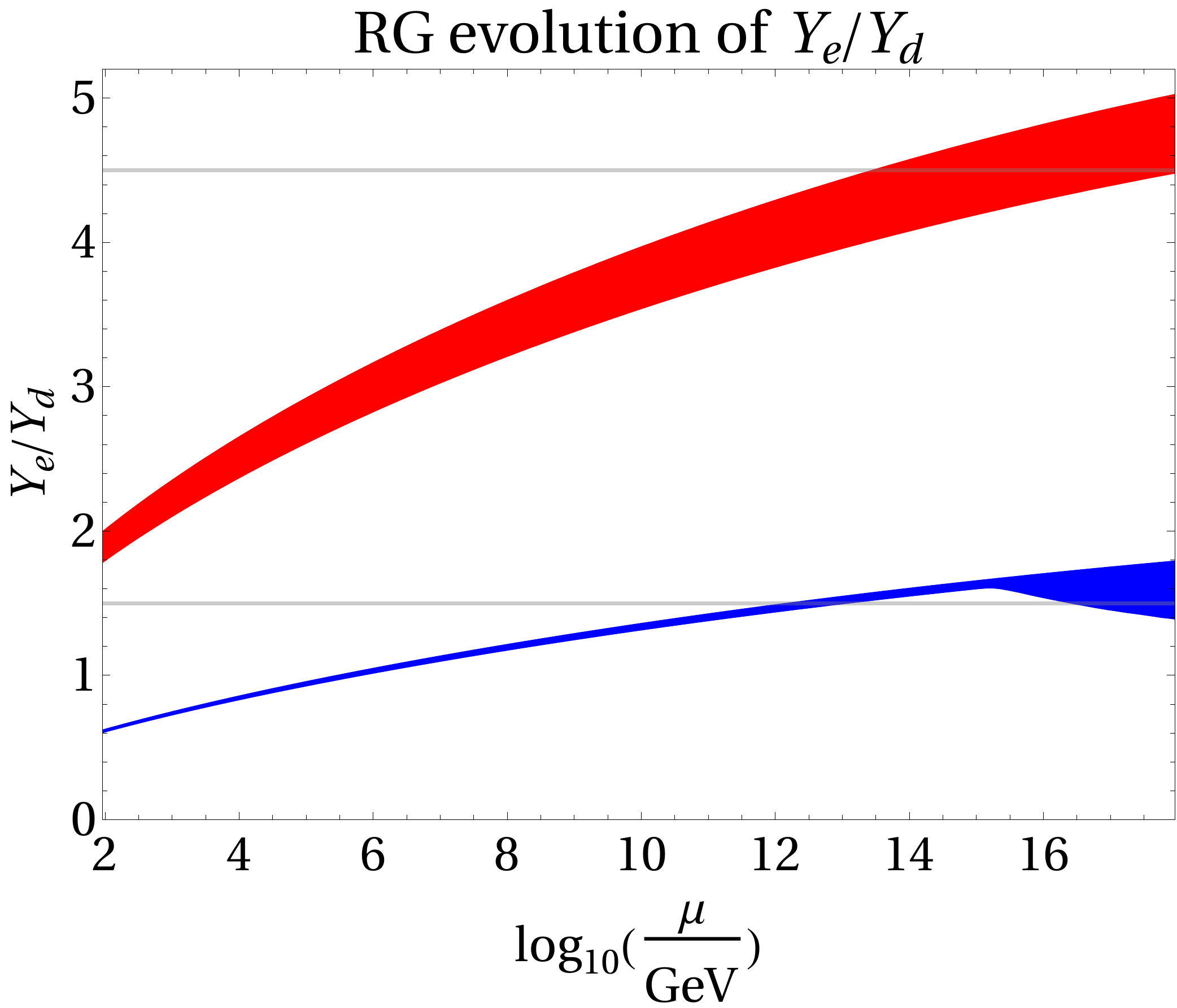}
\caption{The left plot shows an example of the 2-loop RG evolution of the SM gauge couplings, where they meet at $10^{16.3}$~GeV (left) with $y_\tau/y_b=3/2$ and $y_\mu/y_s=9/2$ realized. The right plot shows the estimated SM 2-loop RG evolution of the 1-$\sigma$ range of the ratios $y_\tau/y_b$ and $y_\mu/y_s$ including the effect of the 33 element of the neutrino Yukawa matrix $y_{\nu_3}$, where $y_{\nu_3}\leq 2.8$ was assumed (cf.\ details in the main text and in Appendix~\ref{sec: running of neutrino Yukawa coupling}). The gauge couplings $g_1$, $g_2$ and $g_3$ are colored red, green and blue, respectively, whereas the Yukawa ratios $y_\mu/y_s$ and $y_\tau/y_b$ are colored blue and red. The black curve in the left plot is the unified gauge coupling above the GUT scale. The horizontal grid lines in the left plot indicate the masses of the added scalar fields $\Phi_8$, $\Phi_6$ and $\Phi_{(3,3)}$, while the gird lines in the right plot indicate the values 9/2 and 3/2.}\label{fig: RG evolution of the gauge couplings and Yukawa ratios with non-zero neutrino Yukawa couplings}
\end{figure}

The right plot of Figure~\ref{fig: RG evolution of the gauge couplings and Yukawa ratios with non-zero neutrino Yukawa couplings} shows the estimated 2-loop RG evolution of the ratios $y_\tau/y_b$ and $y_\mu/y_s$ in the SM with an additional contribution of the 33 element of the neutrino Yukawa matrix $y_{\nu_3}$. To draw this plot, we performed an MCMC analysis varying the SM parameters around their central experimental values, taken from~\cite{Antusch:2013jca}, and weighting the points according to the $\chi^2$-function. For the contribution of the 33 element of the neutrino Yukawa matrix $y_{\nu_3}$ we took the light neutrino mass $m_{\nu_3}\approx 50.14\,$meV and computed its RG evolution from the low scale up to the right-chiral neutrino mass scale $M_3$ (which we also varied in the MCMC analysis) according to the following (1-loop) equation~\cite{Antusch:2003kp}:
\begin{equation}
\begin{split}
16\pi^2\mu \frac{d m_{\nu_3}}{d\mu}=[\alpha-3 y_\tau^2\cos^2(\theta_{13}\PMNS)\cos^2(\theta_{23}\PMNS)]m_{\nu_3},
\end{split}
\end{equation}
where $\alpha=-3g_2^2+2(y_\tau^2+y_\mu^2+y_e^2)+6(y_t^2+y_b^2+y_c^2+y_s^2+y_u^2+y_d^2)+\lambda$, $\theta_{13}\PMNS$ and $\theta_{23}\PMNS$ are the PMNS-mixing angles, and $\lambda$ is the SM Higgs self-coupling. At the mass scale of the right-chiral neutrino $M_3$ we determined the 33 element of the neutrino Yukawa matrix $y_{\nu_3}$ with the type I see-saw relation \cite{Minkowski:1977sc}
\begin{equation}m_{\nu_3}=\frac{v^2 y_{\nu_3}^2}{M_3}.
\end{equation} 
Above  the right-chiral neutrino mass $M_3$ we then took the SM contributions as well as the contributions of $y_{\nu_3}$ to the RG evolution into account. We stopped the integration at $\mu=10^{17}\,$GeV or as soon as $y_{\nu_3}\geq 2.8$. The latter condition insures that perturbativity of our models are maintained up to energy scales at least one order of magnitude above the GUT scale, see Appendix~\ref{sec: running of neutrino Yukawa coupling} for details. 

The right plot of figure~\ref{fig: RG evolution of the gauge couplings and Yukawa ratios with non-zero neutrino Yukawa couplings} shows that if the GUT scale is above $10^{16.3}\,$GeV it is possible to realize both GUT scale relations $y_\tau/y_b=3/2$ and $y_\mu/y_s=9/2$ with deviations within the $1\sigma$ uncertainties. Allowing for larger deviations from the experimental best-fit values, an extended range for $\MGUT$ is possible. Furthermore, allowing for larger deviations also the case with very small neutrino Yukawa couplings, with negligible impact on the RG evolution, becomes possible. 
The left plot of figure~\ref{fig: RG evolution of the gauge couplings and Yukawa ratios with non-zero neutrino Yukawa couplings} visualises an example of the RG evolution of the gauge couplings, in which the masses  $\Phi_8$, $\Phi_6$ and $\Phi_{(3,3)}$ were chosen such that the gauge couplings unify at $10^{16.3}\,$GeV, demonstrating that the gauge couplings can indeed meet at a point at which the ratios $y_\tau/y_b=3/2$ and $y_\mu/y_s=9/2$ are both viable.

\section{Nucleon decay fingerprints of two ``toy models"}\label{sec: nucleon decay fingerprints of two toy models}
In the following, we extend our minimal GUT scenario to two ``toy models'' with different Yukawa flavor textures, towards explaining the quark and lepton masses and mixings. Fitting these two models to the low energy data and performing MCMC analyses in order to investigate their predictions for various nucleon decay channels, we show that the ratios between partial nucleon decay rates provide ``fingerprints'' to distinguish between models with different Yukawa flavor structures.

\subsection{Formulation of the ``toy models"}\label{sec: toy models}

We refer to the two considered flavor ``toy models" we consider as model 1 and model 2. They have been discussed in the context of SUSY GUTs in \cite{Antusch:2014poa, Antusch:2013kna}. We assume for both models \textit{single operator dominance} in each entry \cite{Antusch:2012fb, Antusch:2013kna, Antusch:2017ano, Antusch:2018gnu, Antusch:2019avd}, including now also the first family of quarks and leptons. 
For the neutrino sector in model 1 and model 2 we choose the CSD2~\cite{Antusch:2011ic, Antusch:2013wn} and CSD3~\cite{King:2013iva} texture, respectively.  Moreover, we assume that the added scalar fields with intermediate-scale masses $\Phi_8$, $\Phi_6$ and $\Phi_{(3,3)}$ only effect the RG evolution of the gauge couplings, but not the running of the SM Yukawa couplings. This is consistent with the assumption of \textit{single operator dominance}, since, as it will be discussed later, the Higgs representation $\textbf{45}_H$ only interacts with the fermionic representations $\boldsymbol{\bar{5}}_{F2}\boldsymbol{\bar{5}}_{F2}$, but not with any other fermionic representations. \textit{Single operator dominance} then implies that the interactions between the intermediate-scale mass fields $\Phi_8$, $\Phi_6$ and $\Phi_{(3,3)}$ and the SM fermions are up to Clebsch-Gordan (CG) factors equal to the 22 entry of the down-quark Yukawa matrix $(\textbf{Y}_d)_{22}$ which is of the order of $\mathcal{O}(10^{-4})$. Hence, their effect on the running of the SM Yukawa couplings is negligible.

In the two models the Yukawa textures have the following form:
\begin{equation}
\begin{split}
\text{model 1 :}\hspace{2cm}\boldsymbol{Y_{10}}=
\begin{pmatrix}
\star && \star && 0 \\
\star && \star && \star \\ 
0 && \star && \star \\
\end{pmatrix},\hspace{2cm}
\boldsymbol{Y_{\overline{5}}}=
\begin{pmatrix}
0 && \star && 0 \\
\star && \star && 0 \\
0 && 0 && \star
\end{pmatrix},
\\
\text{model 2 :}\hspace{2cm}\boldsymbol{Y_{10}}=
\begin{pmatrix}
\star && \star && \star \\
\star && \star && \star \\ 
\star && \star && \star \\
\end{pmatrix},\hspace{2cm}
\boldsymbol{Y_{\overline{5}}}=
\begin{pmatrix}
\star && 0 && 0 \\
0 && \star && 0 \\
0 && 0 && \star
\end{pmatrix},
\end{split}
\end{equation}
where non-vanishing complex entries are denoted by $\star$. In Table 1 we indicate which SU(5) operator is dominant in which Yukawa entry. The index contraction between different brackets is specified by the SU(5) representation in the index position. As discussed in \cite{Antusch:2014poa, Antusch:2013kna} with these choices of GUT operators one can also get viable quark-lepton Yukawa ratios for the first family. 
\begin{table}[H]
\centering
\begin{tabu}{cll}
\tabucline[1.1pt]{-} \noalign{\vskip 1mm}
entry & model 1 operator & model 2 operator  \\\noalign{\vskip 1mm}
\hline\noalign{\vskip 2mm}
$(\boldsymbol{Y_{10}})_{ij}$  &  $\textbf{10}_{Fi}\textbf{10}_{Fj}\textbf{5}_H$  &  $\textbf{10}_{Fi}\textbf{10}_{Fj}\textbf{5}_H$  \\\noalign{\vskip 2mm}
$(\boldsymbol{Y_{\overline{5}}})_{11}$  &  /  &  $(\textbf{10}_{F1}\boldsymbol{\bar{5}}_H)_{\boldsymbol{5}}(\textbf{24}_H)_{\boldsymbol{\bar{5}}\otimes\textbf{5}}(\textbf{1}_H)_{\boldsymbol{\bar{5}}\otimes\textbf{5}}(\textbf{24}_H)_{\boldsymbol{\bar{5}}\otimes\textbf{5}}(\boldsymbol{\bar{5}}_{F1}\textbf{1}_H)_{\boldsymbol{\bar{5}}}$  \\
\noalign{\vskip 2mm}
$(\boldsymbol{Y_{\overline{5}}})_{12}$  &  
$\textbf{10}_{F1}\boldsymbol{\overline{5}}_{F2}\textbf{5}_H$  &  /  \\\noalign{\vskip 2mm}
$(\boldsymbol{Y_{\overline{5}}})_{21}$  &  
$(\textbf{10}_{F2}\textbf{5}_H)_{\textbf{45}}(\boldsymbol{\overline{5}}_{F1}\textbf{24}_H)_{\boldsymbol{\overline{45}}}$  & 
 /  \\\noalign{\vskip 2mm}
$(\boldsymbol{Y_{\overline{5}}})_{22}$  &  
$(\textbf{10}_{F3}\boldsymbol{\overline{45}}_H)_{\textbf{5}}(\boldsymbol{\overline{5}}_{F3}\textbf{24}_H)_{\boldsymbol{\overline{5}}}$  & 
$(\textbf{10}_{F3}\boldsymbol{\overline{45}}_H)_{\textbf{5}}(\boldsymbol{\overline{5}}_{F3}\textbf{24}_H)_{\boldsymbol{\overline{5}}}$   
\\\noalign{\vskip 2mm}
$(\boldsymbol{Y_{\overline{5}}})_{33}$  &  
$(\textbf{10}_{F3}\boldsymbol{\overline{5}}_H)_{\textbf{45}}(\boldsymbol{\overline{5}}_{F3}\textbf{24}_H)_{\boldsymbol{\overline{45}}}$  &   
$(\textbf{10}_{F3}\boldsymbol{\overline{5}}_H)_{\textbf{45}}(\boldsymbol{\overline{5}}_{F3}\textbf{24}_H)_{\boldsymbol{\overline{45}}}$ 
\\\noalign{\vskip 2mm}
\tabucline[1.1pt]{-} 
\end{tabu}
\caption{Table of operators which dominate the respective Yukawa entry of model 1 and model 2.}
\end{table}
The SM Yukawa matrices stem from SU(5) operators of the form 
\begin{equation}
\begin{split}
\boldsymbol{10}_{Fi}\,\textbf{10}_{Fj}\,\textbf{X}\hspace{5mm}&\supset\hspace{5mm} (\textbf{Y}_u)_{ij},\\
\textbf{10}_{Fi}\,\boldsymbol{\overline{5}}_{Fj}\, \textbf{X}\hspace{5mm}&\supset\hspace{5mm}(\textbf{Y}_d)_{ij},\,(\textbf{Y}_e)_{ij},
\end{split}
\end{equation}
where \textbf{X} represents one or multiple Higgs fields. The fact that in $\textbf{Y}_{\textbf{10}}$ the operator $\textbf{10}_F\textbf{10}_F\textbf{5}_H$ is dominant implies that $\textbf{Y}_u$ is symmetric, i.e.\ $\textbf{Y}_u=\textbf{Y}_u^T$. In principle, also an operator $\textbf{10}_F\textbf{10}_F\textbf{45}_H$ would be allowed by the gauge symmetry. Such an operator would yield antisymmetric contributions to $\textbf{Y}_u$. However, we assume that this operator is absent (or highly suppressed) because of an appropriate family symmetry, which is consistent with the assumption of \textit{single operator dominance} for the entries of $\boldsymbol{Y_{\bar{5}}}$. Since $\textbf{Y}_d$ and $\textbf{Y}_e$ both stem from the same SU(5) operators, their entries are related via the following CG factors \cite{Antusch:2009gu, Antusch:2013rxa}:
\begin{equation}\label{eq: Relation Yd Ye}
\begin{split}
\text{model 1 :}\hspace{2cm}\textbf{Y}_e=
\begin{pmatrix}
0 && 1 && 0 \\
\frac{3}{2} && \frac{9}{2} && 0 \\
0 && 0 && \frac{3}{2}
\end{pmatrix}\cdot \textbf{Y}_d^T ,
\\
\text{model 2 :}\hspace{2cm}\textbf{Y}_e=
\begin{pmatrix}
\frac{4}{9} && 0 && 0 \\
0 && \frac{9}{2} && 0 \\
0 && 0 && \frac{3}{2}
\end{pmatrix}\cdot \textbf{Y}_d^T,
\end{split}
\end{equation}
where by the dot $\cdot$ entry-wise multiplication of matrices is specified.

As mentioned above, we assume the CSD2~\cite{Antusch:2011ic, Antusch:2013wn} and CSD3~\cite{King:2013iva} texture in the neutrino sector of model 1 and model 2, respectively. The two CSD variants each fit to the considered flavor structures in the quark and charged lepton sectors. The texture of $\boldsymbol{Y_{\overline{5}}}$ in model 1 leads to a predicted 1-2 mixing in $\textbf{Y}_d$ and $\textbf{Y}_e$, and may therefore be combined with CSD2 to explain the measured lepton mixing angle $\theta_{13}^\mathrm{PMNS}$ (cf.\ \cite{Antusch:2011ic, Antusch:2013wn}). On the other hand, CSD3 works best with small mixings in $\textbf{Y}_e$, which results from the texture of $\boldsymbol{Y_{\overline{5}}}$ considered in model 2. 

In both ``toy models" only two right-handed neutrinos have been considered, which is the minimal number needed to explain the two observed neutrino mass squared differences. We note that this also covers the case that a third right-handed neutrino exists, but is approximately decoupled, with its contribution to the neutrino mass matrix and to the RG evolution being negligible. We work in the mass basis for the right-handed neutrinos where $\textbf{M}_R$ has the form 
\begin{equation}
\textbf{M}_R=\diag(M_A,M_B).
\end{equation}
\begin{itemize}
\item \textbf{Model 1}: CSD2 comes along in two varieties, denoted by $\phi_{102}$ and $\phi_{120}$, which correspond to the following neutrino Yukawa matrices:
\begin{equation}
\textbf{Y}_\nu^{(102)}=\begin{pmatrix}
0 & b \\ a & 0 \\ -a & 2b
\end{pmatrix},\hspace{2cm}
\textbf{Y}_\nu^{(120)}=\begin{pmatrix}
0 & b \\ a & 2b \\ -a & 0
\end{pmatrix}.
\end{equation}
With the seesaw mechanism these yield the following mass matrices for the left-handed neutrinos:
\begin{equation}\label{eq: left-handed neutrino mass matrix model 1}
\begin{split}
\textbf{M}_\nu^{(102)}=m_a\begin{pmatrix}
\epsilon e^{i\alpha} & 0 & 2\epsilon e^{i\alpha} \\
0 & 1 & -1 \\
2\epsilon e^{i\alpha} & -1 & 1+4\epsilon e^{i\alpha}
\end{pmatrix},\\
\textbf{M}_\nu^{(120)}=m_a\begin{pmatrix}
\epsilon e^{i\alpha} & 2\epsilon e^{i\alpha} & 0 \\
2\epsilon e^{i\alpha} & 1+4\epsilon e^{i\alpha} & -1 \\
0 & -1 & 1
\end{pmatrix},
\end{split}
\end{equation}
where we defined the complex mass parameter $m_a$ as well as the real parameters $\epsilon$ and $\alpha$ as
\begin{equation}\label{eq: def ma epsilon alpha}
m_a=\frac{v^2a^2}{M_a},\hspace{1cm} \epsilon e^{i\alpha}=\frac{a^2}{b^2}\frac{M_B}{M_A}.
\end{equation}
The phase of $m_a$ is unphysical and can be absorbed by the overall phase freedom of $\boldsymbol{\bar{5}}_i$, leaving 3 physical parameters of the left-handed neutrino mass matrix: $m_a$, $\alpha$ and $\epsilon$.
\item \textbf{Model 2}: There are two varieties of the CSD3 which are denoted by $\phi_{113}$ and $\phi_{131}$. The neutrino Yukawa matrices take the following form:
\begin{equation}
\textbf{Y}_\nu^{(113)}=\begin{pmatrix}
0 & b \\ a & b \\ a & 3b
\end{pmatrix}, \hspace{2cm}
\textbf{Y}_{\nu}^{(131)}=\begin{pmatrix}
0 & b \\ a & 3b \\ a & b
\end{pmatrix}.
\end{equation}
The left-handed neutrino mass matrices are then given by:
\begin{equation}\label{eq: left-handed neutrino mass matrix model 2}
\begin{split}
\textbf{M}_\nu^{(113)}=m_a\begin{pmatrix}
\epsilon e^{i\alpha} & \epsilon e^{i\alpha} & 3\epsilon e^{i\alpha} \\
\epsilon e^{i\alpha} & 1+\epsilon e^{i\alpha} & 1+3\epsilon e^{i\alpha} \\
3\epsilon e^{i\alpha} & 1+3\epsilon e^{i\alpha} & 1+9\epsilon e^{i\alpha}
\end{pmatrix}, \\
\textbf{M}_\nu^{(131)}=m_a\begin{pmatrix}
\epsilon e^{i\alpha} & 3\epsilon e^{i\alpha} & \epsilon e^{i\alpha} \\
3\epsilon e^{i\alpha} & 1+9\epsilon e^{i\alpha} & 1+3\epsilon e^{i\alpha} \\
\epsilon e^{i\alpha} & 1+3\epsilon e^{i\alpha} & 1+\epsilon e^{i\alpha}
\end{pmatrix},
\end{split}
\end{equation}
with an analogous definition of the parameters $m_a$, $\epsilon$ and $\alpha$ as in Eq.~\eqref{eq: def ma epsilon alpha}. By the same reasoning as in model 1, the left-handed neutrino mass matrix has 3 physical parameters: $m_a$, $\epsilon$ and $\alpha$.
\end{itemize}

\newpage

\subsection{Numerical procedure}\label{sec: numerical analysis}
\subsubsection{Parametrization of the Yukawa matrices}\label{sec: Parametrization of the Yukawa matrices}
We implement model 1 and model 2 at the GUT scale as described in Section~\ref{sec: toy models}. Because of the relation between $\textbf{Y}_d$ and $\textbf{Y}_e$, according to Eq.~\eqref{eq: Relation Yd Ye} the full Yukawa sector of quarks and charged leptons can be reconstructed by specifying $\textbf{Y}_u$ and $\textbf{Y}_d$. The implementation of the neutrino sector is discussed at the end of this Section.

Since in both models $\textbf{Y}_u$ is a complex symmetric matrix it can be decomposed by a Takagi decomposition
\begin{equation}
\textbf{Y}_u=\textbf{U}\;\textbf{Y}_u^{\text{diag}}\;\textbf{U}^\text{T},
\end{equation}
where $\textbf{U}$ is a unitary matrix and $\textbf{Y}_u^\text{diag}$ is a diagonal matrix with real positive entries.
The unitary matrix \textbf{U} can be parametrized as
\begin{equation}
\textbf{U}=\diag(e^{i\eta_1},e^{i\eta_2},e^{i\eta_3})\;\textbf{U}_{23}(\theta_{23}^{uL},\delta_0)\;\textbf{U}_{13}(\theta_{13}^{uL},\delta)\;\textbf{U}_{12}(\theta_{12}^{uL},0)\;\diag(e^{i\phi_1/2},e^{i\phi_2/2},1).
\end{equation}
Here, we defined
\begin{equation}
\begin{split}
\textbf{U}_{12}(\theta,\delta)=\begin{pmatrix}
\cos\theta  &  \sin\theta\,e^{-i\delta}  &  0  \\
-\sin\theta \,e^{i\delta}  &  \cos\theta  &  0  \\
0  &  0  &  1
\end{pmatrix},
\\
\textbf{U}_{13}(\theta,\delta)=\begin{pmatrix}
\cos\theta  &  0  &  \sin\theta\,e^{-i\delta}  \\
0  &  1  &  0  \\
-\sin\theta \,e^{i\delta}  &  0  &  \cos\theta
\end{pmatrix},
\\
\textbf{U}_{23}(\theta,\delta)=\begin{pmatrix}
1  &  0  &  0  \\
0  &  \cos\theta  &  \sin\theta\,e^{-i\delta}  \\
0  &  -\sin\theta \,e^{i\delta}  &  \cos\theta
\end{pmatrix}.
\end{split}
\end{equation}
As \textbf{U} is a unitary $3\times3$ matrix, it has 9 parameters. These read:
\begin{equation}
\theta_{12}^{uL},\;\theta_{13}^{uL},\;\theta_{23}^{uL},\;\delta,\;\phi_1,\;\phi_2,\;\eta_1,\;\eta_2,\;\eta_3.
\end{equation}
However, only 6 of these parameters are physical, since by a phase redefinition of each representation $\textbf{10}_{Fi}$ we can w.l.o.g. assume that\footnote{The matrix \textbf{U} enters the formulas describing the nucleon decay rates only as part of the products $\textbf{U}^\dagger\textbf{D}_L$ and $\textbf{E}_{L/R}^{\dagger}\textbf{U}$ (cf.\ Section~\ref{sec: nucleon decay rates}). Therefore, the phases $\eta_1$, $\eta_2$ and $\eta_3$ can be chosen such that in the product $\textbf{U}^\dagger\textbf{D}_L$ they cancel exactly with three phases of the matrix $\textbf{U}_L^d$ (and equivalently in the product $\textbf{E}_{L/R}^{\dagger}\textbf{U}$ with three phases of $\textbf{E}_R$). Therefore, the Yukawa matrices $\textbf{Y}_d$ and $\textbf{Y}_e$ only have one complex phase $\varphi$ in model 1 and are real matrices in model 2.}
\begin{equation}
\eta_1=\eta_2=\eta_3=0.
\end{equation}

\begin{itemize}
\item \textbf{Model 1}: In model 1 we use the following parametrization of the down- and up-type Yukawa matrix
\begin{equation}
\begin{split}
\textbf{Y}_d=\begin{pmatrix}
0 & y_{12}^d & 0 \\
y_{21}^d e^{i\varphi} & y_{22}^d & 0 \\
0 & 0 & y_{3}^d
\end{pmatrix},
\\
\textbf{Y}_u=\textbf{U}_1\,\diag(y_1^u,y_2^u,y_3^u)\,\textbf{U}_1^\text{T},
\end{split}
\end{equation}
where 
\begin{equation}
\textbf{U}_1=\textbf{U}_{23}(\theta_{23}^{uL},0)\,\textbf{U}_{13}(\theta_{13}^{uL},\delta)\,\textbf{U}_{12}(\theta_{12},-\pi/2).
\end{equation}
Note that $\theta_{13}^{uL}$ and $\delta$ are not free parameters, but their values depend on the other parameters of $\textbf{Y}_u$. They are chosen such that the condition $(\textbf{Y}_u)_{13}=0$ is satisfied. Furthermore, the CKM parameters are determined by diagonalizing both $\textbf{Y}_u$ and $\textbf{Y}_d$. It has been shown that the choice $\delta_0 \approx -\pi/2$ yields a correct CKM phase \cite{Antusch:2009hq, Antusch:2018gnu}. This can e.g.\ be achieved using a spontaneous CP violation mechanism \cite{Antusch:2011sx, Antusch:2013wn}. Finally, the Yukawa matrix of the charged leptons $\textbf{Y}_e$ is given by Eq.~\eqref{eq: Relation Yd Ye}.

\item \textbf{Model 2}: At the GUT scale the down- and up-quark Yukawa matrices are in model 2 parametrized as
\begin{equation}
\begin{split}
\textbf{Y}_d&=\diag(y_1^d,y_2^d,y_3^d),\\
\textbf{Y}_u&=\textbf{U}_2\,\diag(y_1^u,y_2^u,y_3^u)\,\textbf{U}_2^\text{T},
\end{split}
\end{equation}
with
\begin{equation}
\textbf{U}_2=\textbf{U}_{23}(\theta_{23}^{uL},0)\,\textbf{U}_{13}(\theta_{13}^{uL},\delta)\,\textbf{U}_{12}(\theta_{12}^{uL},0).
\end{equation}
The CKM matrix is fully given by $\textbf{Y}_u$. The charged lepton Yukawa matrix $\textbf{Y}_e$ is determined using Eq.~\eqref{eq: Relation Yd Ye}.
\end{itemize}

The neutrino sector of the models 1 and 2 is considered for two cases separately, namely $a \ll 1$ and $a = {\cal O}(1)$. In both cases we assume that $b \ll 1$,  and w.l.o.g.\ we take $a$ and $b$ positive. Although here we have different textures for the neutrino Yukawa matrix \textbf{Y}$_\nu$, the case $a = {\cal O}(1)$ allows --- similarly as in Section~\ref{sec: Viability of the GUT scale Yukawa ratios} ---  the predicted ratio $y_\tau/y_b$ to be viable for a larger GUT scale through the impact of the neutrino Yukawa couplings on the RG evolution (cf.\ Eq~\eqref{eq: RGE Yd Ye}).

Regarding the running of $\textbf{Y}_d$ and $\textbf{Y}_e$, the case $a \ll 1$ is numerically equivalent to having both right-handed neutrinos already integrated out at the GUT scale $\MGUT$. In this case the left-handed neutrino mass matrix of Eq.~\eqref{eq: left-handed neutrino mass matrix model 1} for model 1 and Eq.~\eqref{eq: left-handed neutrino mass matrix model 2} for model 2, respectively, is implemented at the GUT scale $\MGUT$. 
The case $a = {\cal O}(1)$ and $b \ll 1$ will be treated assuming one right-handed neutrino (the one with couplings $\sim b$) already being integrated out at the GUT scale $\MGUT$ whereas the second one is not. In this case the neutrino Yukawa matrix consists of one column, while the mass matrix of the right-handed neutrinos is $1\times 1$. We implement this case at the GUT scale $\MGUT$ for model 1 and model 2 as follows:
\begin{itemize}
\item \textbf{Model 1}: The neutrino Yukawa matrix $\textbf{Y}_\nu$ reads:
\begin{equation}
\textbf{Y}_\nu=\begin{pmatrix}
0 \\ a \\ -a
\end{pmatrix}.
\end{equation}
Depending on the variety of CSD2, the left-handed neutrino mass matrix $\textbf{M}_\nu$ is at the GUT scale $\MGUT$ given by:
\begin{equation}
\begin{split}
\phi_{102}\,:\hspace{2cm}
\textbf{M}_\nu=m_a\epsilon e^{i\alpha}\begin{pmatrix}
1 & 0 & 2 \\ 0 & 0 & 0 \\ 2 & 0 & 4
\end{pmatrix}, \\
\phi_{120}\,:\hspace{2cm}\textbf{M}_\nu=m_a\epsilon e^{i\alpha}\begin{pmatrix}
1 & 2 & 0 \\ 2 & 4 & 0 \\ 0 & 0 & 0
\end{pmatrix}.
\end{split}
\end{equation}
\item \textbf{Model 2}: The Yukawa matrix of the neutrinos $\textbf{Y}_\nu$ takes the form: 
\begin{equation}
\textbf{Y}_\nu=\begin{pmatrix}
0 \\ a \\ a
\end{pmatrix}.
\end{equation}
The left-handed neutrino mass matrix $\textbf{M}_\nu$ reads for the two different varieties of CSD3:
\begin{equation}
\begin{split}
\phi_{113}\,:\hspace{2cm}
\textbf{M}_\nu=m_a\epsilon e^{i\alpha}\begin{pmatrix}
1 & 1 & 3 \\ 1 & 1 & 3 \\ 3 & 3 & 9
\end{pmatrix}, \\
\phi_{131}\,:\hspace{2cm}\textbf{M}_\nu=m_a\epsilon e^{i\alpha}\begin{pmatrix}
1 & 3 & 1 \\ 3 & 9 & 3 \\ 1 & 3 & 1
\end{pmatrix}.
\end{split}
\end{equation}
\end{itemize}

\subsubsection{Calculation of the nucleon decay rates}\label{sec: nucleon decay rates}
In SU(5) GUTs the gauge bosons mediate nucleon decay via the following well-known dimension six operators \cite{Weinberg:1979sa, Weinberg:1980bf, Weinberg:1981wj, Wilczek:1979hc, Sakai:1981pk}:
\begin{equation}\label{eq: nucleon decay operators EW unbroken}
\mathcal{L}_6=-\textbf{C}_{6L}^{ijkl}\;\epsilon_{\hat{a}\hat{b}\hat{c}}\,\epsilon_{ab} \,\overline{u_{\hat{a}i}^c}\,\gamma^\mu\, Q_{\hat{b}aj}\,\overline{e_k^c}\,\gamma_\mu\, Q_{\hat{c}bl}
-
\textbf{C}_{6R}^{ijkl}\;\epsilon_{\hat{a}\hat{b}\hat{c}}\,\epsilon_{ab}\,\overline{u_{\hat{a}i}^c}\,\gamma^\mu\, Q_{\hat{b}aj}\,\overline{d_{\hat{c}k}^c}\,\gamma_\mu\, L_{bl}+\text{H.c.}\,,
\end{equation}
where $\hat{a}$, $\hat{b}$, $\hat{c}$ are SU(3) indices with $\epsilon_{123}=\epsilon^{123}=1$, $a$, $b$ are SU(2) indices with $\epsilon_{12}=\epsilon^{12}=1$ and $i$, $j$ are family indices, respectively. Moreover, we defined the coefficients 
\begin{equation}
\textbf{C}_{6L}^{ijkl}=\textbf{C}_{6R}^{ijkl}=\frac{g_{\text{GUT}}^2}{2\MGUT^2}\delta^{ij}\delta^{kl}.
\end{equation}

In the GUT scenario described in Section~\ref{sec: gauge coupling unification} with general Yukawa matrices there would be an additional contribution to nucleon decay mediated by the scalar field $\Phi_{(3,3)}$ coming from the interactions $(\textbf{Y}_{QQ})_{fg}Q_fQ_g\Phi_{(3,3)} $ and $(\textbf{Y}_{QL})_{fg}Q_fL_g\Phi_{(3,3)}^*$, with $f,g$ being family indices, and $\textbf{Y}_{QQ}$, $\textbf{Y}_{QL}$ the respective quasi-Yukawa matrices. However, in the two models considered in this thesis we assume that the operator $\textbf{10}_F\textbf{10}_F\textbf{45}_H$ is absent (or highly suppressed, cf.\ Section~\ref{sec: toy models}). This implies that $\textbf{Y}_{QQ}$ vanishes. Therefore, there is no significant contribution to the nucleon decay rates mediated by the scalar sector of our models.

The Yukawa matrices of the quarks and charged leptons can be diagonalized by a singular value decomposition, while the left-handed neutrino mass matrix is diagonalized by a Takagi decomposition:
\begin{equation}\label{eq: singular value decomposition}
\begin{split}
\textbf{Y}_u^{\diag}=\textbf{U}_R^T \textbf{Y}_u \textbf{U}_L,\hspace{2cm} \textbf{Y}_d^{\diag}=\textbf{D}_R^T \textbf{Y}_d \textbf{D}_L, \\
\textbf{Y}_e^{\diag}=\textbf{E}_R^T \textbf{Y}_e \textbf{E}_L,\hspace{2cm} \textbf{M}_\nu^{\diag}=\textbf{N}^T \textbf{M}_\nu \textbf{N},
\end{split}
\end{equation}
where $\textbf{Y}_u^{\diag}$, $\textbf{Y}_d^{\diag}$, $\textbf{Y}_e^{\diag}$ and $\textbf{M}_\nu^{\diag}$ are diagonal matrices with real positive entries, whereas $\textbf{U}_R$, $\textbf{U}_L$, $\textbf{D}_R$, $\textbf{D}_L$, $\textbf{E}_R$, $\textbf{E}_L$ and $\textbf{N}$ are unitary matrices. These unitary matrices also define the transformations to the mass eigenbasis.

If only mass eigenstates of the fermions are included which are lighter than the nucleons the effective operators take the following form in the physical basis in the EW symmetry broken phase of the SM:
\begin{equation}
\begin{split}
\mathcal{L}_\slashed{B}=\frac{1}{16\pi^2}\;\epsilon_{\hat a\hat b\hat c}&\Big(
{\mathbf{\widetilde C}}_{RL}(udue)^{ik}(u_R^{\hat a} d_{Ri}^{\hat b})(u_L^{\hat c}e_{Lk})
+
\mathbf{\widetilde C}_{LR}(udue)^{ik}(u_L^{\hat a} d_{Li}^{\hat b})(u_R^{\hat c}e_{Rk})
\\
&+\mathbf{\widetilde C}_{RL}(udd\nu)^{ijk}(u_R^{\hat a} d_{Ri}^{\hat b})(d_{Lj}^{\hat c}\nu_{Lk})
\Big)+\text{H.c.}\,,
\end{split}
\end{equation}
where $u_i$, $d_i$, $e_i$ and $\nu_i$, $i=1,2,3$, denote the mass eigenstates of the up-type quarks, the down-type quarks, the charged leptons and the neutrinos, whereas the label $L$ respectively $R$ specifies if a state is a left-chiral or a right-chiral spinor. 
Since the heaviest down-type quark and charged lepton are heavier than the nucleons, the index $i$ in $d_i$ and $e_i$ only runs from 1 to 2. Similarly, only the lightest up-type quark is considered in Eq.~\eqref{eq: nucleon decay operators EW broken} and the index $i$ is dropped, because only this state is lighter than the nucleons. Contrarily, all three neutrino states $\nu_i$ are present. Furthermore, the operators $\mathbf{\widetilde C}$ are defined by \cite{FileviezPerez:2004hn}:
\begin{equation}\label{eq: nucleon decay operators EW broken}
\begin{split}
\mathbf{\widetilde C}_{RL}(udue)^{ik}&=\frac{g_{\text{GUT}}^2}{2\MGUT^2}\left((\textbf{U}_R^\dagger \textbf{U}_L)_{11}(\textbf{D}_R^\dagger \textbf{D}_L)_{ik}+(\textbf{U}_R^\dagger \textbf{D}_L)_{1k}(\textbf{E}_R^\dagger \textbf{U}_L)_{i1}\right),
\\
\mathbf{\widetilde C}_{LR}(udue)^{ik}&=\frac{g_{\text{GUT}}^2}{2\MGUT^2}\;(\textbf{U}_R^\dagger \textbf{U}_L)_{11}(\textbf{D}_R^\dagger \textbf{E}_L)_{ki},
\\
\mathbf{\widetilde C}_{RL}(udd\nu)^{ijk}&=\frac{g_{\text{GUT}}^2}{2\MGUT^2}\;(\textbf{U}_R^\dagger \textbf{D}_L)_{1i}(\textbf{D}_R^\dagger \textbf{N})_{jk}.
\end{split}
\end{equation}

The leading-log renormalization of the operators $\textbf{C}_{6L}^{ijkl}$ and $\textbf{C}_{6R}^{ijkl}$ from the GUT scale $\MGUT$ to the the $Z$ boson mass scale $M_Z$ is captured by overall scaling factors described by the coefficients $A_{SL}$ and $A_{SR}$, respectively. They read \cite{Wilczek:1979hc, Buras:1977yy, Ellis:1979hy, Goldman:1980ah}:
\begin{equation}\label{eq: long range factors}
A_{SL(R)}=\prod_{i=1,2,3}\prod_I^{M_Z\leq M_I\leq \MGUT}\left(\frac{\alpha_i(M_{I+1})}{\alpha_i(M_I)}\right)^{\frac{\gamma_{L(R)i}}{\sum_J^{M_Z\leq M_J\leq\MGUT}b_{iJ}}},
\end{equation}
where $\gamma_{L(R)i}=(23(11)/20,\,9/4,\,2)$.

Regarding the running of the dimension six operators $\mathbf{\widetilde C}$ from $M_Z$ to the nucleon mass scale the leading term comes from the gauge coupling of the strong interaction $g_3$. The RGEs of the effective operators $\mathbf{\widetilde C}$ and of the gauge coupling $g_3$ are at 1-loop given by \cite{Nihei:1994tx}
\begin{equation}
\begin{split}
16\pi^2\mu\frac{d\,\mathbf{\widetilde C}^{ijkl}}{d\mu}&=-4\,g_3^2\; \mathbf{\widetilde C}^{ijkl}\,,\\
16\pi^2\mu\frac{dg_3}{d\mu}&=(-11+\frac{2}{3}N_F)g_3^3\,,
\end{split}
\end{equation}
where $N_F$ denotes the number of quarks with a mass below the renormalization scale $\mu$. Again, the running of the dimension six operators is captured by an overall scaling factor.

The partial decay widths of a nucleon $B_i$ into a meson $M_j$ and a lepton $l_k$ are described by the following formula \cite{Goto:1998qg}:
\begin{equation}
\Gamma(B_i\rightarrow M_j l_k)=\frac{m_i}{32\pi}\left(1-\frac{m_j^2}{m_i^2}\right)^2\frac{1}{f_\pi^2}\left(\left|A_L^{ijk}\right|^2+\left|A_R^{ijk}\right|^2\right).
\end{equation}
Here, $m_i$ and $m_j$ are the nucleon and meson masses, while $f_\pi=0.130\,\text{GeV}$ is the pion decay constant (cf.\ PDG \cite{Tanabashi:2018oca}). The amplitudes $A_L^{ijk}$ and $A_R^{ijk}$ depend on the effective operators $\mathbf{\widetilde C}$ at the nucleon mass scale. They are listed in Table 1 of \cite{Goto:1998qg}, where they defined the nucleon mass $m_N$ as the average of the proton mass $m_p$ and the neutron mass $m_n$, while the baryon mass $m_{B'}$ was defined as the average of $m_\Gamma$ and $m_\Sigma$. Moreover, they used the constants from hyperon decay $F\approx 0.463$, $D\approx 0.804$ (cf.\ \cite{Nath:2006ut, Cabibbo:2003cu}), as well as the proton decay matrix element parameters $\alpha\approx 0.0090\,\text{GeV}^3$, $\beta\approx 0.0096\,\text{GeV}^3$ (cf.\ \cite{Nath:2006ut, Tsutsui:2004qc}).

After these considerations 8 proton decay channels as well as 5 neutron decay channels can be calculated. We list these 13 nucleon decay channels together with their experimental bounds in Table \ref{tab: nucleon decay channels experimental bounds}.

\begin{table}[t]
\centering
\begin{tabu}{llll}\tabucline[1.1pt]{-} \noalign{\vskip 2mm}
& decay channel  &  $\tau/\mathcal{B}$ [year] & $\Gamma_{\text{partial}}$ [GeV]  
\\\noalign{\vskip 1mm}\hline\noalign{\vskip 2mm}
Proton: & $p\rightarrow \pi^0\,e^+$  &  $>\,2.4\cdot 10^{34}$  &  $>\,8.7\cdot 10^{-67}$  \\\noalign{\vskip 2mm}
&  $p\rightarrow \pi^0\,\mu^+$  &  $>\,1.6\cdot 10^{34}$  &  $>\,1.3\cdot 10^{-66}$  \\\noalign{\vskip 2mm}
&  $p\rightarrow \eta^0\,e^+$  &  $>\,4.1\cdot 10^{33}$  &  $>\,5.1\cdot 10^{-66}$  \\\noalign{\vskip 2mm}
&  $p\rightarrow \eta^0\,\mu^+$  &  $>\,1.2\cdot 10^{33}$  &  $>\,1.7\cdot 10^{-65}$  \\\noalign{\vskip 2mm}
&  $p\rightarrow K^0\,e^+$  &  $>\,1.1\cdot 10^{33}$  &  $>\,1.9\cdot 10^{-65}$  \\ \noalign{\vskip 2mm}
&  $p\rightarrow K^0\,\mu^+$  &  $>\,1.6\cdot 10^{34}$  &  $>\,1.3\cdot 10^{-66}$  \\\noalign{\vskip 2mm}
&  $p\rightarrow \pi^+\,\overline{\nu}$  &  $>\,2.8\cdot 10^{32}$  &  $>\,7.4\cdot 10^{-65}$  \\\noalign{\vskip 2mm}
&  $p\rightarrow K^+\,\overline{\nu}$  &  $>\,6.6\cdot 10^{33}$  &  $>\,3.2\cdot 10^{-66}$ 
\\\noalign{\vskip 1mm}\hline\noalign{\vskip 2mm}
Neutron:  &  $n\rightarrow \pi^-\,e^+$  &  $>\,2.1\cdot 10^{33}$  &  $>\,1.0\cdot 10^{-65}$  \\\noalign{\vskip 2mm}  
&  $n\rightarrow \pi^-\,\mu^+$  &  $>\,9.9\cdot 10^{32}$  &  $>\,2.1\cdot 10^{-65}$  \\\noalign{\vskip 2mm}
&  $n\rightarrow \pi^0\,\overline{\nu}$  &  $>\,9.9\cdot 10^{32}$  &  $>\,2.1\cdot 10^{-65}$  \\\noalign{\vskip 2mm}
&  $n\rightarrow \eta^0\,\overline{\nu}$  &  $>\,5.6\cdot 10^{32}$  &  $>\,3.7\cdot 10^{-65}$  \\\noalign{\vskip 2mm}
&  $n\rightarrow K^0\,\overline{\nu}$  &  $>\,1.2\cdot 10^{32}$  &  $>\,1.7\cdot 10^{-64}$  \\\noalign{\vskip 1mm}
\tabucline[1.1pt]{-}
\end{tabu}
\caption{Table of experimental bounds on different nucleon decay channels with 90~\% confidence level. The lifetime bound $\tau/\mathcal{B}$, where $\mathcal{B}$ is the branching ratio, as well as the partial decay width $\Gamma_{\text{partial}}$ are listed for each decay channel. The values are from Figure 5-3 of \cite{Brock:2012ogj}, updated with the results from \cite{Mine:2016mxy, Takenaka:2020vqy, Bajc:2016qcc, Miura:2016krn, Abe:2014mwa, 1205.6538}.}\label{tab: nucleon decay channels experimental bounds}
\end{table}

\subsubsection{Input parameters}\label{sec: input parameters}
We now discuss the GUT scale input parameters of our two models and their allowed ranges.
\begin{itemize}
\item \textbf{Model 1}: The input parameters of model 1 are:
\begin{equation}\label{eq: model parameters model 1}
\begin{split}
&g_{\text{GUT}},\;\MGUT,\;M_{\Phi_8},\;M_{\Phi_6},\;M_{\Phi_{(3,3)}},\;
\\
&y_{12}^d,\;y_{21}^d,\;y_{22}^d,\;y_3^d,\;y_1^u,\;y_2^u,\;y_3^u,\;\theta_{12}^{uL},\;\theta_{23}^{uL},\;\varphi,\;\phi_1,\;\phi_2,\hspace{1.6cm}
\\
&a,\;M_A,\;\epsilon,\;\alpha.
\end{split}
\end{equation}
\item \textbf{Model 2}: Model 2 has the following input parameters:
\begin{equation}\label{eq: model parameters model 2}
\begin{split}
&g_{\text{GUT}},\;\MGUT,\;M_{\Phi_8},\;M_{\Phi_6},\;M_{\Phi_{(3,3)}},\;
\\
&y_{1}^d,\;y_{2}^d,\;y_3^d,\;y_1^u,\;y_2^u,\;y_3^u,\;\theta_{12}^{uL},\;\theta_{13}^{uL},\;\theta_{23}^{uL},\;\delta,\;\phi_1,\;\phi_2,\hspace{1.6cm}
\\
&a,\;M_A,\;\epsilon,\;\alpha.
\end{split}
\end{equation}
\end{itemize}
Both models contain 21 real parameters. The five parameters in the first row of Eq.~\eqref{eq: model parameters model 1}, respectively Eq.~\eqref{eq: model parameters model 2}, namely the unified gauge coupling at the GUT scale $g_\text{GUT}$, the GUT scale itself $\MGUT$ and the intermediate-scale masses $M_{\Phi_8},\,M_{\Phi_6}$ and $M_{\Phi_{(3,3)}}$, are used to fit the gauge couplings of the SM. While the Yukawa sector is constructed by the 12 parameters in the second row, the four parameters in the third row construct the neutrino sector. The parameters $\phi_1$ and $\phi_2$ are the so-called GUT phases \cite{Ellis:1979hy, Ellis:2019fwf}. While they do affect the predictions for the nucleon decay rates of the nucleon decay channels, they have no impact on the predictions for the SM fermion masses and mixing angles.

The allowed ranges of the input parameters read:
\begin{equation}
\begin{split}
\MGUT\;&>M_Z,\\
M_{\Phi_8},\,M_{\Phi_6},\,M_{\Phi_{(3,3)}},\;\;M_A\;&\in\;[\MZ,\MGUT],\\
g_{\text{GUT}},\;\;y_{12}^d,\,\,y_{21}^d,\,y_{22}^d,\;\;y_1^d,\;y_2^d,\,y_3^d,\;\;y_1^u,\,y_2^u,\,y_3^u\;&>\;0,\\
\theta_{12}^{u},\,\theta_{13}^u,\,\theta_{13}^u\,&\in\;[0,\pi/2],\\
\varphi\;&\in\;[-\pi,\pi),\\
\delta,\,\phi_1,\,\phi_2,\;\;\alpha\;&\in\;[0,2\pi),\\
\epsilon\;&\in\;[0,1].
\end{split}
\end{equation} 
When fitting the two models we will consider the cases $a \ll 1$ and $a = {\cal O}(1)$. In the former case, the parameters $a$ and $M_A$ are replaced by the parameter $m_a$ of Eq.~\eqref{eq: def ma epsilon alpha}. In the latter case it turns out that values $a\gtrsim 3.5$ would be needed to fully minimize the $\chi^2$-function. However, since we want our models to be renormalizable up to at least one order of magnitude above the GUT scale, we have fixed $a = 2.4$ by hand. The reasoning behind the choice of this value is explained in Appendix \ref{sec: running of neutrino Yukawa coupling}, but we would like to emphasize  that this is not a strict constraint, and other values can be justified as well.

\subsubsection{Observables}
Apart from the nucleon decay rates (cf.~Table~\ref{tab: nucleon decay channels experimental bounds}), the 22 low scale observables of both models consist of the SM gauge couplings, the fermion masses of the up, down, charged lepton, the neutrino mass squared differences as well as the CKM and four of the PMNS parameters:
\begin{equation}
\begin{split}
&g_1,\;g_2,\;g_3,\;\\
&y_u,\;y_c,\;y_t,\;y_d,\;y_s,\;y_b,\;\theta_{12}\CKM,\;\theta_{13}\CKM,\;\theta_{13}\CKM,\;\delta\CKM,\;y_e,\;y_\mu,\;y_\tau,\;\hspace{0.6cm}\\
&\Delta m_{21}^2,\;\Delta m_{31}^2,\;\theta_{12}\PMNS,\;\theta_{13}\PMNS,\;\theta_{23}\PMNS,\;\delta\PMNS.
\end{split}
\end{equation} 

We take the experimental central values and standard deviations at $M_Z$ of the SM gauge and Yukawa observables from \cite{Antusch:2013jca}. The values for the neutrino mass squared differences and the PMNS parameters are taken from NuFIT~v5.0~\cite{Esteban:2020cvm}. For the charged lepton Yukawa couplings we use a standard deviation of 1~\%, to take into account, e.g., the limited precision of the RG evolution.

\subsubsection{Fit to low energy data}
At the GUT scale we implement the parameters of model 1 and model 2, which are given in Eq.~\eqref{eq: model parameters model 1}~and~Eq.~\eqref{eq: model parameters model 2}, respectively, as described in Section~\ref{sec: Parametrization of the Yukawa matrices}. Furthermore, using Eq.~\eqref{eq: nucleon decay operators EW unbroken} we determine the dimension 6 operators at the GUT scale. To calculate the RG evolution of the SM parameters we use the Mathematica package \texttt{REAP} \cite{Antusch:2005gp}. For the dimension 6 operators, on the other hand, the running is computed explicitly according to Eq.~\eqref{eq: long range factors}. For the SM gauge couplings and Yukawa matrices the running is computed at 2-loop, while the running of the right-handed neutrino mass matrix as well as of the effective neutrino mass operator is computed at the 1-loop level. 

After the RG running, the dimension six operators defined in Eq.~\eqref{eq: nucleon decay operators EW broken} are determined at the mass scale of the Z-boson $\MZ$. With the Mathematica package \texttt{ProtonDecay} \cite{Antusch:2020ztu} the partial decay widths of the nucleons given in Table~\ref{tab: nucleon decay channels experimental bounds} are computed. Moreover, at $\MZ$, the Yukawa couplings in the quark and charged lepton sector, the CKM and PMNS parameters as well as the neutrino squared mass differences are calculated.

The statistical analysis of the models is performed as follows:
\begin{itemize}
\item We use the $\chi^2$ function, given by the sum of the individual $\chi_i^2$ of each measured observable $i$,
\begin{equation}
\chi^2(\vec{x})=\sum_i \chi_i^2,
\end{equation}
where $\vec{x}$ is a vector consisting of the input parameters of model 1 or 2, listed in Eq.~\eqref{eq: model parameters model 1} and \eqref{eq: model parameters model 2}, respectively. For the observables $\theta_{23}\PMNS$ and $\delta\PMNS$ we use the exact $\Delta\chi^2$ function provided by NuFIT~5.0~\cite{Esteban:2020cvm}. For all other observables we compute the individual $\chi_i^2$ function as
\begin{equation}
\chi_i^2=\frac{(x_i^{\text{pred}}-x_i^{\text{exp}})^2}{\sigma_{i\pm}^2},
\end{equation}
where $x_i^{\text{pred}}$ is our prediction for the observable $i$ and $x_i^{\text{exp}}$ is its central experimental value, whereas $\sigma_i$ denotes the corresponding standard deviation which may be asymmetric, i.e.\ $\sigma_{i+}$ for $x_i^{\text{exp}}>x_i^{\text{pred}}$ and $\sigma_{i-}$ for $x_i^{\text{exp}}>x_i^{\text{pred}}$, respectively.
\item We use a differential evolution algorithm to calculate the best-fit point of a certain model.
\item Using an adaptive Metropolis-Hastings algorithm \cite{Metropolis-Hastings algorithm} to perform an MCMC analysis we determine the posterior density of the observables of a given model. With flat prior probability distributions starting from the best-fit point we compute 8 independent chains with $1.25\cdot 10^5$ data points giving us a total of $10^6$ points for each dataset.
\end{itemize}

\subsection{Best-fit points and MCMC analyses}\label{sec: results}
After specifying the implementation of the GUT scale parameters for the models 1 and 2 as well as the comparison of the observables to the experimental data in Section~\ref{sec: numerical analysis}, we present in this Section our numerical results. Apart from the predicted nucleon decay rates we are especially interested in the predictions for the masses of the charged leptons and down-type quarks as well as the predictions of the neutrino sector. Since CSD2 and CSD3 both come in two varieties, and we are further interested in the cases $a \ll 1$ and $a = {\cal O}(1)$ as discussed in Section~\ref{sec: input parameters}, we consider four different cases for each of the two models.

\subsubsection{Best-fit points}\label{sec: best-fit points}

\begin{figure}[t]\centering
\includegraphics[width=7cm]{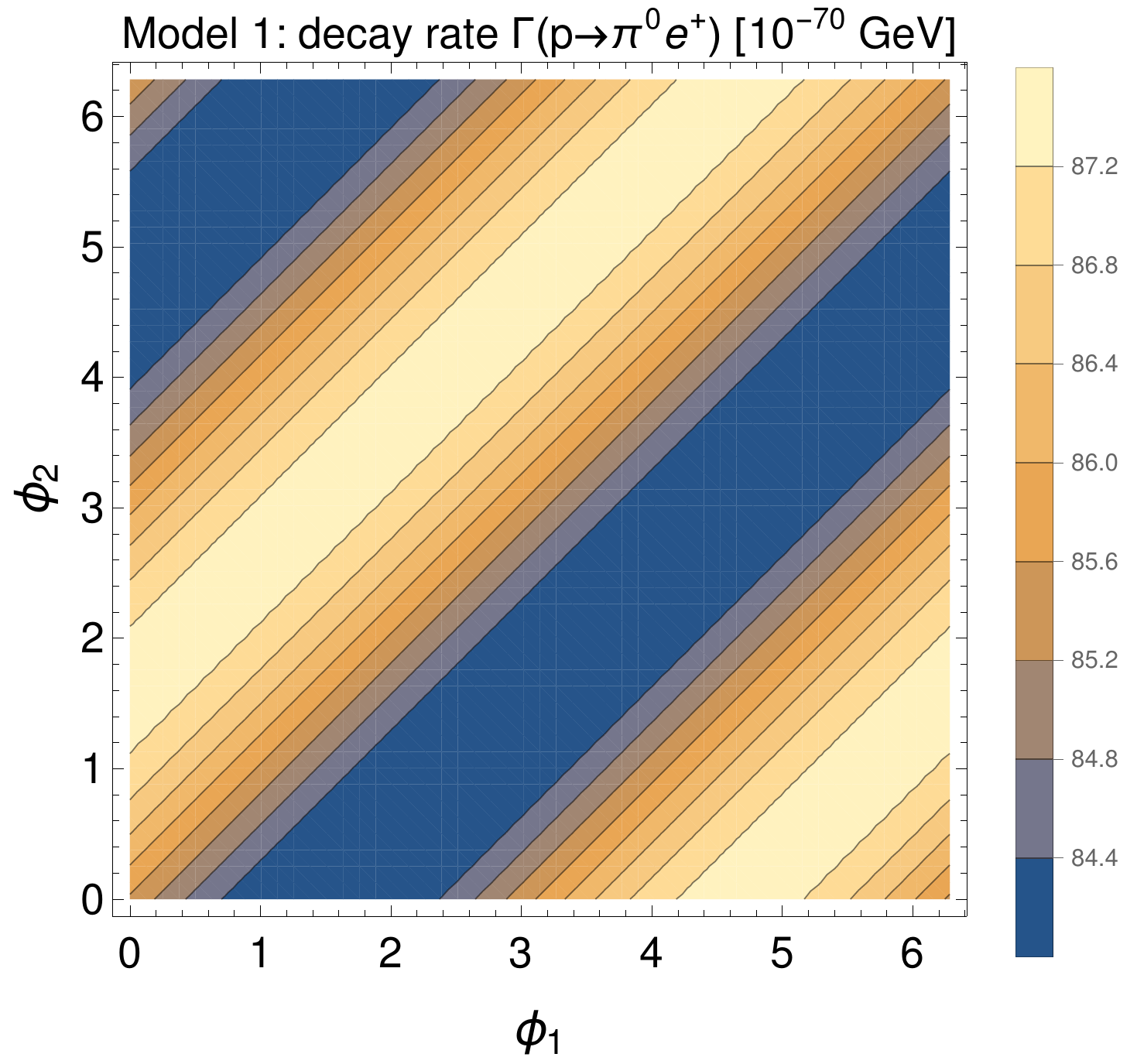}\hspace{6mm}
\includegraphics[width=7cm]{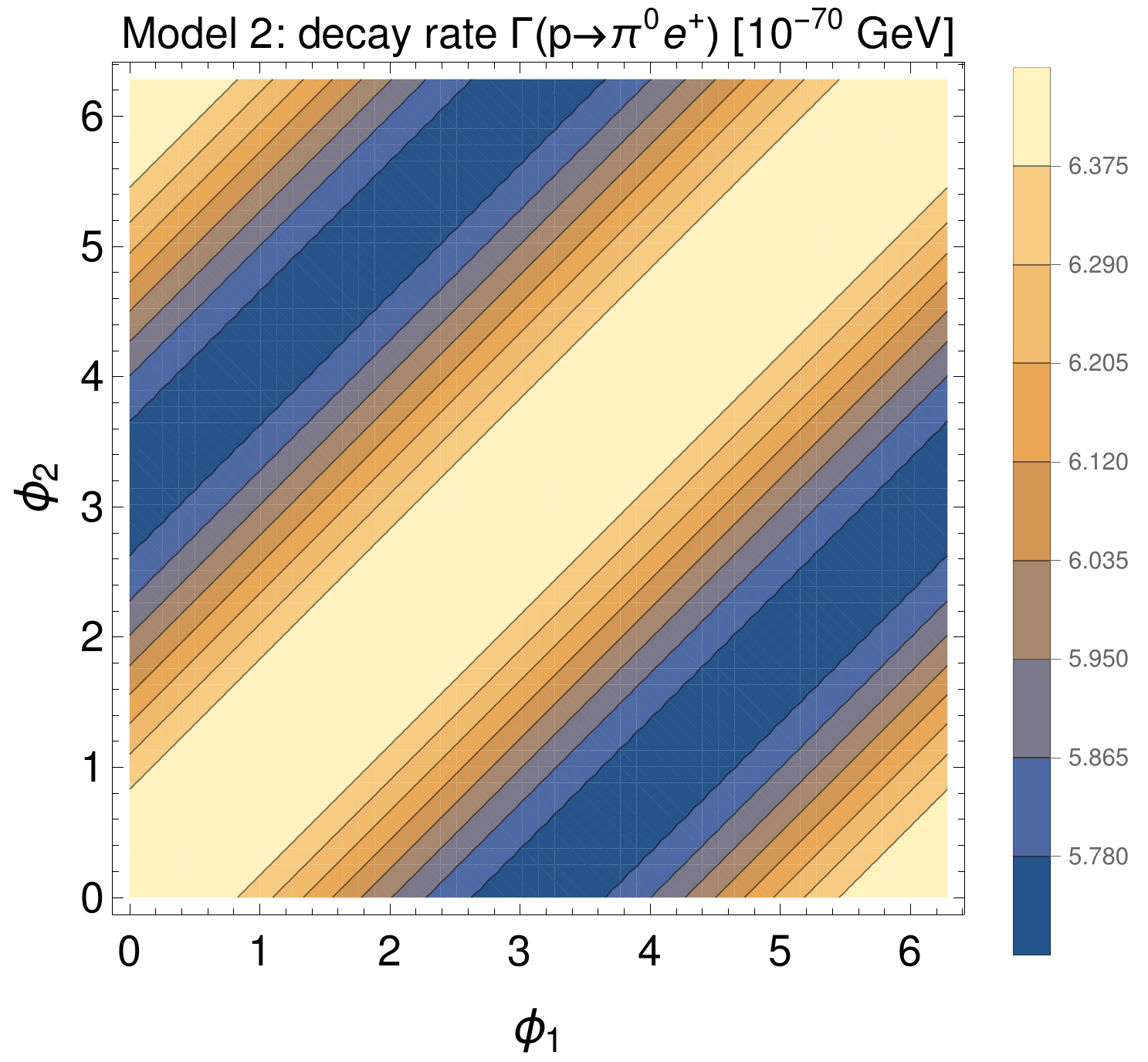}\\\vspace{6mm}

\caption{Contour plots of the decay rate for the dominant proton decay channel $p\rightarrow \pi^0e^+$,
in the $\phi_1$-$\phi_2$ plane, for the best-fit points of model 1 (left) and model 2 (right).}\label{fig: GUT phase dependence of nucleon decay rates}
\end{figure}

The best-fit points for the different cases of each model are obtained by minimizing the $\chi^2$ function. Table~\ref{tab: best-fit points input parameters} shows the values of the input parameters at the GUT scale for the best-fit points, while the total $\chi^2$ as well as the dominant pulls $\chi_i^2$ are listed in Table~\ref{tab: best-fit chi squared}. The most dominant pulls come from the bottom and tau mass. 

Comparing the cases with small ($a \ll 1$) and large ($a = {\cal O}(1)$) neutrino couplings we notice that for $a = {\cal O}(1)$ the GUT scale $\MGUT$ gets lifted which yields smaller predictions for the nucleon decay rates. The reason for this is the following: As discussed in Section~\ref{sec: Viability of the GUT scale Yukawa ratios} a finite neutrino coupling yields better predictions of $y_\tau/y_b$ since it modifies the running of the Yukawa matrices $\textbf{Y}_e$ and $\textbf{Y}_d$ as described in Eq.~\eqref{eq: RGE Yd Ye}. This effect is present in the energy range $M_A\leq\mu\leq\MGUT$. Since the ratio $a^2/M_A$ is fixed by the fit to predict the observed neutrino masses at $M_Z$ and we further fixed $a=2.4$ by hand, the only parameter which is left to increase this effect is the GUT scale $\MGUT$ --- and of course the larger $\MGUT$ the better the predictions of $y_\tau/y_b$. On the other hand, the neutrinoless SM predicts that the ratios $y_\tau/y_b$ is monotonously increasing and $y_\tau/y_b>3/2$ for $\gtrsim10^{14}$~GeV --- hence, the larger the GUT scale $\MGUT$ the worse the prediction for $y_\tau/y_b$. The best-fit point gives then the optimum between these two competing effects. 

For $a \ll 1$ only the second effect is present. Therefore, a smaller GUT scale $\MGUT$ is needed for better predictions of $y_\tau/y_b$. This appears to be in some tension with the limits on the nucleon decay rates, which prefer a higher $\MGUT$. With $a \ll 1$ the pull of the bottom and tau mass is significantly worse than for $a = {\cal O}(1)$. 
We like to note, however, that there are several potential additional effects which were not considered in our minimal model: 
First, as discussed in Section~\ref{sec: gauge coupling unification}, we have assumed that the mass of the second Higgs doublet $h^\perp$ is at the GUT scale $\MGUT$, but it could well also be at intermediate scales. Second, we have not considered possible effects of large GUT-Higgs sector couplings on the running of the Yukawa couplings, but rather assumed them to be $\ll 1$ such that they can be neglected in the RG evolution. Third, we have assumed that all other particles from GUT representations that are not at intermediate scales have masses at the GUT scale. All these effects might change our predictions for $y_\tau/y_b$ and bring the models in even better agreement with the data (potentially also the case with $a \ll 1$).

The different different varieties of CSD2 and CSD3 give different predictions for the neutrino observables. The varieties $\phi_{120}$ and $\phi_{131}$ yield smaller pulls of $\theta_{23}\PMNS$, whereas the pull of $\delta\PMNS$ is smaller for the CSD varieties $\phi_{102}$ and $\phi_{113}$. The MCMC results for the neutrino sector will be discussed in Section~\ref{sec: MCMC}.

The different nucleon decay channels will be compared in Section~\ref{sec: MCMC}. Here, we will discuss the dependence of the dominant decay channels on the GUT phases $\phi_1$ and $\phi_2$. This dependence is of particular interest since the GUT phases do not change the observed fermion masses and mixing angles, but only enter the fit by minimizing the decay rates. Figure~\ref{fig: GUT phase dependence of nucleon decay rates} shows for model 1 (with $\phi_{102}$) and model 2 (with $\phi_{131}$) for $a = {\cal O}(1)$ contour plots for the proton decay channel $p\rightarrow \pi^0e^+$. 
The legend on the right indicates which value is represented by which color --- be aware that in different plots the same color represents different values.  
We note that for other channels, the different models affect the nucleon decay rates with different strength. E.g.\ for $p\rightarrow \pi^+\bar\nu$, in model 1 the size of the effect on the decay rate is similar to the channel $p\rightarrow \pi^0e^+$, while it is strongly suppressed in model 2.

Figure~\ref{fig: GUT phase dependence of nucleon decay rates} also indicates that for the shown decay rate the dependence on the difference of the GUT phases $\phi_1-\phi_2$ dominates over the dependence on the sum $\phi_1+\phi_2$. We have checked that in model 1 this is the case for all nucleon decay channels. In model 2 however, some decay channels also depend mainly on the difference between the GUT phases $\phi_1-\phi_2$, while, as mentioned above, others are almost independent of the GUT phases. Moreover, for all decay channels it holds that the decay rates vary by a factor less than 1.3 for different choices of the GUT phases.

\subsubsection{MCMC analysis}\label{sec: MCMC}
We now use a MCMC method to analyze our models by varying the parameters listed in Eq.~\eqref{eq: model parameters model 1}~and~\eqref{eq: model parameters model 2} around the best-fit point giving us estimates of the posterior densities on the parameter space as described in Section~\ref{sec: numerical analysis}. So we note that also $a$ is varied, but constrained to be below 2.4. In the following, we discuss the results for the charged lepton and down-type quark mass ratios, the PMNS parameters, the GUT scale and the masses of the additional scalar fields as well as the nucleon decay rates.

\begin{enumerate}

\item \textbf{Results for the mass ratios}
\begin{figure}[t]\centering
\includegraphics[width=5cm]{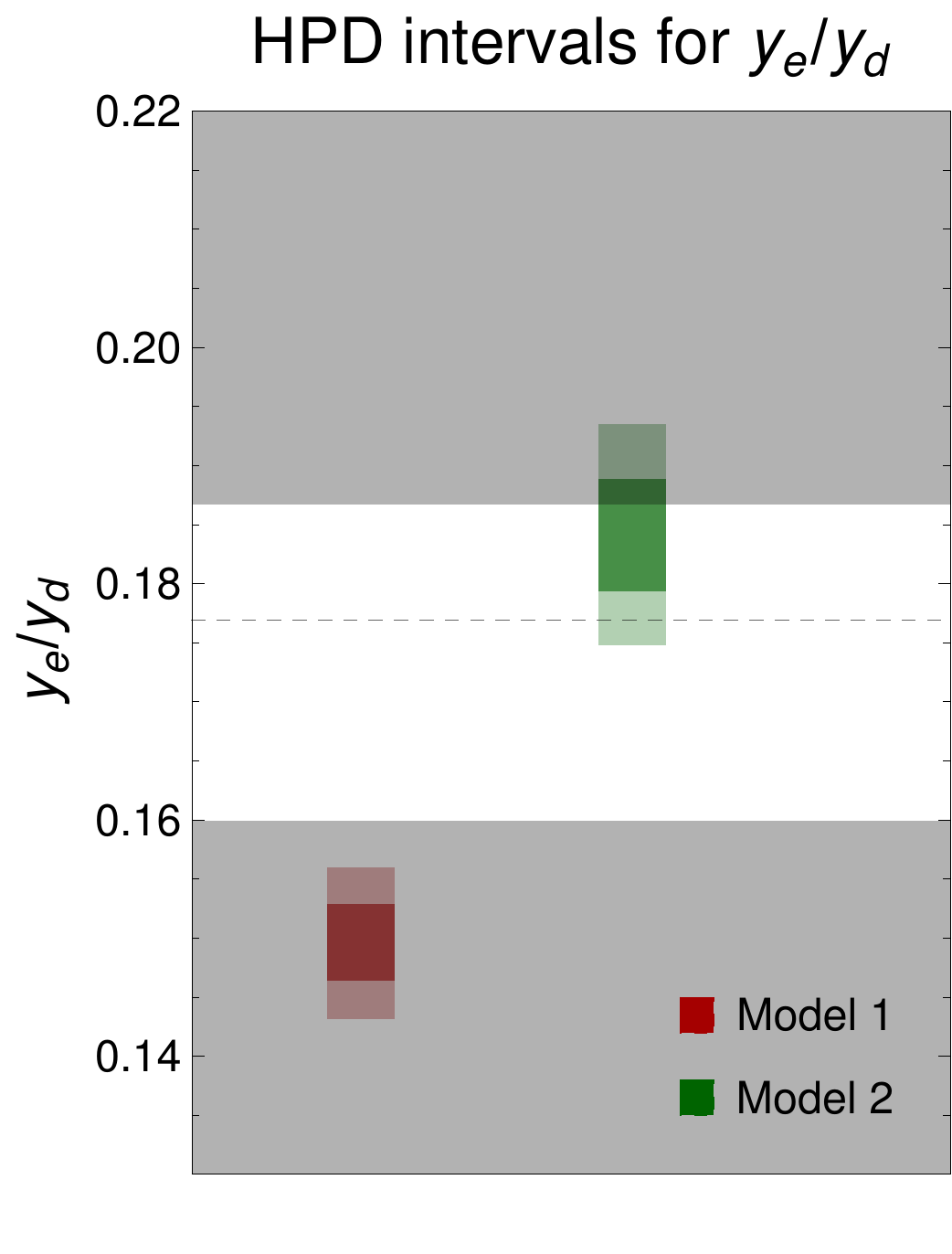}\hspace{6mm}
\includegraphics[width=5cm]{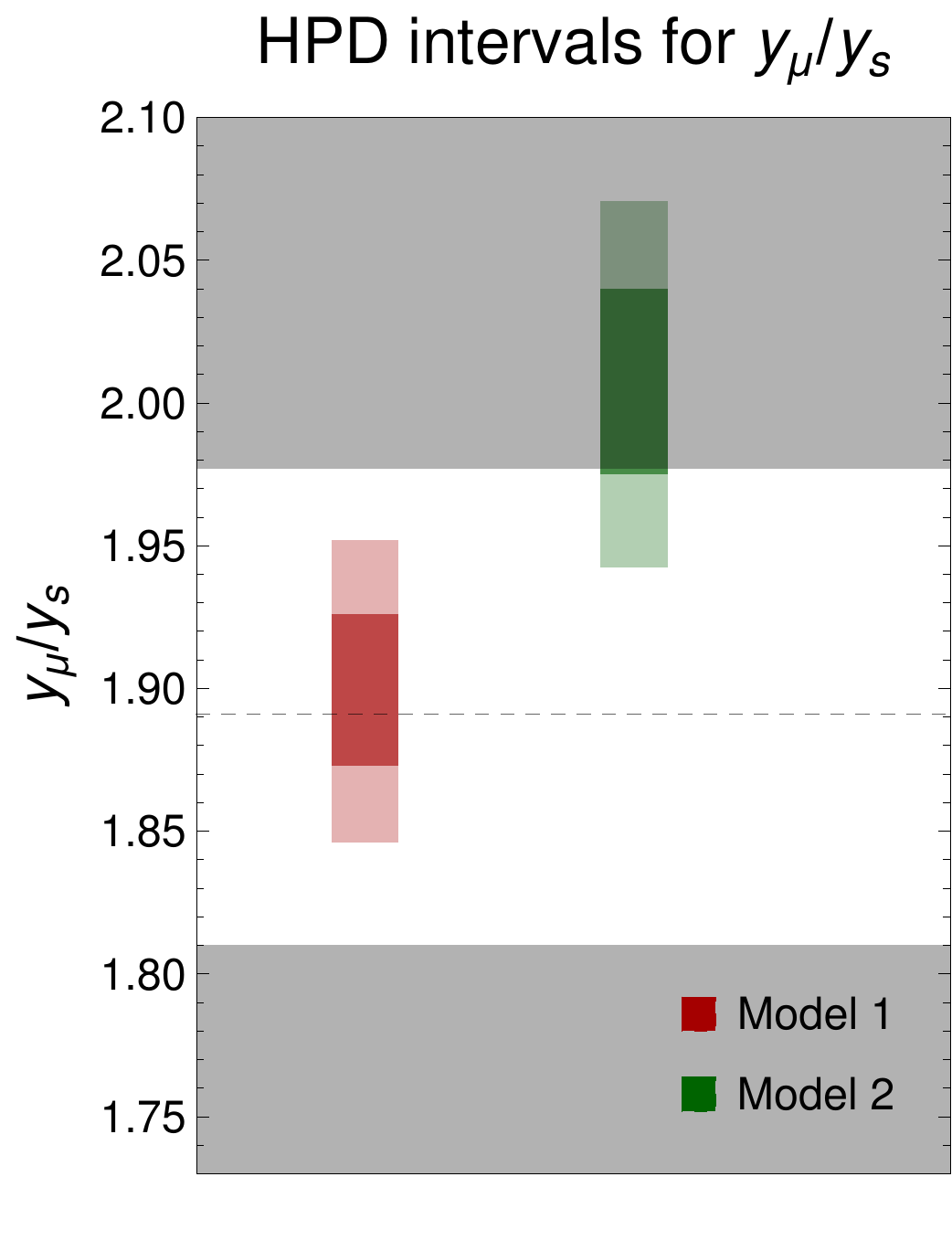}\hspace{6mm}
\includegraphics[width=5cm]{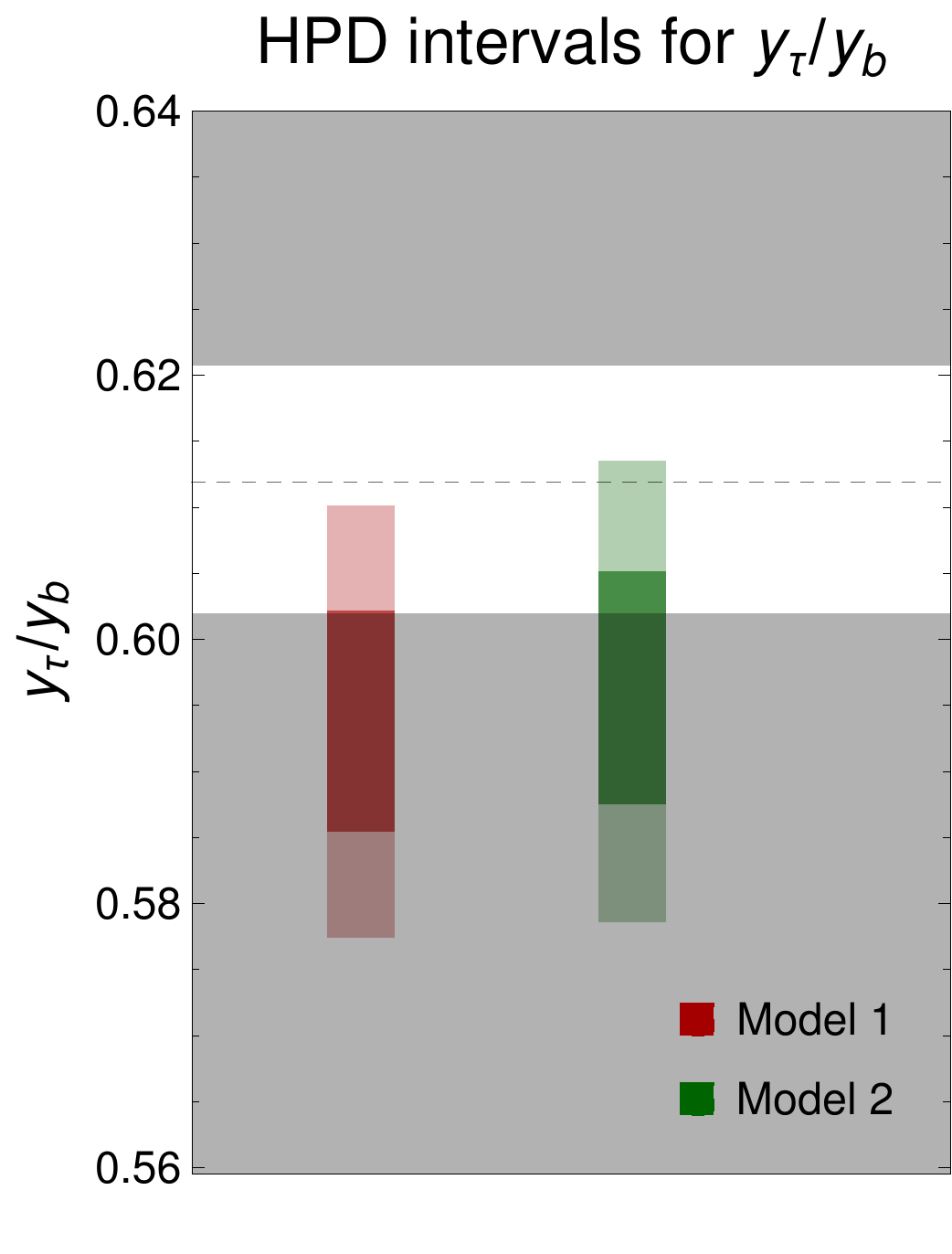}
\caption{The 1-$\sigma$ (dark) and 2-$\sigma$ (light) HPD intervals of $y_e/y_d$, $y_\mu/y_s$ and $y_\tau/y_b$ for model 1 and model 2 with $a = {\cal O}(1)$ and the CSD varieties $\phi_{120}$ and $\phi_{131}$, respectively. The dashed lines represent the current experimental central values, while the grey areas indicate the regions outside the experimental 1-$\sigma$ bounds.}\label{fig: HPD Ye/Yd}
\end{figure}

The 1-$\sigma$ and 2-$\sigma$ HPD intervals of the ratios $y_e/y_d$, $y_\mu/y_s$ and $y_\tau/y_b$ at $\MZ$ for model 1 and model 2 with $a = {\cal O}(1)$ are shown in Figure~\ref{fig: HPD Ye/Yd}. Since the posterior densities of these ratios do approximately not depend on the CSD variety, we show the HPD intervals only for one selected CSD variety for each model, namely $\phi_{120}$ for model 1 and $\phi_{131}$ for model 2. The 1-$\sigma$ and 2-$\sigma$ HPD intervals are indicated by dark and light colors, respectively, and are colored red (green) for model 1 (2). Moreover, indicated by dashed lines  the current experimental central values are shown, while the grey areas represent the regions outside the experimental 1-$\sigma$ bounds. Note, that as discussed in Section~\ref{sec: input parameters}, for the charged lepton Yukawa couplings we use a standard deviation of at least 1\% to take into account theoretical uncertainties, e.g.\ from the limited precision of the RG running. 

Figure~\ref{fig: HPD Ye/Yd} shows that model 1 predicts the ratio $y_e/y_d$ to be a bit smaller than the experimental 1-$\sigma$ range. On the other hand, the prediction by model 2 of this ratio spans from the experimental central value to the upper bound of the experimental 1-$\sigma$ range. The ratio $y_\mu/y_s$ is predicted to be within the 1-$\sigma$ range and around the upper bound of the 1-$\sigma$ range by the models 1 and 2, respectively. Both models predict a large range for the ratio $y_\tau/y_b$ which can (almost) be at the experimental central value.

In summary, our theoretical predictions of the ratios between charged lepton and down-type quark masses are well compatible with the experimental bounds.

\item \textbf{Results for the neutrino sector} 
\begin{figure}[t]\centering
\includegraphics[width=7cm]{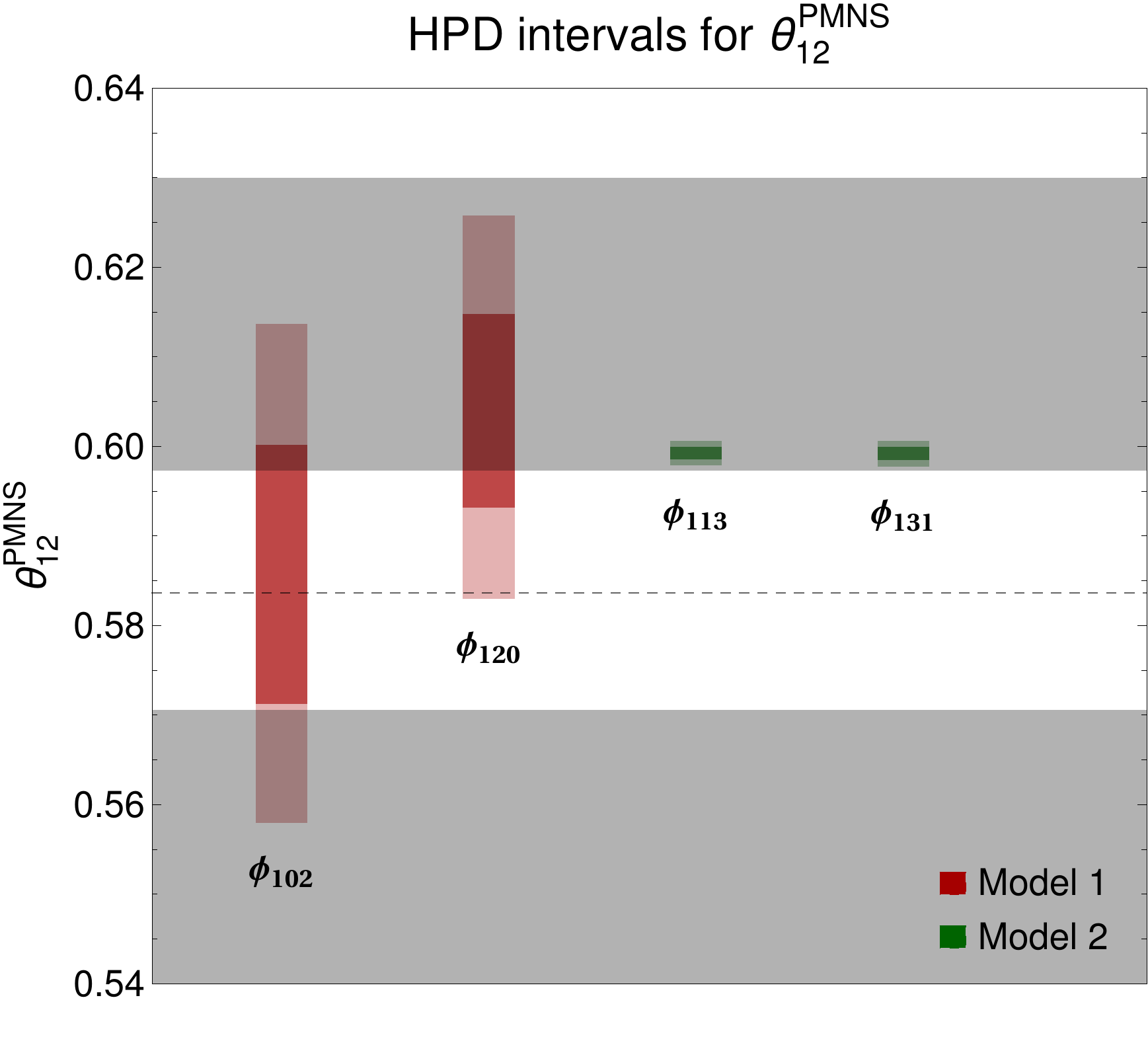}\hspace{6mm}
\includegraphics[width=7cm]{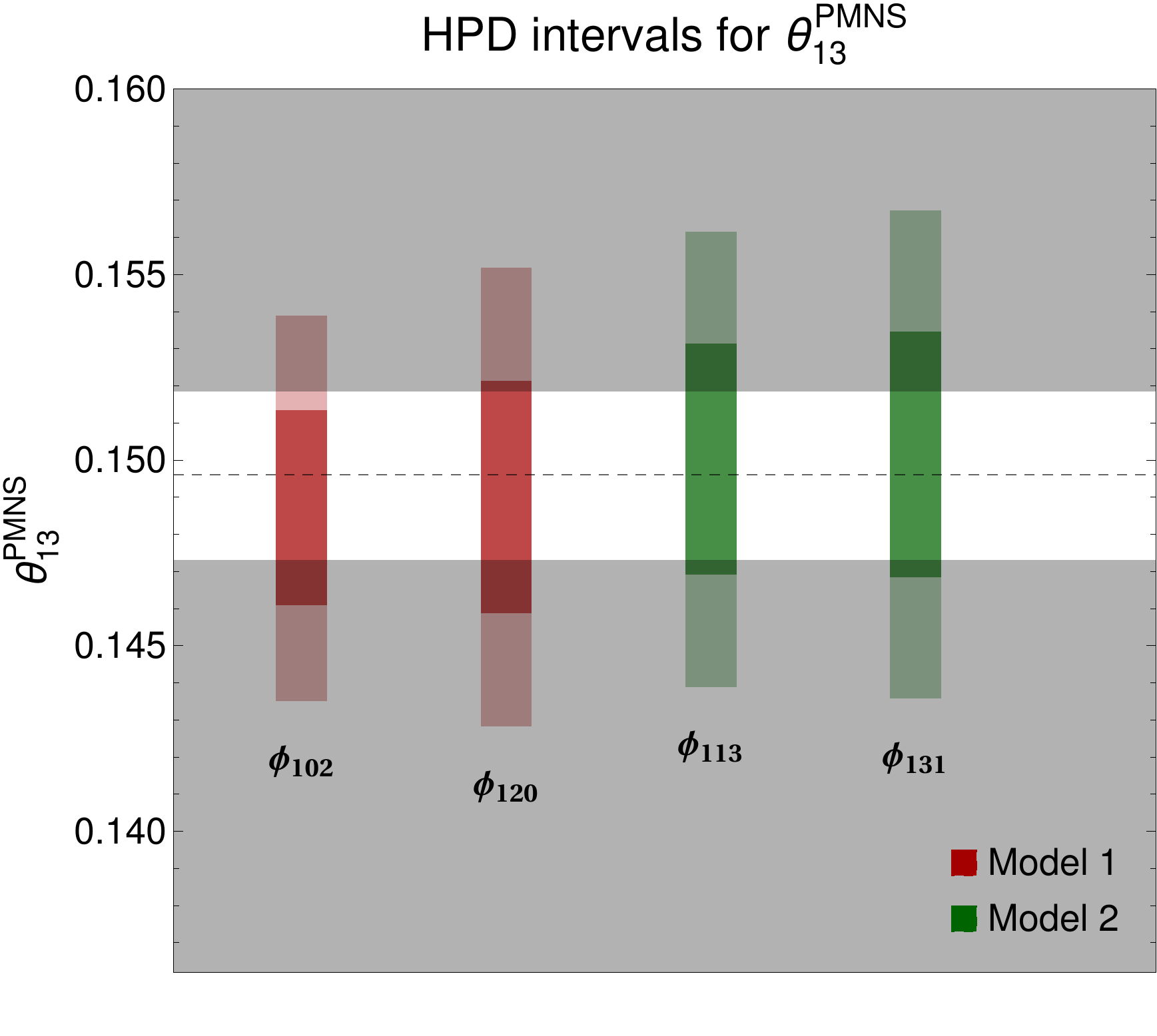}\\\vspace{6mm}

\includegraphics[width=7cm]{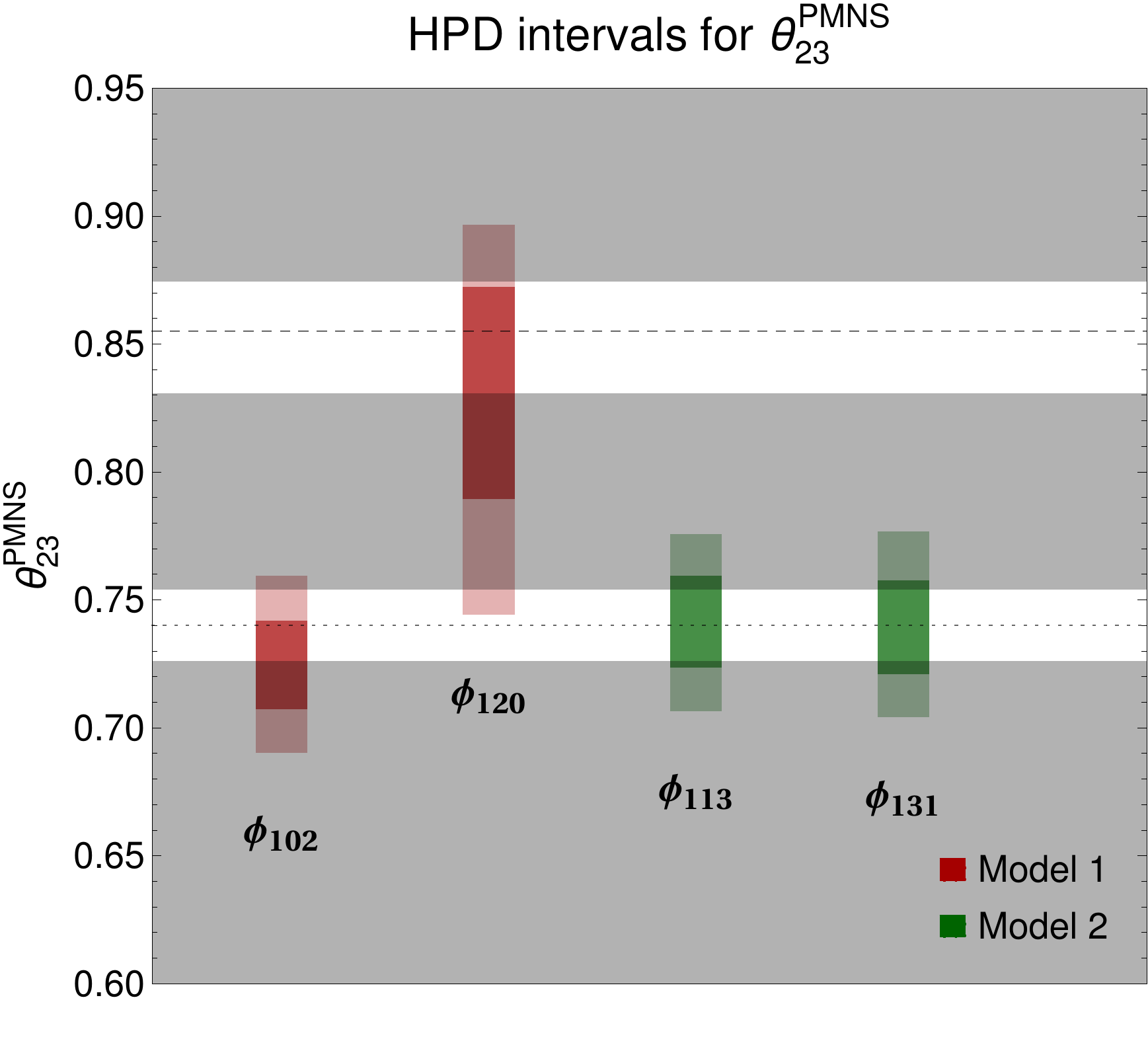}\hspace{9mm}
\includegraphics[width=6.7cm]{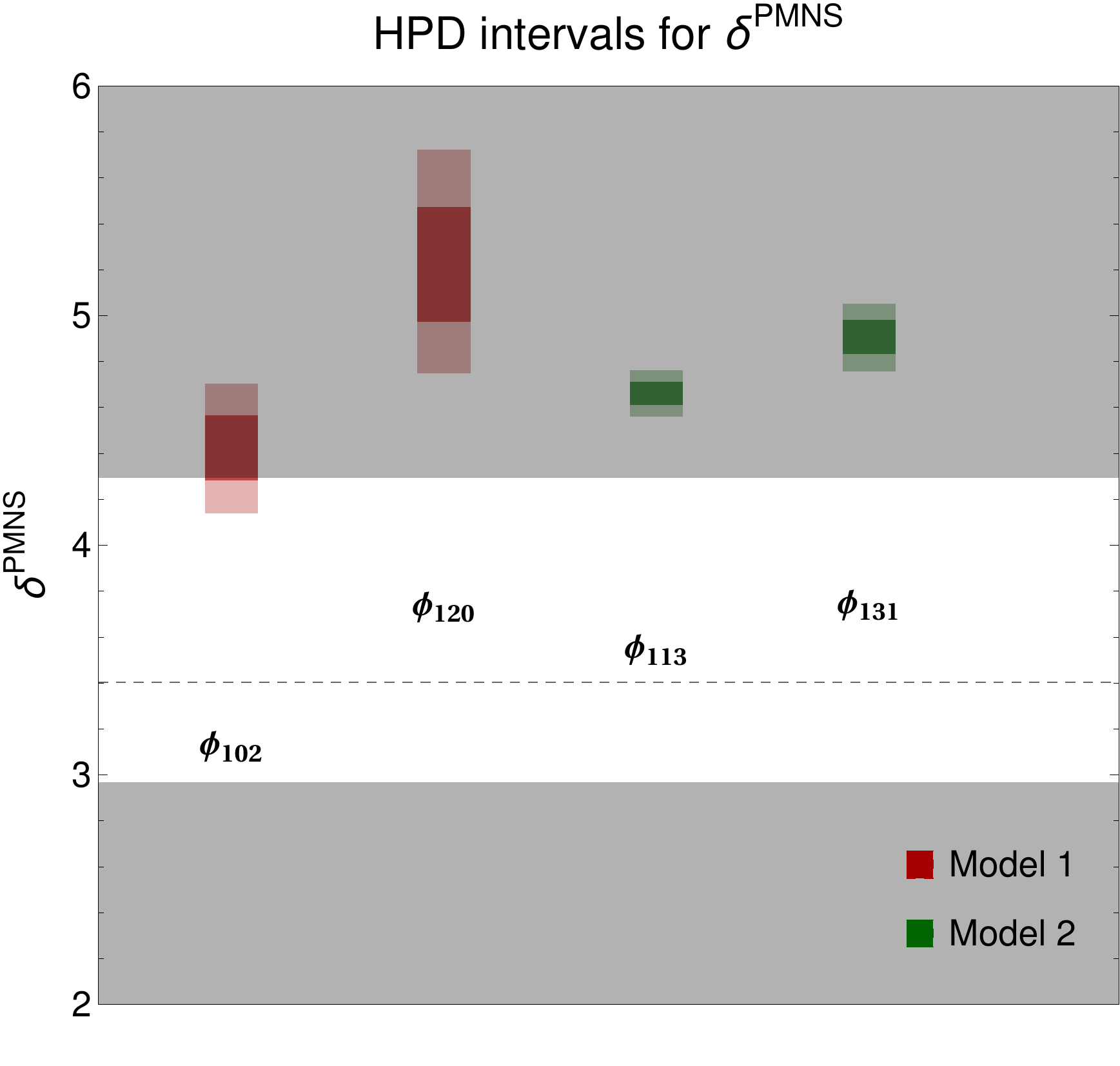}
\caption{The 1-$\sigma$ (dark) and 2-$\sigma$ (light) HPD intervals of the neutrino observables $\theta_{12}\PMNS$, $\theta_{13}\PMNS$, $\theta_{23}\PMNS$ and $\delta\PMNS$ for both CSD varieties of both models with $a = {\cal O}(1)$.  The current experimental central values are indicated by the dashed lines, while the regions outside the experimental 1-$\sigma$ range is represented by the grey areas. The plot for $\theta_{23}\PMNS$ has two white regions representing the two minima of the current experimental $\chi^2$ function. The dotted line indicates the second (local) minimum, whereas the dashed line represents the global minimum.}\label{fig: HPD neutrino}
\end{figure}

Figure~\ref{fig: HPD neutrino} visualizes the 1-$\sigma$ and 2-$\sigma$ HPD intervals of the neutrino observables $\theta_{12}\PMNS$, $\theta_{13}\PMNS$, $\theta_{23}\PMNS$ and $\delta\PMNS$ at $M_Z$, represented by dark and light colors, respectively, for both CSD varieties of both models (with $a = {\cal O}(1)$). The current experimental central values are indicated by dashed lines, whereas the areas outside the experimental 1-$\sigma$ range is represented by the grey areas. Since the $\chi^2$ function of $\theta_{23}\PMNS$ has two minima, its plot contains two white regions. The global minimum of the experimental $\chi^2$ function of $\theta_{23}\PMNS$ is indicated by a dashed line, while the local minimum is represented by a dotted line. 

As it can be seen from Figure~\ref{fig: HPD neutrino}, the predicted range for $\theta_{12}\PMNS$ is much wider in model 1 than in model 2. The reason for this is that while in model 2 this observable is mainly fixed by the CSD3 structure of $\textbf{Y}_\nu$, in model 1 using CSD2 there is also a contribution to $\theta_{12}\PMNS$  from the charged lepton mixing matrix $\textbf{E}_L$ defined in Eq.~\eqref{eq: singular value decomposition}. While for model 1 the predicted 1-$\sigma$ range of $\theta_{12}\PMNS$ has overlap with the current experimental 1-$\sigma$ range, for model 2 $\theta_{12}\PMNS$ is predicted to be slightly above the current experimental 1-$\sigma$ range. For both models the predicted 1-$\sigma$ range of $\theta_{13}\PMNS$ approximately coincides with the current experimental 1-$\sigma$ range. Concerning $\theta_{23}\PMNS$ in model 1 the predicted range depends on the choice of the CSD2 variety. For $\phi_{102}$ it overlaps with the second minimum of the experimental $\chi^2$ distribution, while for $\phi_{120}$ it overlaps with the first minimum, but since the prediction in this case is not sharp it almost spans to the second minimum. In model 2, both CSD3 varieties can yield a prediction of $\theta_{23}\PMNS$ matching the local minimum (with lower value of $\theta_{23}\PMNS$). Regarding $\delta\PMNS$, while for model 1 the case $\phi_{102}$ yields a predicted 1-$\sigma$ range approximately overlapping  the current experimental 1-$\sigma$ range, for $\phi_{120}$ the predicted 1-$\sigma$ range lies above the upper bound of the experimental 1-$\sigma$ range. Furthermore, both models predict $\delta\PMNS$ to be at or above the current experimental 1-$\sigma$ range. In model 2, both cases of the CSD3 variety predict $\delta\PMNS$ to be somewhat above the current experimental 1-$\sigma$ range. Note however that at higher $\sigma$-level the experimental range becomes very wide, so model 2 is not excluded by this discrepancy.

\item \textbf{Results for the GUT scale and the masses of the added scalar fields}
\begin{figure}[t]
\centering
\includegraphics[width=7cm]{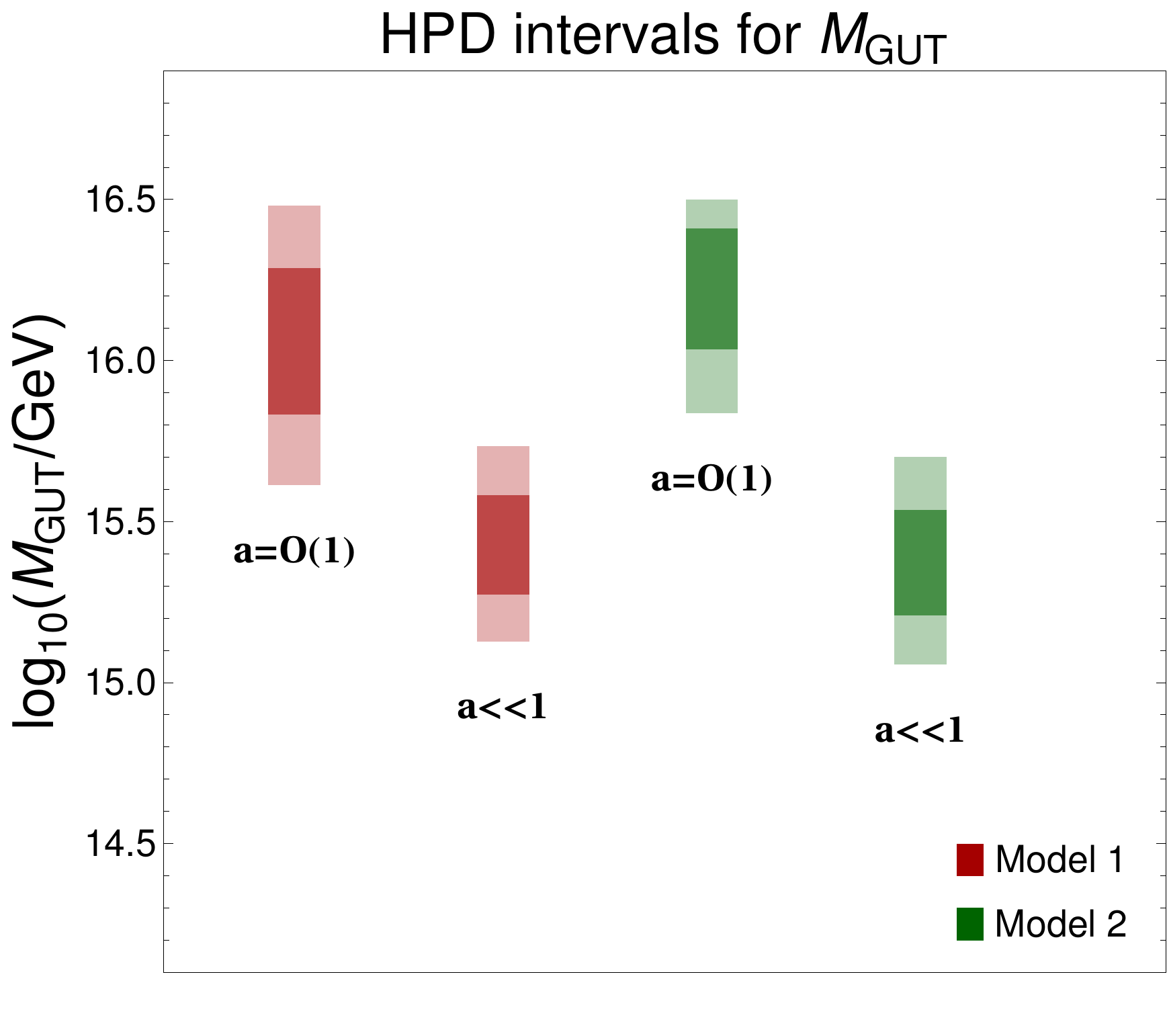}\hspace{6mm}
\includegraphics[width=7cm]{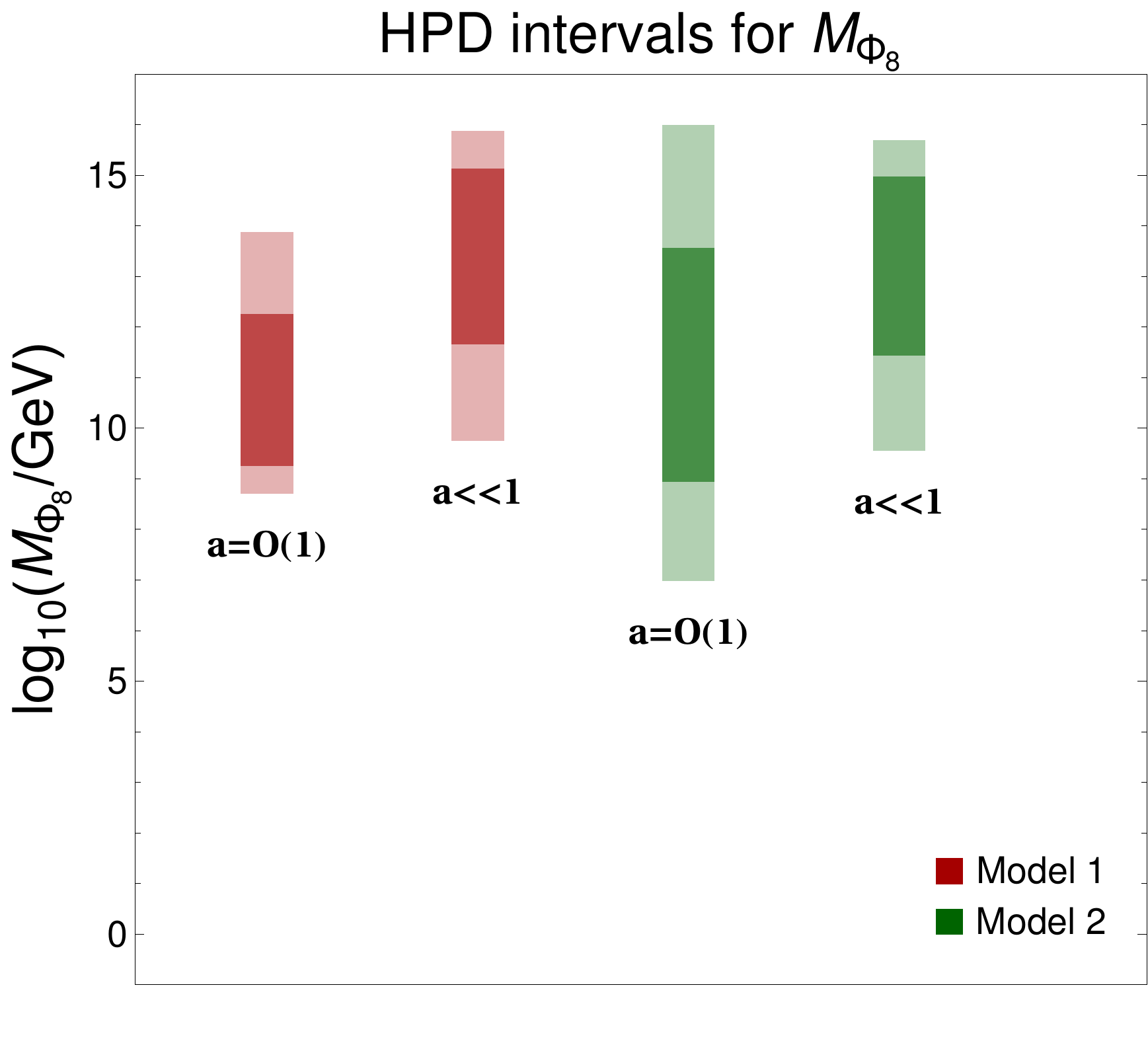}\\\vspace{6mm}
\includegraphics[width=7cm]{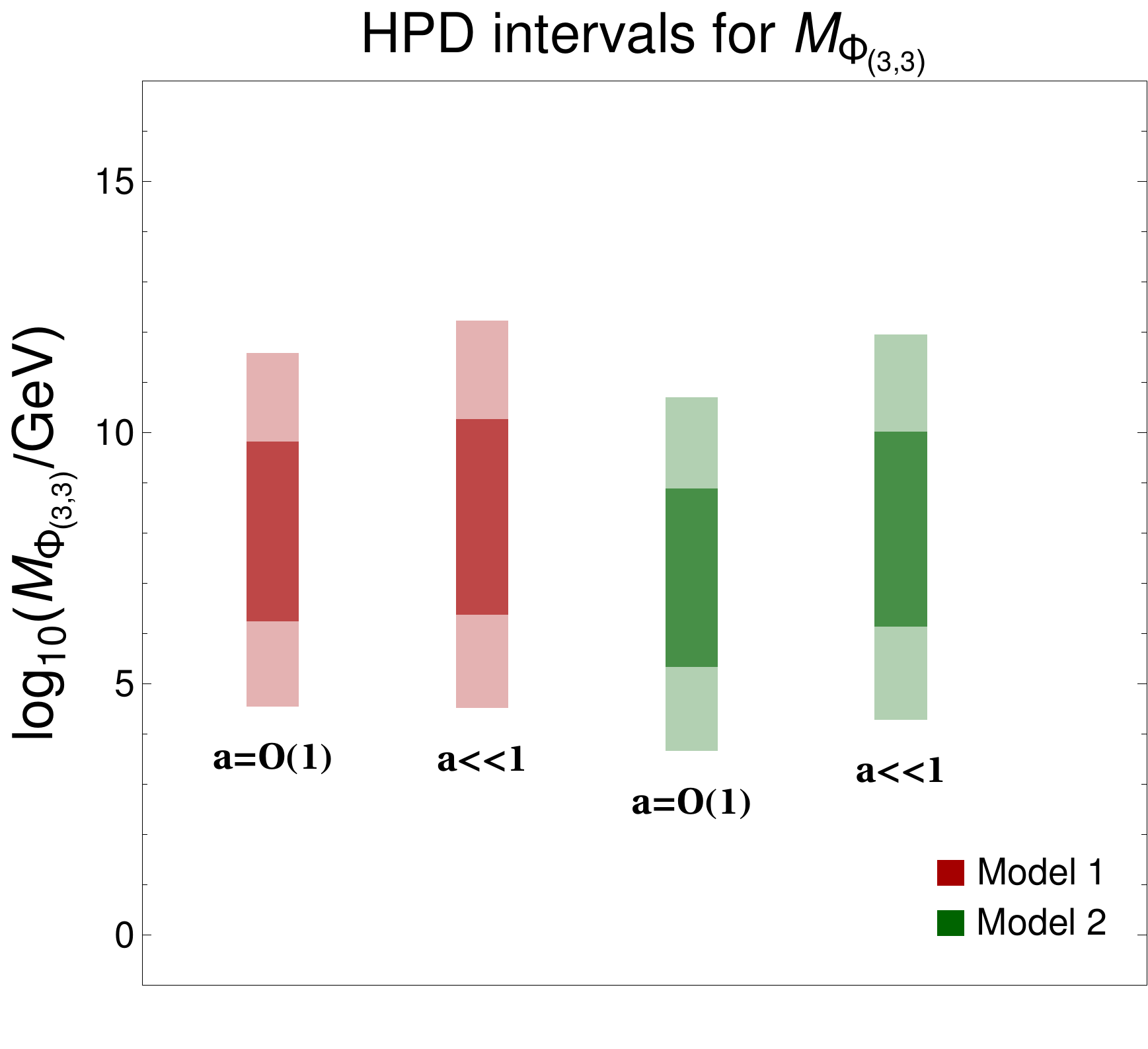}\hspace{6mm}
\includegraphics[width=7cm]{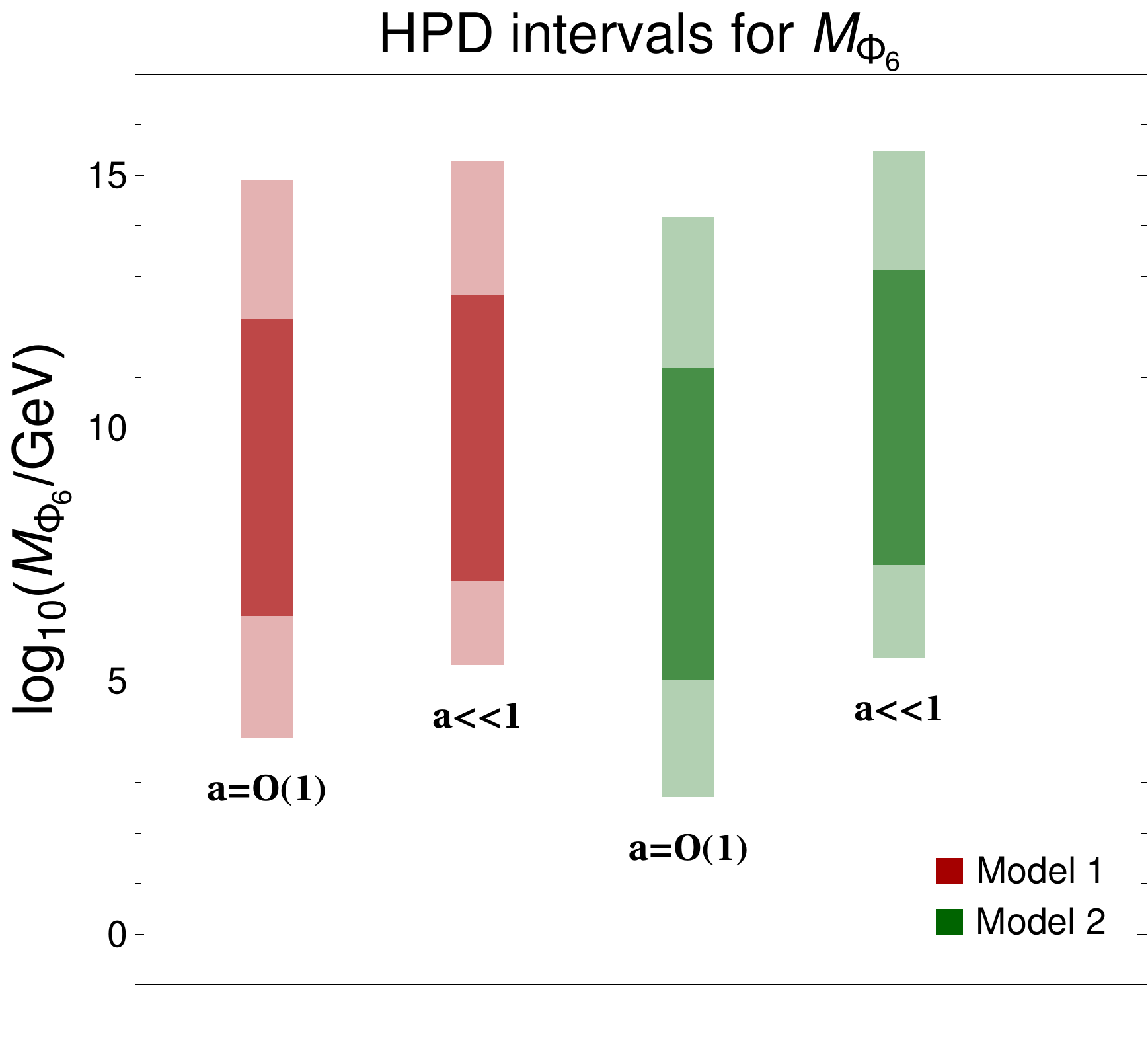}
\caption{The 1-$\sigma$ (dark) and 2-$\sigma$ (light) HPD intervals of the GUT scale $\MGUT$, and the masses of the added scalar fields $M_{\Phi_8}$, $M_{\Phi_{(3,3)}}$, $M_{\Phi_6}$.}\label{fig: HPD masses parameters}
\vspace{12mm}
\includegraphics[width=5cm]{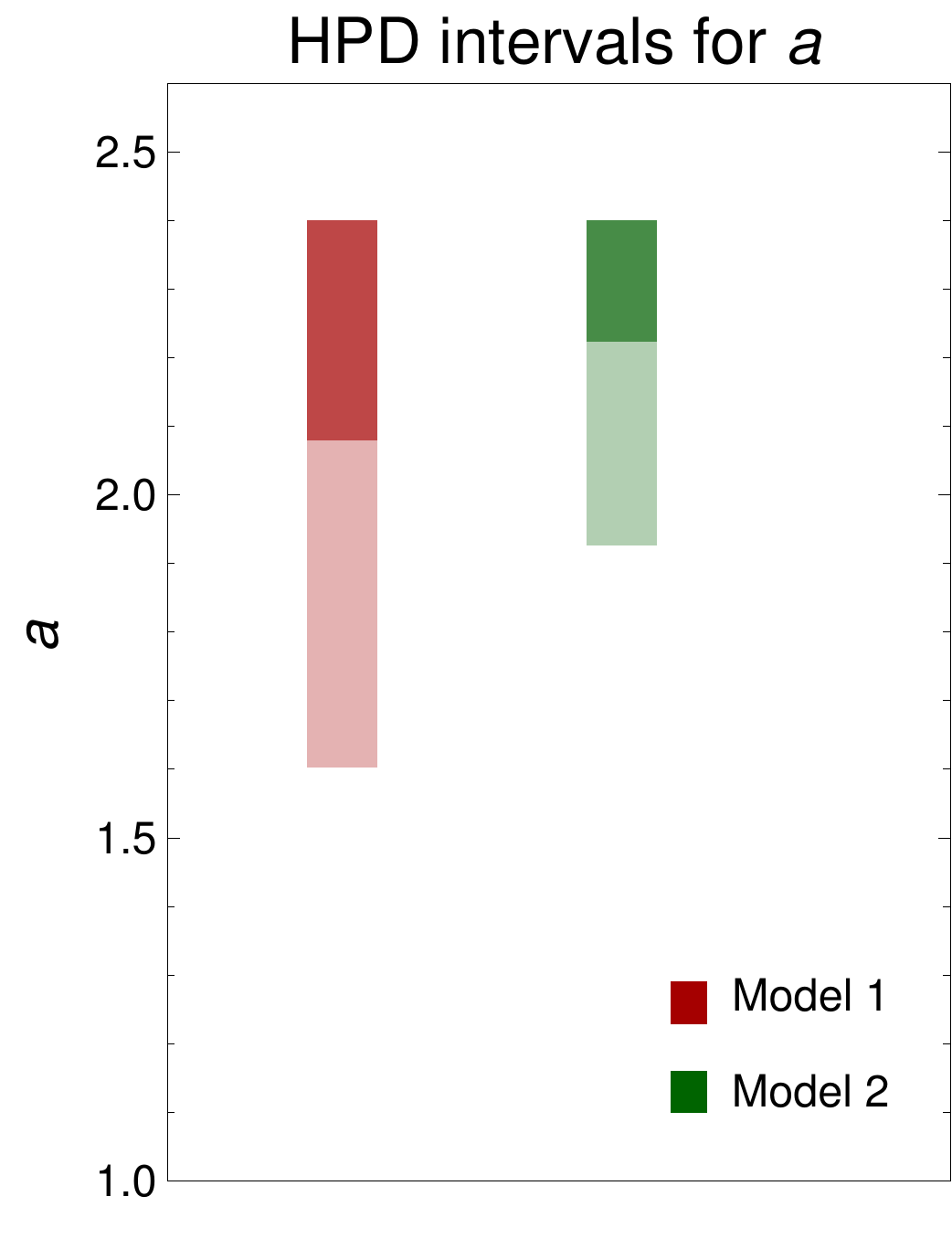}\hspace{2.4cm}
\includegraphics[width=5cm]{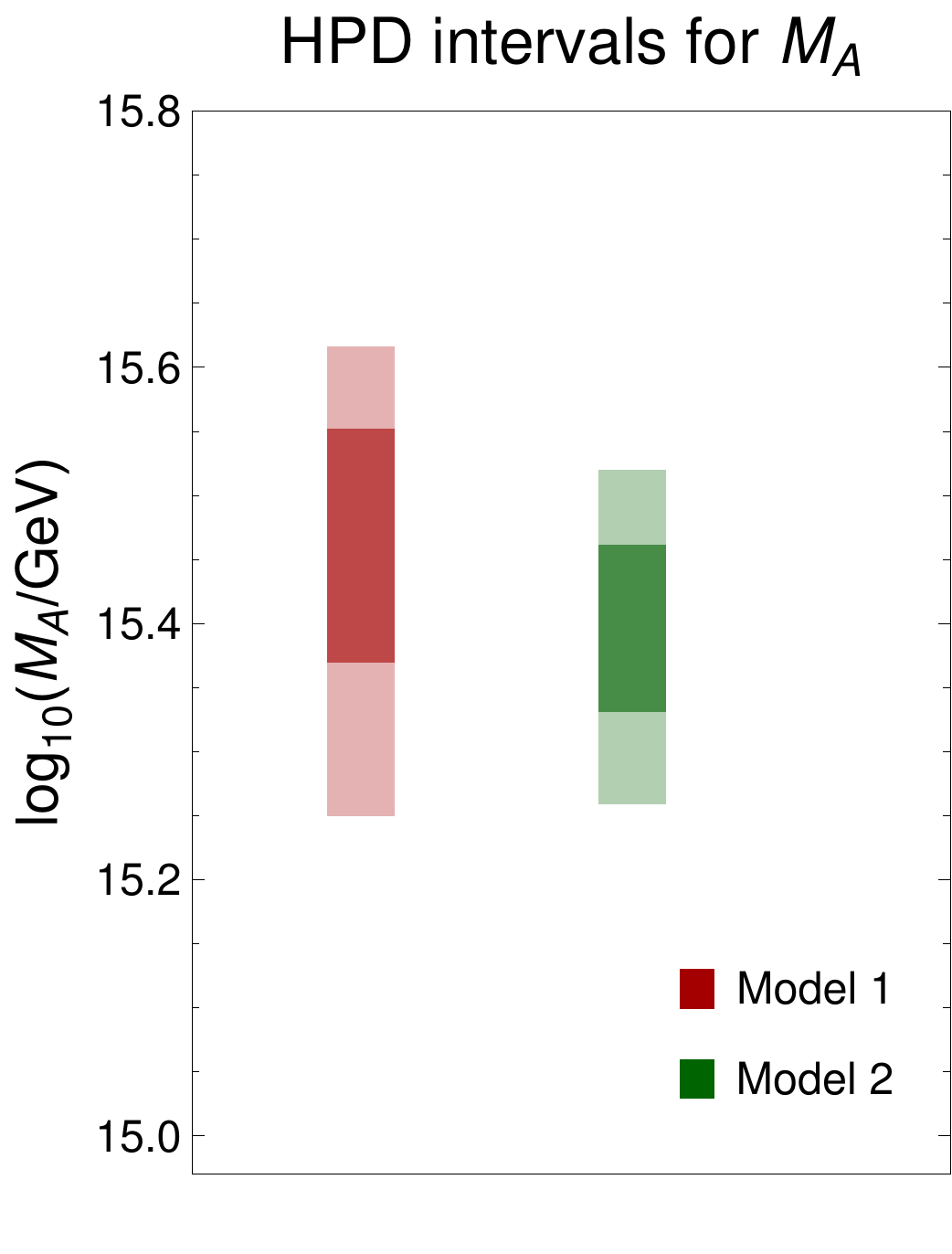}
\caption{The 1-$\sigma$ (dark) and 2-$\sigma$ (light) HPD intervals of the GUT scale value of the neutrino coupling $a$, and the right-handed neutrino mass $M_A$.}\label{fig: HPD neutrino parameter}

\end{figure}

Figure \ref{fig: HPD masses parameters} shows the 1-$\sigma$ (dark) and 2-$\sigma$ (light) HPD intervals of the GUT scale $\MGUT$ and of the masses of the added scalar fields $M_{\Phi_8}$, $M_{\Phi_6}$ and $M_{\Phi_{(3,3)}}$ for model 1 (red) and 2 (green). For both models the cases $a \ll 1$ and $a = {\cal O}(1)$ are shown. While for model 1 we chose the CSD2 variety $\phi_{120}$, for model 2 the results for the CSD3 variety $\phi_{131}$ are shown. 

From Figure \ref{fig: HPD masses parameters} it can be clearly seen that the GUT scale $\MGUT$ is predicted to be significantly higher in the case $a = {\cal O}(1)$ compared to the case $a \ll 1$ for both models. This reinforces the results we got for the best-fit points (cf.\ Section~\ref{sec: best-fit points}). The reason for this is --- as discussed in Section~\ref{sec: best-fit points} --- that a higher GUT scale leads to better predictions for $y_\tau/y_b$ in the case $a = {\cal O}(1)$. On the other hand, there is no large difference in the prediction of the GUT scale between the two models if the same case is considered. 

For the masses of the added scalar fields we get relatively similar predictions for both cases and for both models. Clearly, the masses of the extra particles can be above the reach of present and planned particle collider experiments, but discoveries could also be possible. Comparing the 2-$\sigma$ HPD results of these masses from the 2-loop calculation with the allowed ranges from the 1-loop estimation we did in Section~\ref{sec: gauge coupling unification}, we notice that for $M_{\Phi_{(3,3)}}$, in the case of the 2-loop calculation, larger values are allowed than for the 1-loop estimation. 

\item \textbf{Results for the neutrino parameters}

The 1-$\sigma$ (dark) and 2-$\sigma$ (light) HPD intervals of the neutrino coupling $a$ and the right-handed neutrino mass $M_A$ at the GUT scale $\MGUT$ for model 1 (red) and model 2 (green) for the case $a = {\cal O}(1)$ and the CSD variety $\phi_{120}$ (model 1), respectively $\phi_{131}$ (model 2) are shown in Figure~\ref{fig: HPD neutrino parameter}.

In both models the neutrino coupling $a$ is bounded from the top by 2.4 since we set this bound by hand, as discussed in Section~\ref{sec: input parameters}. The lower bound is in model 1 predicted to be smaller than in model 2. At 1-$\sigma$ (2-$\sigma$) it is given by 2.08 (1.60) in model 1, respectively 2.22 (1.93) in model 2.

The prediction right-handed neutrino mass $M_A$ is similar in both models, but in model 1 the upper limit of the 1-$\sigma$ and 2-$\sigma$ regions are larger than in model 2.

\item \textbf{Results for the nucleon decay rates}

We now compute the HPD intervals of the 13 different nucleon decay channels listed in Table~\ref{tab: nucleon decay channels experimental bounds}. These predictions are approximately independent of the choice of the CSD2 or CSD3 variety. We therefore choose the variety $\phi_{120}$ ($\phi_{131}$) for model 1 (2). The main contribution to the theoretical uncertainty comes from the GUT scale $\MGUT$ since the decay rates are proportional to the inverse of the GUT scale to the fourth power $\Gamma \propto (\MGUT)^{-4}$. However, the ratios of specific decay channels are independent on the GUT scale $\MGUT$. We will therefore also analyze these quantities to discriminate between the two models. 

Figure~\ref{fig: HPD nucleon decay rates}~and~\ref{fig: HPD ratios nucleon decay rates} show the 1-$\sigma$ (dark) and 2-$\sigma$ (light) HPD intervals of the various decay channels and selected ratios of specific decay channels, respectively. For model 1 (2) the HPD intervals are colored red (green). In Figure~\ref{fig: HPD nucleon decay rates} the current experimental bounds are indicated by blue lines.

Figure~\ref{fig: HPD nucleon decay rates} shows that none of the two models is excluded by the current experimental bounds on the decay rates of the different nucleon decay channels. 
In both models the dominant proton decay channel is $p\rightarrow \pi^0e^+$, while the dominant neutron decay channel is $n\rightarrow \pi^-e^+$. With the future sensitivity of $2.0\cdot 10^{-67}\,\GeV$ for the lifetime of the decay channel $p\rightarrow \pi^0e^+$ at HyperKamiokande after 15 years of runtime \cite{Abe:2018uyc}, part of the predicted 2-$\sigma$ region will be tested. Further channels close to the experimental bound are $p\rightarrow K^0\mu^+$, $p\rightarrow K^+\bar\nu$ and $p\rightarrow \pi^0\mu^+$. The decays channels with a pion in the final state as well as the ones with a kaon always dominate that of the eta meson. Figure~\ref{fig: HPD nucleon decay rates} further indicates that for a given decay channel the 2-$\sigma$ HPD intervals always overlap. Also, the predicted ranges are not sharp --- the 2-$\sigma$ HPD range spans over 3 to 4 orders of magnitude for each decay channel. The reason for this is the dependence of the decay rates on the inverse of the GUT scale to the fourth power. Therefore, with a measurement of a single decay channel alone it is not possible to discriminate between the two models. Instead, the ratios between different decay channels are needed.

To draw Figure~\ref{fig: HPD ratios nucleon decay rates} we first computed for each point in the MCMC dataset the ratio and only then determined the HPD intervals. There are more than 30 ratios with non-overlapping 2-$\sigma$ HPD intervals. On the other hand, there are several decay widths for which an exact ratio is predicted because of the form of the low-energy effective Lagrangian. The following ratios have in every SU(5) GUT the exact value $1/2$:
\begin{equation}\label{eq: fixed ratios 1}
\frac{\Gamma_{p\rightarrow \pi^0e^+}}{\Gamma_{n\rightarrow \pi^-e^+}},
\hspace{10mm}
\frac{\Gamma_{p\rightarrow \pi^0\mu^+}}{\Gamma_{n\rightarrow \pi^-\mu^+}},
\hspace{10mm}
\frac{\Gamma_{n\rightarrow \pi^0\bar\nu}}{\Gamma_{p\rightarrow \pi^+\bar\nu}}.
\end{equation}
Moreover, since in this paper we only consider gauge boson mediated nucleon decay and neglect any contributions to the decay widths from the scalar fields, the following ratios are further fixed in our two models:
\begin{equation}\label{eq: fixed ratios 2}
\frac{\Gamma_{p\rightarrow \pi^0e^+}}{\Gamma_{p\rightarrow \eta^0e^+}},
\hspace{10mm}
\frac{\Gamma_{p\rightarrow \pi^0\mu^+}}{\Gamma_{p\rightarrow \eta^0\mu^+}},
\hspace{10mm}
\frac{\Gamma_{p\rightarrow \pi^+\bar\nu}}{\Gamma_{n\rightarrow \eta^0\bar\nu}}.
\end{equation}
Their exact value depends on the coupling constants for the interaction between baryons and mesons $F\approx0.463$ and $D\approx0.804$ (cf.\ \cite{Nath:2006ut, Cabibbo:2003cu}) and is given by:
\begin{equation}
\frac{(1+F+D)^2}{3(-\frac{1}{3}+F-\frac{1}{3}D)^2}\approx197.66\,.
\end{equation}

\begin{figure}[H]\centering
\includegraphics[width=16.5cm]{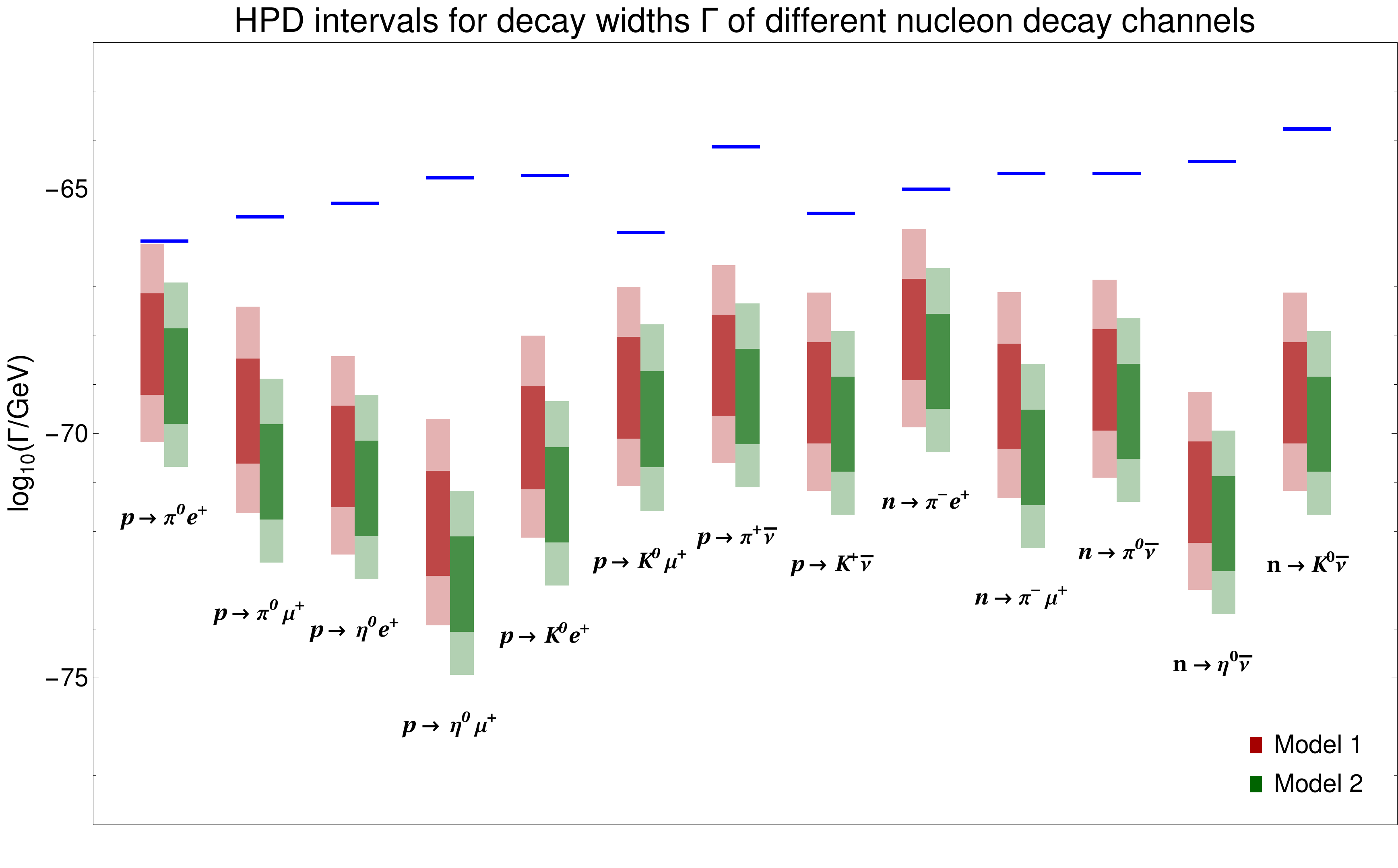}
\caption{The 1-$\sigma$ (dark) and 2-$\sigma$ (light) HPD intervals of the nucleon decay rates for model 1 and model 2 with $a = {\cal O}(1)$ and the CSD varieties $\phi_{120}$ and $\phi_{131}$, respectively. The experimental bounds are represented by the blue lines.\vspace{6mm}}\label{fig: HPD nucleon decay rates}
\includegraphics[width=16.5cm]{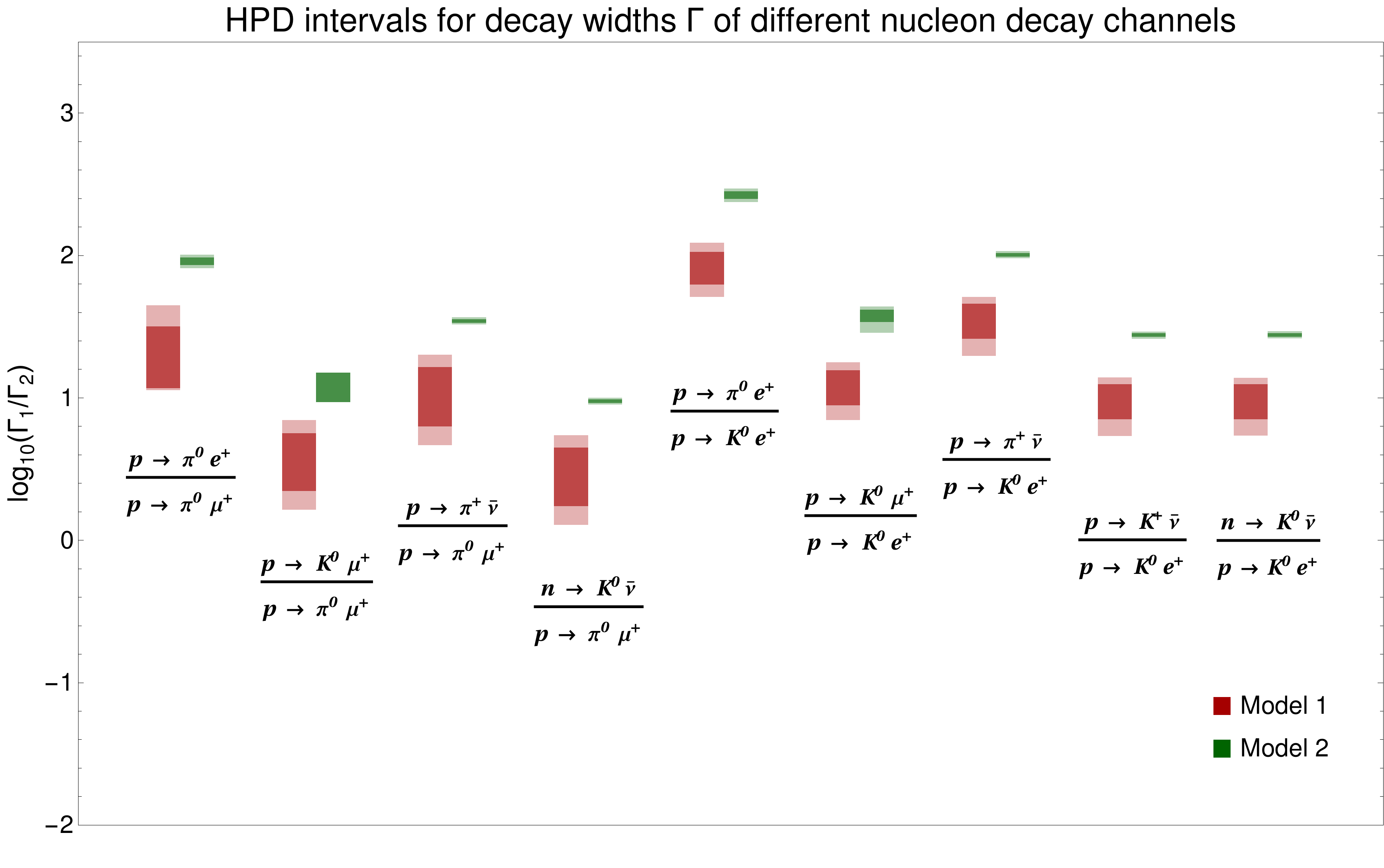}
\caption{The 1-$\sigma$ (dark) and 2-$\sigma$ (light) HPD intervals of selected ratios of nucleon decay rates for model 1 and model 2 with $a = {\cal O}(1)$ and the CSD varieties $\phi_{120}$ and $\phi_{131}$, respectively.}\label{fig: HPD ratios nucleon decay rates}
\end{figure}

Finally, by combining ratios from Eq.~\eqref{eq: fixed ratios 1} with the ratios listed in Eq.~\eqref{eq: fixed ratios 2} further fixed ratios can be constructed.

In Figure~\ref{fig: HPD ratios nucleon decay rates} we picked for each fixed ratio defined in Eq.~\eqref{eq: fixed ratios 1}~and~\eqref{eq: fixed ratios 2} one representative giving us nine independent ratios with non-overlapping 2-$\sigma$ HPD intervals. All other ratios with non-overlapping 2-$\sigma$ HPD intervals can then be reconstructed from the ones given in Figure~\ref{fig: HPD ratios nucleon decay rates} using Eq.~\eqref{eq: fixed ratios 1}~and~\eqref{eq: fixed ratios 2}. Measuring one such ratio already allows to discriminate between the two models. On the other hand, by experimentally determining multiple ratios a model can be excluded. 

The decay channels which are predicted to be close to the experimental bounds deliver candidates for the most interesting ratios. Furthermore, upcoming experiments are more probable to measure ratios close to $1:1$  than ratios in which the enumerator and denominator differ by several orders of magnitude. Therefore, the most interesting ratios are the ones where one of the three decay channels $p\rightarrow \pi^0\mu^+$, $n\rightarrow \pi^-\mu^+$ and $p\rightarrow K^0e^+$ is compared to a decay channel with a pion in the final state. 

Note that despite the different construction of the neutrino sector in the two models, all ratios between two decay channels both having neutrinos in the final states have overlapping 2-$\sigma$ intervals, and therefore do not allow to discriminate between the different models. The reason for this is that if neutrinos are in the final state, we sum over the neutrino flavor states, and thus the flavor information is lost.

\end{enumerate}

\section{Summary and conclusions}\label{sec:conclusions}

In this paper we investigated a novel non-supersymmetric SU(5) GUT scenario, where the masses of the third and second family down-type quarks and charged leptons both stem dominantly from single joint GUT operators (i.e.\ satisfying \textit{single operator dominance}), leading to the GUT scale predictions $y_\tau/y_b = 3/2$ and $y_\mu/y_s = 9/2$. 

To realize this scenario, we have extended the Georgi-Glashow SU(5) model by a 45-dimensional GUT-Higgs representation, such that the GUT scale predictions $y_\tau/y_b = 3/2$ and $y_\mu/y_s = 9/2$ can be realized from joint effective GUT operators, and the gauge couplings can meet at the high scale $M_\mathrm{GUT}$ when components of GUT representations have masses at intermediate scales. We have further added SU(5) singlet fermion representations to explain the observed neutrino masses via the type I seesaw mechanism. 

We found that the GUT scale predictions $y_\tau/y_b = 3/2$ and $y_\mu/y_s = 9/2$ emerging from this scenario can indeed be compatible with the low energy experimental data on the quark and charged lepton masses, with the best-fit to the data arising when renormalization group effects from the neutrino Yukawa matrix have an impact on the RG running of the charged fermion Yukawa couplings (cf.\ Figure \ref{fig: RG evolution of the gauge couplings and Yukawa ratios with non-zero neutrino Yukawa couplings}). 

We then extended this minimal scenario to two ``toy models'' towards explaining quark and lepton masses and mixings with different flavor structure. 
We performed MCMC analyses to confront the two ``toy models" with the available experimental data, with results presented and discussed in Section~\ref{sec: nucleon decay fingerprints of two toy models}. Regarding the nuclear decay rates, we highlighted that if several nucleon decay channels are observed in the future, the ratios between their partial decay rates (cf.\ Figure~\ref{fig: HPD ratios nucleon decay rates}) can serve as ``fingerprints'', allowing to separate between GUT models with different flavor structure.   

In summary, non-supersymmetric SU(5) GUTs realizing the GUT scale predictions $y_\tau/y_b = 3/2$ and $y_\mu/y_s = 9/2$ and neutrino masses with \textit{single operator dominance}  by adding an extra scalar 45-plet and fermionic singlets, can be interesting rather minimal and predictive scenarios of Grand Unification. For the future, it will be interesting to further extend them to fully realistic flavor models, and to study them in a cosmological context. As we have demonstrated, the predictions for the nucleon decay rates in GUT flavor models can provide useful ``fingerprints'' that might help testing such models.

\textcolor{black}{\section*{Acknowledgements}}
\noindent This work has been supported by the Swiss National Science Foundation.

\appendix
\appendixpage

\section{Renormalization group evolution of \textbf{Y}$_\nu$ above $\MGUT$}\label{sec: running of neutrino Yukawa coupling}
As we have seen in Section~\ref{sec: results} large neutrino couplings $a = {\cal O}(1)$ are helpful to improve the fit to the low energy experimental data on charged fermion masses. But since we assume that our Yukawa entries are generated by \textit{single operator dominance} (cf.\ Section~\ref{sec: toy models}) in the respective entry, assuming that higher order effective operators are suppressed, we have to require that our models remain perturbative some range above the GUT scale $\MGUT$.

In the following we want to estimate up to which energy scale $\mu$ above the GUT scale $\MGUT$ this is the case for our two ``toy models". For this we compute an approximation of the 1-loop RGE of the neutrino Yukawa matrix $\textbf{Y}_\nu$ above $\MGUT$, focussing on the contribution from $\textbf{Y}_\nu$ and the gauge coupling $g$. 

In the following $f,g=1,2,3$ are family indices, $i,j=1,\dots 5$ are fundamental SU(5) indices and $A,B=1,\dots,24$ are adjoint SU(5) indices, respectively. Moreover, $\xi$ is the gauge fixing parameter, $M_{\textbf{5}_H}$ is the mass of the scalar field $\textbf{5}_H$, $\textbf{T}^A$ are the SU(5) generators. The wave function renormalization constants $Z_i=1+\delta Z_i$ $(i\in\lbrace \textbf{1}_F, \mathbf{\bar 5}_F,\textbf{5}_H\rbrace)$ are defined in the usual way.

\subsection{Relevant Feynman rules}
We first list the relevant Feynman rules which are needed to compute the counter terms.

\begin{enumerate}
\item \textbf{Propagator}
\begin{equation}
\begin{split}
\vcenter{\hbox{\begin{tikzpicture}
\begin{feynman}
\vertex(a1){\(\textbf{1}_{Fi}\)};
\vertex[right=3cm of a1](a5){\(\textbf{1}_{Fj}\)};
\diagram* {
(a1)--[
] (a5),
};
\end{feynman}
\end{tikzpicture}}}\hspace{5mm} &=\hspace{3mm} \frac{i\slashed p+M_i}{p^2-M_i^2+i\epsilon}\delta_{ij}\\
\vcenter{\hbox{\begin{tikzpicture}
\begin{feynman}
\vertex(a1){\(\mathbf{\bar{5}}_{Fia}\)};
\vertex[right=3cm of a1](a5){\(\mathbf{\bar{5}}_{Fjb}\)};
\diagram* {
(a1)--[fermion, 
] (a5),
};
\end{feynman}
\end{tikzpicture}}}\hspace{5mm} &=\hspace{3mm} \frac{i\slashed p}{p^2+i\epsilon}\delta_{ji}\\
\vcenter{\hbox{\begin{tikzpicture}
\begin{feynman}
\vertex(a1){\(\mathbf{5}_{Ha}\)};
\vertex[right=3cm of a1](a5){\(\mathbf{5}_{Hb}\)};
\diagram* {
(a1)--[charged scalar, 
] (a5),
};
\end{feynman}
\end{tikzpicture}}}\hspace{5mm} &=\hspace{3mm} \frac{i}{p^2-M_{\textbf{5}_H}^2+i\epsilon}\delta_{ba}\\
\vcenter{\hbox{\begin{tikzpicture}
\begin{feynman}
\vertex(a1){\(X^A_{\mu}\)};
\vertex[right=3cm of a1](a5){\(X^B_{\nu}\)};
\diagram* {
(a1)--[boson, 
] (a5),
};
\end{feynman}
\end{tikzpicture}}}\hspace{5mm} &=\hspace{3mm} \frac{i(-\eta^{\mu\nu}+(1-\xi)\frac{p^\mu p^\nu}{p^2})}{p^2+i\epsilon}\delta_{BA}
\end{split}
\end{equation}

\item \textbf{Yukawa interaction}
\begin{equation}
\begin{split}
\vcenter{\hbox{\begin{tikzpicture}
\begin{feynman}
\vertex(a1){\(\textbf{1}_{Fj}\)};
\vertex[below=2.5cm of a1] (a3){\(\mathbf{\bar{5}}_{Fia}\)};
\vertex at ($(a1)!0.5!(a3)+(1.5cm,0)$) (a2);
\vertex[right=1.5cm of a2](a4){\(\mathbf{5}_{Hb}\)};
\diagram* {
(a3)--[fermion] (a2)--[] (a1),
(a4)--[charged scalar] (a2)
};
\end{feynman}
\end{tikzpicture}}}\hspace{1cm} &=\hspace{3mm} -i (\textbf{Y}_\nu)_{ji}\delta_{ba}
\end{split}
\end{equation}

\item \textbf{Gauge boson interactions}
\begin{equation}
\begin{split}
\vcenter{\hbox{\begin{tikzpicture}
\begin{feynman}
\vertex(a1){\(\mathbf{\bar 5}_{Fjb}\)};
\vertex[below=2.5cm of a1] (a3){\(\mathbf{\bar{5}}_{Fia}\)};
\vertex at ($(a1)!0.5!(a3)+(1.5cm,0)$) (a2);
\vertex[right=1.5cm of a2](a4){\(X^A_\mu\)};
\diagram* {
(a3)--[fermion] (a2)--[fermion] (a1),
(a2)--[boson] (a4)
};
\end{feynman}
\end{tikzpicture}}}\hspace{1cm} &=\hspace{3mm} \frac{i}{2}g\delta_{ij}\lambda^A_{ba} \gamma_\mu\\
\vcenter{\hbox{\begin{tikzpicture}
\begin{feynman}
\vertex(a1){\(\mathbf{5}_{Hb}\)};
\vertex[below=2.5cm of a1] (a3){\(\mathbf{5}_{Ha}\)};
\vertex at ($(a1)!0.5!(a3)+(1.5cm,0)$) (a2);
\vertex[right=1.5cm of a2](a4){\(X^A_\mu\)};
\diagram* {
(a3)--[charged scalar] (a2)--[charged scalar] (a1),
(a2)--[boson] (a4)
};
\end{feynman}
\end{tikzpicture}}}\hspace{1cm} &=\hspace{3mm} -\frac{i}{2}g(p^\mu+q^\mu)\lambda^A_{ba} \gamma_\mu\\
\end{split}
\end{equation}

\item \textbf{Counter terms of the two-point function}
\begin{equation}
\begin{split}
\vcenter{\hbox{\feynmandiagram[horizontal=a to b]{
a[particle=\(\textbf{1}_{Fi}\)]--[] c --[with arrow={0.1cm},segment length=0.5cm] d [crossed dot]--[]e--[with arrow={0cm}]b[particle=\(\textbf{1}_{Fj}\)],
};}}\hspace{5mm} &=\hspace{3mm} i\slashed{p}(\delta Z_{\textbf{1}_F})_{ji}-i(\delta Z_{\textbf{M}_R} \textbf{M}_R)_{ji} \\
\vcenter{\hbox{\feynmandiagram[horizontal=a to b]{
a[particle=\(\mathbf{\bar{5}}_{Fia}\)]--[] c --[with arrow={0.1cm},segment length=0.5cm] d [crossed dot]--[]e--[with arrow={0cm}]b[particle=\(\mathbf{\bar{5}}_{Fjb}\)],
};}}\hspace{5mm} &=\hspace{3mm} i\slashed{p}(\delta Z_{\mathbf{\bar{5}}_F})_{ji}\delta_{ba}\\
\vcenter{\hbox{\feynmandiagram[horizontal=a to b]{
a[particle=\(\mathbf{5}_{Ha}\)]--[scalar] c --[scalar,with arrow={0.1cm},segment length=0.5cm] d [crossed dot]--[scalar]e--[scalar,with arrow={0cm}]b[particle=\(\mathbf{5}_{Hb}\)],
};}}\hspace{5mm} &=\hspace{3mm} i(p^2\delta Z_{\textbf{5}_H}-\delta M_{\textbf{5}_H}^2)
\end{split}
\end{equation}

\item \textbf{Yukawa counter term}
\begin{equation}
\begin{split}
\vcenter{\hbox{\begin{tikzpicture}
\draw (1.5,-1.25) circle (4pt);
\draw (1.5,-1.25) node[cross]{};
\begin{feynman}
\vertex(a1){\(\textbf{1}_{F}\)};
\vertex[below=2.5cm of a1] (a3){\(\mathbf{\bar{5}}_{F}\)};
\vertex at ($(a1)!0.5!(a3)+(1.5cm,0)$) (a2);
\vertex[right=1.5cm of a2](a4){\(\mathbf{5}_{H}\)};
\diagram* {
(a3)--[fermion] (a2)--[] (a1),
(a4)--[charged scalar] (a2)
};
\end{feynman}
\end{tikzpicture}}}\hspace{1cm} &=\hspace{3mm} -i (\textbf{Y}_\nu \delta Z_{\textbf{Y}_\nu})_{ji}\delta_{ba}
\end{split}
\end{equation}

\end{enumerate}

\subsection{Calculation of the counter terms}
Using the Feynman rules listed in the previous section we now calculate the relevant counterterms in the MS-scheme.
\begin{enumerate}
\item \textbf{Wave function and mass of right-handed neutrino singlets}
\begin{equation}
\begin{split}
\vcenter{\hbox{\feynmandiagram[horizontal=a to b]{
a[particle=\(\textbf{1}_{F}\)]--[] c --[with arrow={0.1cm},segment length=0.5cm] d [blob]--[]e--[with arrow={0cm}]b[particle=\(\textbf{1}_{F}\)],
};}}
\equiv&
\vcenter{\hbox{\begin{tikzpicture}
\begin{feynman}
\vertex(a1){\(\textbf{1}_{F}\)};
\vertex[right=2.5cm of a1](a5){\(\textbf{1}_{F}\)};
\diagram* {
(a1)--[
] (a5),
};
\end{feynman}
\end{tikzpicture}}}
+
\vcenter{\hbox{\begin{tikzpicture}
\begin{feynman}
\vertex(a1){\(\textbf{1}_{F}\)};
\vertex[right=4cm of a1](a5){\(\textbf{1}_{F}\)};
\vertex[right=1.5cm of a1](a2);
\vertex[right=2.5cm of a1](a3);
\diagram* {
(a1)--[ 
] (a2) --[charged scalar, half left, edge label=\(\textbf{5}_{H}\)](a3),
(a2) --[fermion, half right, edge label'=\(\mathbf{\bar 5}_{F}\)](a3)--[](a5)
};
\end{feynman}
\end{tikzpicture}}}
\\
&+
\vcenter{\hbox{\feynmandiagram[horizontal=a to b]{
a[particle=\(\textbf{1}_{F}\)]--[] c --[with arrow={0.1cm},segment length=0.5cm] d [crossed dot]--[]e--[with arrow={0cm}]b[particle=\(\textbf{1}_{F}\)],
};}}
\overset{!}=
\text{UV finite}
\end{split}
\end{equation} 

\begin{equation}
\Rightarrow \delta Z_{\textbf{1}_F}= -\frac{1}{16\pi^2}\left(\frac{5}{2}\textbf{Y}_\nu \textbf{Y}_\nu^\dagger\right)\frac{1}{\epsilon}
\end{equation}

\item \textbf{Wave function of $\mathbf{\bar{5}}_{F}$}
\begin{equation}
\begin{split}
\vcenter{\hbox{\feynmandiagram[horizontal=a to b]{
a[particle=\(\mathbf{\bar{5}}_{F}\)]--[] c --[with arrow={0.1cm},segment length=0.5cm] d [blob]--[]e--[with arrow={0cm}]b[particle=\(\mathbf{\bar{5}}_{F}\)],
};}}
\equiv&
\vcenter{\hbox{\begin{tikzpicture}
\begin{feynman}
\vertex(a1){\(\mathbf{\bar{5}}_{F}\)};
\vertex[right=2.5cm of a1](a5){\(\mathbf{\bar{5}}_{F}\)};
\diagram* {
(a1)--[fermion 
] (a5),
};
\end{feynman}
\end{tikzpicture}}}
+
\vcenter{\hbox{\begin{tikzpicture}
\begin{feynman}
\vertex(a1){\(\mathbf{\bar{5}}_{F}\)};
\vertex[right=4cm of a1](a5){\(\mathbf{\bar{5}}_{F}\)};
\vertex[right=1.5cm of a1](a2);
\vertex[right=2.5cm of a1](a3);
\diagram* {
(a1)--[ 
] (a2) --[charged scalar, half left, edge label=\(\textbf{5}_{H}\)](a3),
(a2) --[fermion, half right, edge label'=\(\mathbf{1}_{F}\)](a3)--[](a5)
};
\end{feynman}
\end{tikzpicture}}}
\\
&+
\vcenter{\hbox{\begin{tikzpicture}
\begin{feynman}
\vertex(a1){\(\mathbf{\bar{5}}_{F}\)};
\vertex[right=4cm of a1](a5){\(\mathbf{\bar{5}}_{F}\)};
\vertex[right=1.5cm of a1](a2);
\vertex[right=2.5cm of a1](a3);
\diagram* {
(a1)--[fermion 
] (a2) --[boson, half left, edge label=\(\textbf{5}_{H}\)](a3),
(a2) --[fermion, half right, edge label'=\(\mathbf{\bar{5}}_{F}\)](a3)--[fermion](a5)
};
\end{feynman}
\end{tikzpicture}}}
\\
&+
\vcenter{\hbox{\feynmandiagram[horizontal=a to b]{
a[particle=\(\mathbf{\bar{5}}_{F}\)]--[] c --[with arrow={0.1cm},segment length=0.5cm] d [crossed dot]--[]e--[with arrow={0cm}]b[particle=\(\mathbf{\bar{5}}_{F}\)],
};}}
\overset{!}{=}
\text{UV finite}
\end{split}
\end{equation}

\begin{equation}
\Rightarrow \delta Z_{\mathbf{\bar{5}}_F}= -\frac{1}{16\pi^2}\left(\frac{1}{2}\textbf{Y}_\nu \textbf{Y}_\nu^\dagger + 6 \xi g^2\right)\frac{1}{\epsilon}
\end{equation}

\item \textbf{Wave function and mass of $\mathbf{5}_{H}$}
\begin{equation}
\begin{split}
\vcenter{\hbox{\feynmandiagram[horizontal=a to b]{
a[particle=\(\mathbf{5}_{H}\)]--[scalar] c --[scalar,with arrow={0.1cm},segment length=0.5cm] d [blob]--[scalar]e--[scalar,with arrow={0cm}]b[particle=\(\mathbf{5}_{H}\)],
};}}
\equiv&
\vcenter{\hbox{\begin{tikzpicture}
\begin{feynman}
\vertex(a1){\(\mathbf{5}_{H}\)};
\vertex[right=2.5cm of a1](a5){\(\mathbf{5}_{H}\)};
\diagram* {
(a1)--[charged scalar 
] (a5),
};
\end{feynman}
\end{tikzpicture}}}
+
\vcenter{\hbox{\begin{tikzpicture}
\begin{feynman}
\vertex(a1){\(\mathbf{5}_{H}\)};
\vertex[right=4cm of a1](a5){\(\mathbf{5}_{H}\)};
\vertex[right=1.5cm of a1](a2);
\vertex[right=2.5cm of a1](a3);
\diagram* {
(a1)--[charged scalar 
] (a2) --[half left, edge label=\(\textbf{1}_{F}\)](a3),
(a3) --[fermion,half left, edge label=\(\mathbf{\bar{5}}_{F}\)](a2),(a3)--[charged scalar](a5)
};
\end{feynman}
\end{tikzpicture}}}
\\
&+
\vcenter{\hbox{\begin{tikzpicture}
\begin{feynman}
\vertex(a1){\(\mathbf{5}_{H}\)};
\vertex[right=4cm of a1](a5){\(\mathbf{5}_{H}\)};
\vertex[right=1.5cm of a1](a2);
\vertex[right=2.5cm of a1](a3);
\diagram* {
(a1)--[charged scalar 
] (a2) --[boson, half left, edge label=\(X\)](a3),
(a2) --[charged scalar, half right, edge label'=\(\mathbf{5}_{H}\)](a3)--[charged scalar](a5)
};
\end{feynman}
\end{tikzpicture}}}
\\
&+
\vcenter{\hbox{\feynmandiagram[horizontal=a to b]{
a[particle=\(\mathbf{5}_{H}\)]--[scalar] c --[scalar,with arrow={0.1cm},segment length=0.5cm] d [crossed dot]--[scalar]e--[scalar,with arrow={0cm}]b[particle=\(\mathbf{5}_{H}\)],
};}}
\overset{!}{=}
\text{UV finite}
\end{split}
\end{equation}

\begin{equation}
\Rightarrow \delta Z_{\textbf{5}_H}= -\frac{1}{16\pi^2}\left(\text{Tr}(\textbf{Y}_\nu^\dagger\textbf{Y}_\nu)+(6\xi-18)g^2\right)\frac{1}{\epsilon}
\end{equation}

\item \textbf{Yukawa vertex}
\begin{equation}
\begin{split}
\vcenter{\hbox{\begin{tikzpicture}
\draw[pattern=north west lines] (1.5,-1.25) circle (10pt);
\begin{feynman}
\vertex(a1){\(\textbf{1}_{F}\)};
\vertex[below=2.5cm of a1] (a3){\(\mathbf{\bar{5}}_{F}\)};
\vertex at ($(a1)!0.5!(a3)+(1.5cm,0)$) (a2);
\vertex[right=1.5cm of a2](a4){\(\mathbf{5}_{H}\)};
\diagram* {
(a3)--[fermion] (a2)--[] (a1),
(a4)--[charged scalar] (a2)
};
\end{feynman}
\end{tikzpicture}}} 
\equiv&
\vcenter{\hbox{\begin{tikzpicture}
\begin{feynman}
\vertex(a1){\(\textbf{1}_{F}\)};
\vertex[below=2.5cm of a1] (a3){\(\mathbf{\bar{5}}_{F}\)};
\vertex at ($(a1)!0.5!(a3)+(1.5cm,0)$) (a2);
\vertex[right=1.5cm of a2](a4){\(\mathbf{5}_{H}\)};
\diagram* {
(a3)--[fermion] (a2)--[] (a1),
(a4)--[charged scalar] (a2)
};
\end{feynman}
\end{tikzpicture}}}
+
\vcenter{\hbox{\begin{tikzpicture}
\begin{feynman}
\vertex(a1){\(\textbf{1}_{F}\)};
\vertex[below=2.5cm of a1] (a3){\(\mathbf{\bar{5}}_{F}\)};
\vertex at ($(a1)!0.5!(a3)+(1.5cm,0)$) (a2);
\vertex[right=1.5cm of a2](a4){\(\mathbf{5}_{H}\)};
\vertex[right=0.4cm of a2](a5);
\vertex at ($(a2)!0.225!(a3)$) (a6);
\diagram* {
(a3)--[fermion] (a2)--[] (a1),
(a4)--[charged scalar] (a2),
(a6)--[boson, half right, edge label'=\(X\)] (a5),
};
\end{feynman}
\end{tikzpicture}}}
\\
&+
\vcenter{\hbox{\begin{tikzpicture}
\begin{feynman}
\vertex(a1){\(\textbf{1}_{F}\)};
\vertex[below=3cm of a1] (a3){\(\mathbf{\bar{5}}_{F}\)};
\vertex at ($(a1)!0.5!(a3)+(1.5cm,0)$) (a2);
\vertex[right=1.5cm of a2](a4){\(\mathbf{5}_{H}\)};
\vertex at ($(a2)!0.4!(a1)$) (a5);
\vertex at ($(a2)!0.4!(a3)$) (a6);
\diagram* {
(a3)--[fermion] (a6)--[edge label'=\(\mathbf{1}_F\)] (a2) --[fermion, edge label'=\(\mathbf{\bar{5}}_F\)] (a5) --[](a1),
(a4)--[charged scalar] (a2),
(a5)--[charged scalar, half right, edge label'=\(\textbf{5}_H\)] (a6),
};
\end{feynman}
\end{tikzpicture}}}
+
\vcenter{\hbox{\begin{tikzpicture}
\draw (1.5,-1.25) circle (4pt);
\draw (1.5,-1.25) node[cross]{};
\begin{feynman}
\vertex(a1){\(\textbf{1}_{F}\)};
\vertex[below=2.5cm of a1] (a3){\(\mathbf{\bar{5}}_{F}\)};
\vertex at ($(a1)!0.5!(a3)+(1.5cm,0)$) (a2);
\vertex[right=1.5cm of a2](a4){\(\mathbf{5}_{H}\)};
\diagram* {
(a3)--[fermion] (a2)--[] (a1),
(a4)--[charged scalar] (a2)
};
\end{feynman}
\end{tikzpicture}}}
\\
&\overset{!}{=}\text{UV finite}
\end{split}
\end{equation}

\begin{equation}
\Rightarrow \delta Z_{\textbf{Y}_\nu}= -\frac{1}{16\pi^2}6\xi g^2\frac{1}{\epsilon}
\end{equation}

\end{enumerate}

\subsection{RGEs of $\textbf{Y}_\nu$ and $g$ above the GUT scale at 1-loop}
\begin{figure}
\centering
\includegraphics[width=9cm]{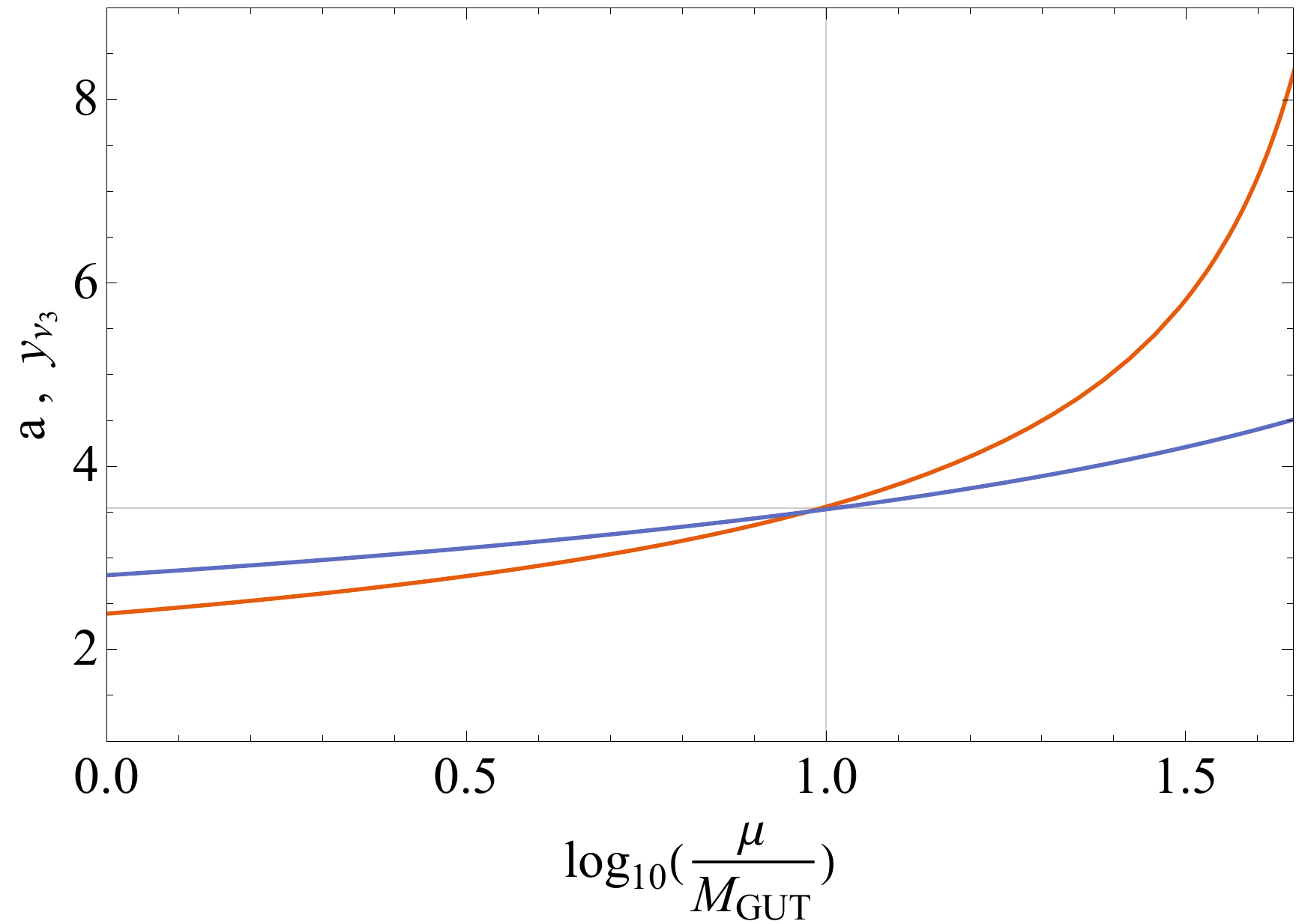}
\caption{RG evolution of the neutrino coupling $y_{\nu_3}$ (blue) for the case of a neutrino matrix dominated by its 33 element, respectively $a$ (orange) for the case of CSD2 or CSD3. For the GUT scale values $y_{\nu_3}=2.8$, $a=2.4$ and $g=0.6$ were chosen. The vertical grey line indicates where the neutrino couplings get larger than $\sqrt{4\pi}$.}\label{fig: running of a}
\end{figure}

We now calculate the RGE for the running of the Neutrino Yukawa matrix above the GUT scale. We first note that   
in our case the bare neutrino Yukawa matrix $\textbf{Y}_{\nu B}$ is related to the renormalized neutrino Yukawa coupling $\textbf{Y}_\nu$ by the following formula
\begin{equation}
\textbf{Y}_{\nu B}=Z_{\textbf{1}_F}^{-\frac{1}{2}}\textbf{Y}_\nu Z_{\textbf{Y}_\nu}\mu^{\frac{\epsilon}{2}}Z_{\mathbf{\bar{5}}_F}^{-\frac{1}{2}}Z_{\textbf{5}_H}^{-\frac{1}{2}}.
\end{equation}
Therefore, using the formulae in Eq.~(18) from \cite{Antusch:2001ck}, we find
\begin{equation}
\mu\frac{d \textbf{Y}_\nu}{d\mu}=\frac{1}{(4\pi)^2}\textbf{Y}_\nu\left\lbrace3\textbf{Y}_\nu^\dagger\textbf{Y}_\nu+\text{Tr}(\textbf{Y}_\nu^\dagger\textbf{Y}_\nu)-9g^2\right\rbrace.
\end{equation}
Turning to the RGE for $g$, with the Dynkin indices of the representations \textbf{1}, $\mathbf{\bar{5}}$, $\textbf{10}$, \textbf{24}, and \textbf{45} given by 0, $\frac{1}{2}$, $\frac{3}{2}$, $5$, and $12$, respectively, we obtain the 1-loop running of the unified gauge coupling. It is given by
\begin{equation}
\mu\frac{dg}{d\mu}=\frac{1}{(4\pi)^2}\frac{19}{2}g^3.
\end{equation}
Having determined approximations of the RGEs we can now, for given GUT scale values of $\textbf{Y}_\nu$ and $g$, estimate up to which scale the models are still perturbative. Since in the RGEs terms of loop order $n$ are suppressed by a factor of $\frac{1}{(4\pi)^{2n}}$ and each parameter $p$ enters with an exponent of maximal power $2n$, we define a theory to be perturbative as long as all parameters are smaller than $\sqrt{4\pi}$. Then loop terms of order $n$ are except of their prefactor suppressed by a factor of $(4\pi)^{-n}$. 
In our work, we use the requirement that our models should remain perturbative for at least one order of magnitude above the GUT scale. With the gauge coupling $g\approx 0.6$ and the neutrino couplings $a = {\cal O}(1)$, $b \ll 1$ we then find an upper bound for $a$ of 2.38, for the cases of CSD2 or CSD3 considered in Section~\ref{sec: nucleon decay fingerprints of two toy models}. On the other hand, for the case of a neutrino Yukawa matrix which is dominated by its 33 element $y_{\nu_3}$ (cf. Section~\ref{sec: A minimal non-SUSY}), we find an upper bound for $y_{\nu_3}$ of 2.81. Figure \ref{fig: running of a} shows the running of the neutrino coupling $a$ respectively $y_{\nu_3}$ above the GUT scale, if the GUT scale values of $a$ respectively $y_{\nu_3}$ are set to 2.4 respectively 2.8 and if the GUT scale value of $g$ is set to 0.6.

Because of the above reasoning we used for the estimated 2-loop RG evolution of the 1-$\sigma$ range of the ratios $y_\tau/y_b$ and $y_\mu/y_s$ the bound $y_{\nu_3}\leq 2.8$ (cf.\ Section~\ref{sec: Viability of the GUT scale Yukawa ratios}). Similarly, we used in the minimization of the $\chi^2$-function (cf.\ Section~\ref{sec: best-fit points}) the fixed value $a=2.4$ and for the MCMC analysis (cf.\ Section~\ref{sec: MCMC}) the range $0\leq a\leq 2.4$. We note that this is just one possible way to ensure perturbativity and not necessarily a strict constraint.

\begin{table}[p]
\centering
\renewcommand{\arraystretch}{1.5}
\begin{tabu}{|c|cc|cc|cc|cc|}\cline{1-9}
\multirow{ 3}{*}{}& \multicolumn{4}{|c|}{\textbf{Model 1}}&\multicolumn{4}{c|}{\textbf{Model 2}}
\\\cline{2-9}
& \multicolumn{2}{|c|}{$a\ll 1$}&\multicolumn{2}{c|}{$a = \mathcal{O}(1)$}&\multicolumn{2}{|c|}{$a\ll 1$}&\multicolumn{2}{c|}{$a = \mathcal{O}(1)$}
\\\cline{2-9} 
& $\phi_{102}$ & $\phi_{120}$ & $\phi_{102}$ & $\phi_{120}$ & $\phi_{113}$ & $\phi_{131}$ & $\phi_{113}$ & $\phi_{131}$ \\\hline
$g_{\text{GUT}}\;/\;10^{-1}$ &  5.64 & 5.64  & 5.81  & 5.80  & 5.66  & 5.66  & 5.87  & 5.94  \\
$\log_{10}(\MGUT\;/\;\GeV)$ & $15.2$ & $15.2$  & 16.0  & 16.1  & 15.2  & 15.2  & 16.1  &  16.2 \\
$\log_{10}(M_{\Phi_8}\;/\;\GeV)$ & $15.2$  & $15.2$  & 11.4  &  10.5 & 15.2  & 15.2  & 11.0  & 11.2  \\
$\log_{10}(M_{\Phi_6}\;/\;\GeV)$ &  8.21 & 8.22  & 8.38  &  8.42 & 7.96  & 7.96  & 7.82  & 7.85  \\
$\log_{10}(M_{\Phi_{(3,3)}}\;/\;\GeV)$ & 8.21  & 8.22  & 8.99  & 9.73  & 7.96  & 7.96  & 9.66  &  9.16 \\
$y_1^u \;/\;10^{-6}$ &  3.11 & 3.13  & 3.13  & 3.14  & 3.13  & 3.27  & 3.13  & 3.27  \\
$y_2^u\;/\; 10^{-3}$ & 1.53  & 1.53  & 1.56  & 1.54  & 1.51  & 1.59  & 1.56  & 1.56  \\
$y_3^u\;/\;10^{-1}$ & 4.68  & 4.68  & 4.73  & 4.71  &  4.65 & 4.90  & 4.81  & 4.86  \\
$y_{11}^d\;/\;10^{-6}$ & -  & -  & -  & -  & 6.19  & 6.19  & 6.61  & 6.90  \\
$y_{12}^d\; /\;10^{-5}$ & 3.50  & 3.51  & 3.01  & 3.11  & -  & -  & -  & -  \\
$y_{21}^d\; /\; 10^{-5}$ & 3.05  & 3.04  & 3.69  & 3.77  & -  & -  & -  & -  \\
$y_{22}^d \;/\; 10^{-4}$ & 1.28  & 1.28  & 1.29  & 1.31 & 1.29  & 1.29  & 1.31  & 1.32  \\
$y_{33}^d  \;/\; 10^{-3}$ & 6.43  & 6.43  & 6.49  & 6.58  & 6.41  & 6.41  & 6.51  & 6.52  \\
$\theta_{12}^{uL}\;/10^{-1}$ & 0.870  & 0.867  & 8.66  & 8.67  & 2.27  & 2.27  & 2.27  &  2.27 \\
$\theta_{13}^{uL}$ & -  & -  & -  & -  & 4.12  & 4.12  & 4.15  & 4.14  \\
$\theta_{23}^{uL}\;/\;10^{-2}$ & 4.76  & 4.76  & 4.78  & 4.78  & 4.7  & 4.77  & 4.78  &  4.79 \\
$\delta$ & -  & -  & -  & -  & 1.28  & 1.28  & 1.28  & 1.28  \\
$\varphi_{12}$ & 5.04  & 4.92  & 4.97  & 4.79  & -  & -  & -  & -  \\
$a$ & -  & -  & 2.4  & 2.4  & -  & -  & 2.4  & 2.4  \\
$\log_{10}(M_A\;/\;\GeV)$ & -  & -  & 15.5  & 15.5  & -  & -  & 15.5  & 15.4  \\
$m_a\;/\;10^{-15}$ & 2.39  & 2.30  & -  & -  & 2.62  & 2.63  & -  & -  \\
$\epsilon\;/\;10^{-1}$ & 1.22  & 1.32  & 2.36  & 2.86  & 1.01  & 1.00  & 2.01  & 2.10  \\
$\alpha$ & 1.54  & 5.02  & 1.60  & 5.58  & 2.09  & 4.18  & 2.06  & 4.29  \\
\hline 
\end{tabu}
\caption{The GUT scale input parameters of the best-fit points of the models 1 and 2.}\label{tab: best-fit points input parameters}
\end{table}

\begin{table}
\centering
\renewcommand{\arraystretch}{1.5}
\begin{tabu}{|c|cc|cc|cc|cc|}\cline{1-9}
\multirow{ 3}{*}{}& \multicolumn{4}{|c|}{\textbf{Model 1}}&\multicolumn{4}{c|}{\textbf{Model 2}}
\\\cline{2-9}
& \multicolumn{2}{|c|}{$a\ll 1$}&\multicolumn{2}{c|}{$a = \mathcal{O}(1)$}&\multicolumn{2}{|c|}{$a\ll 1$}&\multicolumn{2}{c|}{$a = \mathcal{O}(1)$}
\\\cline{2-9}
& $\phi_{102}$ & $\phi_{120}$ & $\phi_{102}$ & $\phi_{120}$ & $\phi_{113}$ & $\phi_{131}$ & $\phi_{113}$ & $\phi_{131}$ \\\hline

$\chi^2$ & 21.1 & 21.9 & 20.7 & 14.4 & 25.8 & 25.1 & 14.8 & 9.64 \\
$\chi_{g_1}^2$ & 1.13 & 1.15 & 1.95 & 1.85 & 0.60 & 0.60 & 1.42 & 1.56 \\
$\chi_{g_2}^2$ & 0.36 & 0.37 & - & - & 0.57 & 0.57 & - & - \\
$\chi_{g_3}^2$ & 3.25 & 3.22 & 0.84 & 0.56 & 2.22 & 2.22 & 0.60 & 0.13 \\
$\chi_{y_t}^2$ & 0.83 & 0.82 & - & - & 0.84 & 0.84 & - & - \\
$\chi_{y_d}^2$ & 0.91 & 0.92 & 0.29 & 0.11 & 2.20 & 2.20 & 0.95 & - \\
$\chi_{y_s}^2$ & 0.01 & 0.01 & 1.27 & 1.60 & 0.72 & 0.72 & 0.33 & 0.17 \\
$\chi_{y_b}^2$ & 5.45 & 5.44 & 2.81 & 1.21 & 5.63 & 5.63 & 2.12 & 1.23 \\
$\chi_{y_\tau}^2$ & 6.00 & 6.01 & 2.71 & 1.29 & 6.19 & 6.19 & 1.21 & 0.97 \\
$\chi_{\theta_{12}\PMNS}^2$ & - & 0.24 & 0.24 & 2.58 & 1.31 & 1.31 & 1.34 & 1.30 \\
$\chi_{\theta_{13}\PMNS}^2$ & - & - & 1.71 & 0.12 & - & - & - & - \\
$\chi_{\theta_{23}\PMNS}^2$ & 0.64 & - & 2.20 & 0.35 & 2.03 & 2.50 & 0.99 & 0.75 \\
$\chi_{\delta\PMNS}^2$ & 1.28 & 2.37 & 1.48 & 2.97 & 1.80 & 2.00 & 1.83 & 2.33 \\
$\chi_{\;\Gamma(p\rightarrow \pi^0e^+)}^2$ & 1.15 & 1.15 & - & - & 0.19 & 0.19 & - & - \\
$\chi_{\;\Gamma(p\rightarrow K^0\mu^+)}^2$ & - & - & - & - & 0.18 & 0.18 & - & - \\
$\chi_{\;\Gamma(n\rightarrow \pi^-e^+)}^2$ & - & - & - & - & 0.57 & 0.57 & - & - \\
\hline 
\end{tabu}
\caption{The total $\chi^2$ and the dominant pulls $\chi_i^2$ of the best-fit points of the models 1 and 2.}\label{tab: best-fit chi squared}
\end{table}

\end{document}